\documentclass[11pt]{article}
\pdfoutput=1
\usepackage[centertags]{amsmath}
\usepackage{amsthm}
\usepackage[square, comma, sort&compress,numbers]{natbib}
\usepackage{array,multirow}
\numberwithin{equation}{section}
\usepackage{amssymb,amsfonts}
\usepackage{graphicx}
\usepackage{color}
\usepackage{mathtools,bm}
\usepackage{accents}
\usepackage{physics}
\usepackage{authblk}
\usepackage[
colorlinks=true
,urlcolor=blue
,anchorcolor=blue
,citecolor=blue
,filecolor=blue
,linkcolor=blue
,menucolor=blue
,linktocpage=true
,pdfproducer=medialab
,pdfa=true
]{hyperref}

\usepackage[title,titletoc]{appendix}

\usepackage{epsfig}
\usepackage{bbold}

\usepackage{caption}
\usepackage{subcaption}
\usepackage{wrapfig}
\usepackage{float}
\usepackage{soul}
\usepackage{tkz-euclide}
\usepackage{braket}
\usepackage{tikz,pgf}
\usetikzlibrary{shapes}
\usetikzlibrary{calc}
\usetikzlibrary{decorations.pathmorphing}
\usetikzlibrary{decorations.pathreplacing,shapes.misc}
\usetikzlibrary{positioning}
\usetikzlibrary{arrows}
\usetikzlibrary{decorations.markings}
\usetikzlibrary{shadings}

\usetikzlibrary{intersections}

\newtheorem{prop}{Proposition}[section]
\newtheorem{lemma}{Lemma}[section]

\newcommand{\Ker}{\mathop{\mathrm{Ker}}}

\def\cD{\mathcal{D}}
\def\cE{\mathcal{E}}

\def\cG{\mathcal{G}}
\def\cH{\mathcal{H}}

\def\cK{\mathcal{K}}

\def\cM{\mathcal{M}}
\def\cN{\mathcal{N}}
\def\cO{\mathcal{O}}
\def\cP{\mathcal{P}}

\def\cR{\mathcal{R}}
\def\cS{\mathcal{S}}

\numberwithin{equation}{section} \makeatletter
\@addtoreset{equation}{section}

\newcommand{\ga}{\alpha}
\newcommand{\gad}{{\dot{\alpha}}}
\newcommand{\gb}{\beta}
\newcommand{\gbd}{{\dot{\beta}}}
\renewcommand\gg{\gamma}
\newcommand{\ggd}{{\dot{\gamma}}}
\newcommand{\gd}{\delta}

\newcommand{\Ad}{{\bar{A}}}

\newcommand{\Cd}{{\bar{C}}}

\newcommand{\Hd}{{\bar{H}}}

\newcommand{\Md}{{\bar{M}}}

\newcommand{\ad}{{\bar{a}}}
\newcommand{\bd}{{\bar{b}}}
\newcommand{\cd}{{\bar{c}}}

\newcommand{\kd}{{\bar{k}}}
\newcommand{\md}{{\bar{m}}}

\newcommand{\pd}{{\bar{p}}}

\newcommand{\ud}{{\bar{u}}}
\newcommand{\vd}{{\bar{v}}}

\newcommand{\yd}{{\bar{y}}}

\newcommand{\lp}{\left(}
\newcommand{\rp}{\right)}

\newcommand{\floor}[1]{\left\lfloor #1 \right\rfloor}
\newcommand{\ceil}[1]{\left\lceil #1 \right\rceil}

\makeatletter
\newcommand{\pushright}[1]{\ifmeasuring@#1\else\omit\hfill$\displaystyle#1$\fi\ignorespaces}
\newcommand{\pushleft}[1]{\ifmeasuring@#1\else\omit$\displaystyle#1$\hfill\fi\ignorespaces}
\makeatother

\usepackage{jheppub}

\makeatletter
\def\@fpheader{\vspace{-.1cm}}
\makeatother

\title{\centering{Bilinear higher-spin currents in the unfolded formalism}}

\author{Yu. A. Tatarenko}

\affiliation{
I.E. Tamm Department of Theoretical Physics, \\
P.N. Lebedev Physical Institute, Leninsky ave. 53, 119991 Moscow, Russia}
\affiliation{
Moscow Institute of Physics and Technology, \\
Institutsky lane 9, 141700, Dolgoprudny, Moscow region, Russia}

\emailAdd{tatarenko.iua@phystech.edu}

\abstract{The classification of primary non-trivial bilinear currents in the 4d higher-spin theory is obtained. It is interpreted in terms of the known classification of cubic Lagrangian vertices in the 4d higher-spin theory. It is shown that some currents join to the action with the tail of higher-derivative terms (as it happens with the Fradkin-Vasiliev $2-s-s$ vertex). The analysis is based on the $\sigma_-$-cohomology technique of the unfolded formalism.}

\begin{document}
\allowdisplaybreaks

\maketitle
\flushbottom
\newpage

\section{Introduction}\label{sec:introduction}

The equations of motion and the action of free massless higher-spin (HS) fields were found by Fronsdal \cite{Fronsdal:1978rb}. Originally, he proposed to construct an interacting HS theory starting from the free action and obtaining interaction vertices order-by-order via the Noether procedure, but however, this program was not realized. Nevertheless, many results in studying of HS interactions were achieved by other methods. For instance, the so-called \emph{unfolded formalism} allowed Vasiliev to construct the system of the nonlinear HS equations on $AdS_4$ (and next on $AdS_d$) background \cite{Vasiliev:1990en, Vasiliev:1992av, Vasiliev:2003ev}. The most of other approaches are focused on the action. So, Fradkin and Vasiliev found an example of action consistent at the cubic order \cite{Fradkin:1987ks}. In works of A.~Bengtsson, I.~Bengtsson, Brink, Linden and Metsaev \cite{Bengtsson:1983pd, Bengtsson:1986kh, Metsaev:2005ar,Metsaev:2007rn, Metsaev:2018xip} the full classification of cubic vertices in HS theory on $d$-dimensional Minkowski space and on $AdS_4$ was obtained in the \emph{lightcone gauge}. The covariant analogue of this classification for bosons on flat background was constructed by Manvelyan, Mkrtchyan and Ruhl \cite{Manvelyan:2010jr} and for bosons in $AdS$ (in TT-gauge) by Joung and Taronna \cite{Joung:2011ww}. See also \cite{Vasiliev:2011knf} where the cubic vertices in the $AdS_d$ HS theory were constructed in the unfolded formalism.

In the 2nd perturbation order, currents that are present in the RHS of dynamical equations are connected with the cubic part of the action as follows:
\begin{equation}\label{gen_J_def}
    J := \fdv{S^3}{\phi}\,.
\end{equation}
The currents have to obey the conservation law, which we write schematically as
\begin{equation}\label{gen_cons_law}
    \partial J \approx 0\,,
\end{equation}
where $\partial$ stands for a differential operator (divergence in the simplest case), $\approx$ denotes weak (on-shell) equality. Obviously, $J$ in \eqref{gen_J_def} can be derivative of another current $\Tilde J$, and so the conservation law for $J$ can be a differential consequence of the conservation law for $\Tilde J$. Conserved currents that are not derivatives of other currents are called \emph{primary}. Another special case is when the weak equality in \eqref{gen_cons_law} becomes the strong equality, i.e., when the conservation law is satisfied identically. Such currents in \cite{Spirin:2024zgy} were called \emph{trivial} in certain sense. Note that the trivial currents not always can be removed by a local field redefinition. For example, the currents produced by the so-called \emph{Born-Infeld-type} vertices made of HS Weyl tensors are trivial in these terms.

From the vertex perspective, the notion of triviality of currents used here is close to the \emph{deformational triviality} notion: deformationally trivial (or also \emph{Abelian} \cite{Vasiliev:2011knf}) vertices are vertices not deforming the gauge transformation law, or, equivalently, off-shell gauge invariant vertices. Since the conservation law \eqref{gen_cons_law} can be derived from the Noether identity, deformationally trivial vertices lead to trivial currents. The deformationally trivial vertices are of special importance beyond the 3rd perturbation order, since, according to \cite{Joung:2019wbl, Fredenhagen:2019hvb, Fredenhagen:2019lsz}, in the $d$-dimensional flat or $AdS$ HS theory all the \emph{on-shell} gauge invariant vertices are in fact \emph{off-shell} gauge invariant, i.e., deformationally trivial.

In the recent paper \cite{Tatarenko:2024csa} we have calculated the bilinear HS currents in the 4d Vasiliev theory and have discussed their relation to the known cubic vertices, with the conclusion that the currents are in one-to-one correspondence with the vertices. However there are several questions remain, namely, which of these currents are primary and which are not, which are trivial and which are not, and whether the currents not corresponding to the vertices exist?

To answer these questions, in this paper we present the full classification of the non-trivial primary bilinear conserved HS currents in the 4d HS theory (in what follows words ``primary" and ``non-trivial" will be implicit). This classification is obtained by methods of the unfolded formalism, so it inherits the features of this formalism: covariance, coordinate and gauge invariance and independence of local field redefinitions (i.e., the currents in our classification cannot be removed by any gauge transformation or local field redefinition). But our approach does not allow to construct currents explicitly, in the form $J = \sum a_{mn} (\partial^m\phi) (\partial^n \phi)$. We parameterize the currents by three spins (spin of the current and spins of two fields inside it) and the number of derivatives. As is shown below, and as was expected in view of vertex classifications \cite{Bengtsson:1983pd, Bengtsson:1986kh, Metsaev:2005ar,Metsaev:2007rn,Manvelyan:2010jr,Joung:2011ww,Metsaev:2018xip}, these values determine the current uniquely (up to complex conjugation).

The rest of the paper is organized as follows. The general concept of the approach is described in Section \ref{sec:general_idea}, where also some aspects of the unfolded formalism are recalled. In Section \ref{sec:definitions} we specify the mathematical objects generally defined in Section \ref{sec:general_idea}, for the case of 4d HS theory. Section \ref{sec:calculations} contains the description of calculation method and technical details; some details and intermediate expressions are put into Appendices \ref{app:calculation_details} and \ref{app:on_rank-two_dynamical_equations}. The results are discussed in Section \ref{sec:discussion}. In Section \ref{sec:conclusion} we summarize our results and draw a conclusion. The useful technical formulae are collected in Appendix \ref{app:technical_formulae}. The reader not interested in technical details can skip Section \ref{sec:calculations} and move directly to Section \ref{sec:discussion}.

\section{General idea}\label{sec:general_idea}

In this section we explain the general aspects of the method we use and recall the underlying basics of the unfolded formalism (see \cite{Vasiliev:2007yc} for a detailed introduction).

Consider a dynamical system with dynamical fields $\omega^I$ taking values in $\Lambda^\bullet \otimes V$, where $\Lambda^\bullet$ is an exterior algebra on a manifold $\cM$ and $V$ is a $\mathbb{Z}_+$-graded module of some (Lie or associative) algebra $A$ with the grading operator $\cG^I{}_J$. Let the dynamical equations of this system have the form
\begin{equation}\label{gen_eqn}
    \cD^I{}_J \omega^J = 0\,,
\end{equation}
where $\cD$ is a nilpotent differential operator associated with background connection 1-forms $\Omega^I$:
\begin{equation}\label{gen_D}
    \cD := \dd + (\text{polylinear in $\Omega^I$ terms})\,,\qquad
    \cD^2 = 0\,.
\end{equation}
(Here $\dd:=dx^\mu \pdv{}{x^\mu}$ is de Rham differential.)
Equations of the form \eqref{gen_eqn} are called free \emph{unfolded equations}. System \eqref{gen_eqn} is invariant under gauge transformations
\begin{equation}\label{gen_gauge}
    \delta_\epsilon \omega^I := \cD^I{}_J \epsilon^J\,.
\end{equation}
Note that equations \eqref{gen_eqn} may content so-called \emph{constraints}, i.e., equations that express some components of $\omega^I$ via derivatives of others. Those fields in $\omega^I$ which are fixed by constraints are called \emph{auxiliary}.

Let $\cD$ have the following structure
\begin{equation}\label{gen_D_struct}
    \cD = D_0 + \sigma_- + \sigma_+\,,
\end{equation}
where all the space-time derivatives are hidden in $D_0$ and at the same time $D_0$ preserves the grading, $\sigma_-$ decreases it by 2\footnote{This is more convenient choice when working with both bosons and fermions.} and $\sigma_+$ is a sum of operators of non-negative grading. From the nilpotency of $\cD$ it follows that $\sigma_-^2 = 0$, so one can define cohomology groups of $\sigma_-$ as usual $H^p(\sigma_-):=(\Ker \sigma_-/ \Im \sigma_-) \cap (\Lambda^p \otimes V)$. Due to this fact the following lemma takes place (for the proof see \cite{Shaynkman:2000ts}).
\begin{lemma}\label{lemma:gen_sigma-}
    For a $p$-form part of $\omega^I$:
    \begin{itemize}
        \item differential gauge symmetries are in $H^{p-1}(\sigma_-)$;
        \item dynamical (i.e., not auxiliary) fields are in $H^p(\sigma_-)$;
        \item differential dynamical equations (i.e., not constraints) are in $H^{p+1}(\sigma_-)$.
    \end{itemize}
\end{lemma}
This lemma can be adapted to the case with $\cD$ containing terms with higher degree of $\Omega^I$, i.e., when equation \eqref{gen_eqn} mixes $p$-forms and $q$-forms in $\omega^I$ with different $p$ and $q$. In this case, when the form degree of $\omega^I$ is uncertain, the subspaces of $H(\sigma_-) := \Ker \sigma_-/ \Im \sigma_-$ that correspond to differential gauge parameters, dynamical fields and dynamical equations will be called spaces of \emph{gauge-like}, \emph{field-like} and \emph{equation-like} $\sigma_-$-cohomology, respectively.

Each dynamical field comes into the unfolded equations \eqref{gen_eqn} with the chain of \emph{descendants}, i.e., auxiliary fields that are expressed via derivatives of the given dynamical field. By construction, if the degree of the dynamical field is $G$, its descendants have degrees $G+2,\,G+4,\,\dots$ -- the descendant with degree $G+2k$ has at most $k$ derivatives of the dynamical field.

One can make system \eqref{gen_eqn} self-interacting by inserting in its RHS $\Upsilon^I$ being formal power series in background and dynamical fields, i.e., in $\Omega^I$ and $\omega^I$. To preserve consistency of the equations one has to set
\begin{equation}\label{gen_Bianchi_id}
    \cD^I{}_J \Upsilon ^J = 0\,.
\end{equation}
As discussed in \cite{Gelfond:2003vh}, this condition is equivalent to the conservation law for the current in the dynamical equations. The role of the conserved current is played by the part of $\Upsilon$ that belongs to equation-like $H(\sigma_-)$, according to Lemma \ref{lemma:gen_sigma-}.

Let us consider the lowest interacting order, i.e., the case with $\Upsilon$ bilinear in the dynamical fields. Then equality \eqref{gen_Bianchi_id} has to be a consequence of that dynamical fields inside $\Upsilon$ obey the free equations of motion \eqref{gen_eqn}. This fact can be reinterpreted in terms of the \emph{rank-two fields} \cite{Gelfond:2003vh}. The rank-two fields $J^{IJ}$ are the fields on $\cM$ taking values in $\Lambda^\bullet \otimes V \otimes V$ and obeying \emph{rank-two equations}
\begin{equation}\label{gen_rk2eqn}
    (\cD^{(2)})^{IJ}{}_{KL} J^{KL} = 0\,,
\end{equation}
where the operator $\cD^{(2)}$ is constructed from $\cD$ such that $J^{KL} = \omega^K\omega^L$ (the wedge symbols are omitted in this paper) satisfies these equations.\footnote{Strictly speaking, $(\cD^{(2)})^{IJ}{}_{KL} := \dd\, \delta^I{}_K \delta^J{}_L + (\cD-\dd)^I{}_K \delta^J{}_L + (\cD-\dd)^J{}_L \delta^I{}_K$.} In these terms, $\omega^I$ are \emph{rank-one fields} and equations \eqref{gen_eqn} are \emph{rank-one equations}.

In the 4d HS theory, $\cD^{(2)}$ has the structure similar to \eqref{gen_D_struct} (if one extends the grading operator $\cG$ to the space $V \otimes V$ using Leibniz rule):
\begin{equation}\label{gen_D2_struct}
    \cD^{(2)} = D^{(2)}_0 + \sigma^{(2)}_- + \sigma^{(2)}_+\,.
\end{equation}
Therefore, one can apply Lemma \ref{lemma:gen_sigma-} to this case and conclude that dynamical rank-two fields belong to $\sigma^{(2)}_-$-cohomology. Since $\Upsilon$ is constructed of the the tensor product of two rank-one fields, which is particular rank-two field realization, the equation \eqref{gen_Bianchi_id} is satisfied by virtue of \eqref{gen_rk2eqn}. Thus, as \eqref{gen_Bianchi_id} expresses the conservation law, the conserved bilinear currents are described by the dynamical rank-two fields or, equivalently, by some elements of $\sigma^{(2)}_-$-cohomology.

It is worth emphasizing that the conserved currents are not stated to be in one-to-one correspondence with $H(\sigma^{(2)}_-)$ (and it is not so in the case of the 4d HS theory considered below). In particular, the relevant elements of $H(\sigma^{(2)}_-)$ must obey the first order differential equations, because the conservation laws are such. This can be controlled by the grading: as in the unfolded equations \eqref{gen_rk2eqn} only term with the space-time derivatives (associated with $D^{(2)}_0$, see \eqref{gen_D2_struct}) preserves grading, the dynamical equations are of the first order iff the $\sigma^{(2)}_-$-cocycles corresponding to the dynamical fields and their dynamical equations have the same degree. Of course, not every first-order differential equation can be interpreted as a conservation law, but, as we will see, in the case of the 4d HS theory such peculiar rank-two fields do not appear.

Concluding, the basic idea of the bilinear HS currents classification method used in this paper is as follows. At first, we find the cohomology groups of $\sigma^{(2)}_-$, and then select those field-like cocycles that contribute via \eqref{gen_rk2eqn} to the equation-like cocycles of the same degree. Such selected cocycles obey differential equations matching the conservation laws thus being related to bilinear HS currents.

\section{Definitions}\label{sec:definitions}

\paragraph{Spinor conventions}

In the 4d HS theory we deal with the two-component (Weyl) spinors and use the following index conventions:
\begin{align}
    &u_\ga = u^\gb \epsilon_{\gb\ga}\,,\ u^\ga = \epsilon^{\ga\gb}u_\gb\,,
        &&\ud_\gad = \ud^\gbd \epsilon_{\gbd\gad}\,,\ \ud^\gad = \epsilon^{\gad\gbd}u_\gbd\,;\\
    &\epsilon_{\gb}{}^\ga = \delta^\ga_\gb\,,
        &&\epsilon_{\gbd}{}^\gad = \delta^\gad_\gbd\,;
\end{align}
and a short-hand notation for spinor contractions:
\begin{align}
    &(ab) := a_\ga b^\ga\,,
        &&(\bar{a}\bar{b}) := \bar{a}_\gad \bar{b}^\gad\,.
\end{align}
Also we introduce operators of differentiation with respect to spinor variables $y^\ga$, $(y_n)^\ga$ and $\yd^\gad$, $(\yd_n)^\gad$ ($n = 1,\,2,\,3,\,\dots$ is an additional index) acting on the whole expression on the right:
\begin{align}\label{p_def}
    &p_\ga := -i\pdv{}{y^\ga}\,,
        &&\pd_\gad := -i\pdv{}{\yd^\gad}\,;\\
    &(p_n)_\ga := -i\pdv{}{(y_n)^\ga}\,,
        &&(\pd_n)_\gad := -i\pdv{}{(\yd_n)^\gad}\,.
\end{align}
With the help of \eqref{p_def} we construct the operators counting spinor variables' degree:
\begin{align}
    &\cN: = -i(yp)\,,
        &&\Bar{\cN}: = -i(\yd\pd)\,;\\
    &\cN_n: = -i(y_np_n)\,,
        &&\Bar{\cN}_n: = -i(\yd_n\pd_n)\,.
\end{align}
\paragraph{HS algebra}

In the 4d HS theory, the dynamical fields take values in the so-called \emph{HS algebra} (see review \cite{Vasiliev:1999ba} and references therein) generated by a pair of commuting spinor variables $Y^A=(y^\ga, \yd^\gad)$ and \emph{Klein operators} $K=(k,\kd)$ such that:
\begin{subequations}\label{Klein}
\begin{align}
    &\{k,y^\ga\} = 0 = \{\kd,\yd^\gad\}\,,
        &&[k,\yd^\gad] = 0 = [\kd,y^\ga]\,,\\
    &[k, \kd] = 0\,,
        &&k k = \kd \bar k  = 1\,.
\end{align}
\end{subequations}

Elements of HS algebra are formal power series in its generators:
\begin{align}\label{f_m,n}
    f(Y; K) := \sum_{m,n=0}^{\infty}\sum_{i,j=0,1} \frac{1}{m! n!}
        f^{ij}_{\ga(m)\ \gad(n)} k^i \kd^j y^{\ga(m)}\yd^{\gad(n)}
        \equiv \sum_{m,n=0} f_{m,n}(Y; K)\,,
\end{align}
where $\alpha(n)$ denotes $n$ symmetrized indices ($\alpha(n) := \ga_1\ga_2\dots\ga_n$ and $y^{\ga(n)} := y^{\ga_1}\dots y^{\ga_n}$) and $f_{m,n}(Y; K)$ is a homogeneous polynomial of degrees $m$ and $n$ in $y$ and $\yd$, respectively.

The HS algebra is equipped with the \emph{star product} defined as
\begin{equation}\label{star_product}
    f(Y; K) \star g(Y; K)
        := \int dU dV f(Y + U; K)e^{i(uv) + i(\ud\vd) }g(Y + V; K)\,.
\end{equation}
Here the integration measure is normalised so that $1\star 1=1$.

\paragraph{Background connection}

The background connection 1-form is the $AdS_4$-connection which is expressed via the vierbein $h^{\ga\gad}$ and the Lorentz connection $\varpi_{\ga \gb},\,\bar{\varpi}_{\gad \gbd}$ as follows
\begin{equation}\label{Omega_def}
    \Omega(Y) := 
        -\frac i4 (
            \varpi_{\ga \gb} y^\ga y^\gb + 
            2h_{\ga \gad} y^\ga \yd^\gad + 
            \bar{\varpi}_{\gad \gbd} \yd^\gad \yd^\gbd
        )\,.
\end{equation}
The background covariant derivative is
\begin{equation}\label{D}
    D_\Omega := \dd + [\Omega,\bullet]_\star\,,\qquad D_\Omega^2 = 0 \Leftrightarrow \dd \Omega + \Omega \star \Omega = 0\,.
\end{equation}

We introduce basic 2- and 3-forms constructed from the vierbein:
\begin{gather}\label{basic_frame_forms}
    H_{\ga\gb} := h_{\ga \ggd} h_\gb{}^\ggd\,,\qquad
        \Hd_{\gad\gbd} := h_{\gg \gad} h^\gg{}_\gbd\,;\\
    \mathcal{H}_{\ga \gad} := h_{\ga\gbd}H^\gbd{}_\gad =- h_{\gb\gad}H^\gb{}_\ga\,.
\end{gather}
In the sequel we will mostly use the following index-free notation
\begin{gather}\label{indexless_notation}
    \varpi(u,v) := \frac12 \varpi_{\ga\gb} u^\ga v^\gb\,,\qquad
        \bar{\varpi}(\ud,\vd) := \frac12 \bar{\varpi}_{\gad\gbd} \ud^\gad \vd^\gbd\,;\\
    h(u,\ud) := h_{\ga\gad} u^\ga \ud^\gad\,,\\
        H(u,v) := \frac 12 H_{\ga\gb}\, u^\ga v^\gb\,,\qquad
            \Hd(\ud,\vd) := \frac 12 \Hd_{\gad\gbd}\, \ud^\gad \vd^\gbd\,,\\
                \cH(u,\ud) := \cH_{\ga\gad} u^\ga \ud^\gad\,.
\end{gather}

\paragraph{Rank-one fields and equations}

The dynamical fields in the 4d HS theory are 1-forms $\omega(Y; K)$ and 0-forms $C(Y; K)$. They depend on the Klein operators differently:
\begin{align}\label{wC_dyn_cond}
    &\omega(Y; -K) = \omega(Y; K)\,,
    &&C(Y; -K) = -C(Y; K)\,.
\end{align}
As a result $D_\Omega$ acts on these fields in different ways:
\begin{align}
    D_\Omega \omega(Y; K) &= \big( D_L + ih(y,\pd) + ih(p,\yd) \big)\omega(Y; K)\,,\label{D^adj}\\
    D_\Omega C(Y; K) &= \big( D_L - ih(y,\yd) - ih(p,\pd) \big) C(Y; K)\,,\label{D^tw}
\end{align}
where the Lorentz covariant derivative is
\begin{equation}\label{D_L_def}
    D_L := \dd + 2i\varpi(y,p) + 2i\bar{\varpi}(\yd,\pd)\,.
\end{equation}
In the sequel the following definition of the Lorentz derivative components $D(a,\ad)$ will be useful\footnote{See also Appendix \ref{app:technical_formulae} for its commutation relations.}:
\begin{equation}\label{D_expansion}
    D_L f(Y) \equiv h^{\ga\gad}D_{\ga\gad} f(Y) \equiv -h(p_1,\pd_1) D(y_1,\yd_1) f(Y)\,.
\end{equation}

However, $D_\Omega$ is not a direct analogue of the $\cD$ from \eqref{gen_eqn} since equations of motion of $\omega$ and $C$ contain a bilinear in $\Omega$ term gluing $\omega$ and $C$:
\begin{align}
    D_\Omega \omega(Y; K) &=\label{omega_eq_free}
        \Upsilon(\Omega, \Omega, C)\,,\\
    D_\Omega C(Y; K) &=\label{C_eq_free}0\,,
\end{align}
where according to the \emph{First on-mass shell theorem} \cite{Vasiliev:1986td,Vasiliev:1988sa}
\begin{align}
    \Upsilon(\Omega, \Omega, C) =
        -\frac i2 \bar{\eta} H(p,p) C(y, 0; K) \kd - \frac i2 \eta \Hd(\pd,\pd) C(0, \yd; K) k\,.\label{WWC_res}
\end{align}
Here $\eta$ is an arbitrary complex number.

Equations \eqref{omega_eq_free}, \eqref{C_eq_free} can be split into independent subsystems with fixed spin corresponding to the eigenvalues of the \emph{spin operator}
\begin{align}\label{S_def}
    &\cS\, \omega(Y; K) := \frac 12 (\cN + \Bar{\cN} + 2)\, \omega(Y; K)\,,
    &&\cS\, C(Y; K) := \frac 12 |\cN - \Bar{\cN}|\, C(Y; K)\,,
\end{align}

The rank-one equations \eqref{omega_eq_free}, \eqref{C_eq_free} can be represented in a way analogous to \eqref{gen_eqn}, \eqref{gen_D} if one chooses the grading operator $\cG$ as
\begin{align}\label{G_def}
    &\cG \,\omega(Y; K) := |\cN - \Bar{\cN}|\, \omega(Y; K)\,,
    &&\cG\, C(Y; K) :=  (\cN + \Bar{\cN})\, C(Y; K)\,,
\end{align}
and introduces the combined rank-one field (here and in what follows it is assumed that $\theta(0) = 0$)
\begin{equation}\label{w_def}
    w(Y; K) := \omega(Y; K) + \theta(\cN - \bar{\cN} + 1) C(Y; K)\kd + \theta(\bar{\cN} - \cN + 1) C(Y; K)k\,.
\end{equation}
Then, the rank-one equations take the form
\begin{align}\label{rk1eq}
    (D_L + \Sigma^{(1)}_- + \Sigma^{(1)}_+) w(Y; K) =0
\end{align}
with
\begin{align}
\begin{split}\label{Sigma1_-}
    \Sigma^{(1)}_- :=
        &\big[
            ih(p,\yd)\theta(\cN-\bar{\cN}-1) +
            ih(y,\pd)\theta(\bar{\cN}-\cN-1)
        \big]
        \theta(2\cS - \cG)
        +\\+
        &\big[
            - ih(p,\pd) +
            \frac i2 \bar{\eta} H(p,p) \delta_{\bar{\cN},0} +
            \frac i2 \eta \Hd(\pd,\pd) \delta_{\cN,0}
        \big]
        \theta(\cG - 2\cS + 1)\,,
\end{split}\\
\begin{split}\label{Sigma1_+}
    \Sigma^{(1)}_+ :=
        &\big[
            ih(p,\yd) \theta(\bar{\cN}-\cN+2) +
            ih(y,\pd) \theta(\cN-\bar{\cN}+2)
        \big]
        \theta(2\cS - \cG)
        -\\-
        &ih(y,\yd) \theta(\cG - 2\cS + 1)\,.
\end{split}
\end{align}
In the equation \eqref{rk1eq} $D_L$ plays the role of $D_0$ from \eqref{gen_D_struct}, while $\Sigma^{(1)}_\pm$ correspond to $\sigma_\pm$.

It is worth emphasizing the difference between the $AdS$ and Minkowski setups. In formulae above it is assumed that the background space is $AdS_4$ with $\Lambda = -1$. If one restores $\Lambda$ in \eqref{rk1eq}, it appears as a factor in front of $\Sigma^{(1)}_+$. Hence, in flat limit the $\Sigma^{(1)}_+$-term vanishes. But it obviously does not affect $\Sigma^{(1)}_-$ and its rank-two counterpart, and therefore our analysis covers simultaneously $AdS$ and Minkowski cases.

In the sequel the following definition is used. Consider the rank-one field, defined by the formula \eqref{w_def}, with some spin \eqref{S_def} and degree \eqref{G_def} values, denoted as $s$ and $G$, respectively. Then its differential form degree is 1 if $G < 2s$, and 0 otherwise. The notion of a rank-one field can be simply generalized by stating that the form degree is $q$ if $G < 2s$, and $q-1$ otherwise. Operator $\Sigma^{(1)}_-$ is assumed to be the same for all $q$. We will denote by $H^{q-(q-1)}(\Sigma^{(1)}_-)$ the subspace of $H(\Sigma^{(1)}_-)$ which elements have form degree $q\geqslant 1$ if $G < 2s$, and $q-1$ otherwise. If $q=0$, the corresponding rank-one fields with $G \geqslant 2s$ do not exist, so in this case we will write $H^{0-\text{x}}(\Sigma^{(1)}_-)$. In these terms, the gauge-like, the field-like and the equation-like cohomology are $H^{0-\text{x}}(\Sigma^{(1)}_-)$, $H^{1-0}(\Sigma^{(1)}_-)$ and $H^{2-1}(\Sigma^{(1)}_-)$, correspondingly.

\paragraph{Rank-two fields and equations}

Rank-two fields \cite{Gelfond:2003vh} generalize the bilineals in the rank-one fields:
\begin{equation}\label{rk2_types}
    J(Y_1; Y_2) :=
        J^{CC}(Y_1; Y_2) + J^{\omega C}(Y_1; Y_2) + J^{C\omega}(Y_1; Y_2) + J^{\omega\omega}(Y_1; Y_2)\,.
\end{equation}
Here $J^{CC}$ is a 0-form, $J^{\omega C}$ and $J^{C\omega}$ are 1-forms, and $J^{\omega\omega}$ is 2-form; Klein operators are implicit. It is convenient to represent $J(Y_1; Y_2)$ as a sum of irreducible components
\begin{equation}\label{irr_comp_def}
    J(Y_1; Y_2) =
    \sum_{a,\ad,b,\bd,c,\cd} J_{a,\ad;\;b,\bd;\;c,\cd}(Y_1; Y_2)\,,
\end{equation}
where it is used an extension of the definition \eqref{f_m,n}:
\begin{gather}\label{f_abc}
    \mathbf{f}_{a,\ad;\;b,\bd;\;c,\cd}(Y_1; Y_2) :=
    \cK_{a,\ad;\;b,\bd;\;c,\cd}\ f_{a+b,\ad+\bd}(Y)\,,\\
    \cK_{a,\ad;\;b,\bd;\;c,\cd}:=
    (py_1)^a(py_2)^b(y_1y_2)^c(\pd\yd_1)^\ad(\pd\yd_2)^\bd(\yd_1\yd_2)^\cd\,.\label{K_abc}
\end{gather}

The rank-two grading operator is
\begin{equation}\label{G2_def}
    \cG = \cG_1 + \cG_2\,,
\end{equation}
\begin{align}\label{G12_def}
    &\cG_1 J^{XY} :=
    \begin{cases}
        |\cN_1 - \bar{\cN}_1|\,, &\text{if }X=\omega\,,\\
        \cN_1 + \bar{\cN}_1\,, &\text{if }X=C\,;
    \end{cases}
    &&\cG_2 J^{XY} :=
    \begin{cases}
        |\cN_2 - \bar{\cN}_2|\,, &\text{if }Y=\omega\,,\\
        \cN_2 + \bar{\cN}_2\,, &\text{if }Y=C\,.
    \end{cases}
\end{align}

The rank-two equations\footnote{Note that, strictly speaking, $J^{\omega C}(Y_1; Y_2)$ corresponds to $\omega(Y_1)(\theta(\cN_2 - \bar{\cN}_2 + 1) C(Y_2)\kd + \theta(\bar{\cN}_2 - \cN_2 + 1) C(Y_2)k)$ not just to $\omega(Y_1)C(Y_2)$. The similar holds for other components of $J$. This is why the RHS of the equations does not contain explicit Klein operators.}:
\begin{align}
    &(D_L - ih(y_1,\yd_1) - ih(y_2,\yd_2) - ih(p_1,\pd_1) - ih(p_2,\pd_2))J^{CC}(Y_1; Y_2) = 0\,;\\
    &(D_L + ih(y_1,\pd_1) - ih(y_2,\yd_2) + ih(p_1,\yd_1) - ih(p_2,\pd_2))J^{\omega C}(Y_1; Y_2) 
        =\nonumber\\&\hspace{3cm}=
        - \frac i2 \bar{\eta} H(p_1,p_1) J^{CC}(y_1,0; Y_2)
            - \frac i2 \eta \Hd(\pd_1,\pd_1) J^{CC}(0,\yd_1; Y_2)\,;\\
    &(D_L - ih(y_1,\yd_1) + ih(y_2,\pd_2) - ih(p_1,\pd_1) + ih(p_2,\yd_2))J^{C\omega}(Y_1; Y_2) 
        =\nonumber\\&\hspace{3cm}=
        - \frac i2 \bar{\eta} H(p_2,p_2) J^{CC}(Y_1; Y_2,0)
            - \frac i2 \eta \Hd(\pd_2,\pd_2) J^{CC}(Y_1; 0,\yd_2)\,;\\
    &(D_L - ih(y_1,\pd_1) + ih(y_2,\pd_2) - ih(p_1,\yd_1) + ih(p_2,\yd_2))J^{\omega\omega}(Y_1; Y_2) 
        =\nonumber\\&\hspace{3cm}=
        + \frac i2 \bar{\eta} H(p_2,p_2) J^{\omega C}(Y_1; Y_2,0)
            + \frac i2 \eta \Hd(\pd_2,\pd_2) J^{\omega C}(Y_1; 0,\yd_2)
        -\nonumber\\&\hspace{3cm}\quad
        - \frac i2 \bar{\eta} H(p_1,p_1) J^{C\omega}(y_1,0; Y_2)
            - \frac i2 \eta \Hd(\pd_1,\pd_1) J^{C\omega}(0,\yd_1; Y_2)\,.
\end{align}
These equations can be represented in the form $(D_L + \Sigma^{(2)}_- + \Sigma^{(2)}_+)J(Y_1; Y_2) = 0$ with $\Sigma^{(2)}_-$ defined as follows, which is suitable for the rank-two fields \eqref{rk2_types} generalized to arbitrary form degree:
\begin{align}\label{Sigma2_-}
\begin{split}
    \Sigma^{(2)}_- =
        &\big[
            ih(p_1,\yd_1)\theta(\cN_1-\bar{\cN}_1-1) +
            ih(y_1,\pd_1)\theta(\bar{\cN}_1-\cN_1-1)
        \big]
        \theta(2\cS_1 - \cG_1)
        +\\+
        &\big[
            - ih(p_1,\pd_1) +
            \frac i2 \bar{\eta} (-1)^q H(p_1,p_1) \delta_{\bar{\cN}_1,0} +
            \frac i2 \eta (-1)^q \Hd(\pd_1,\pd_1) \delta_{\cN_1,0}
        \big]
        \times\\&\hspace{2cm}\times\theta(\cG_1 - 2\cS_1 + 1)
        +\\+
        &\big[
            ih(p_2,\yd_2)\theta(\cN_2-\bar{\cN}_2-1) +
            ih(y_2,\pd_2)\theta(\bar{\cN}_2-\cN_2-1)
        \big]
        \theta(2\cS_2 - \cG_2)
        +\\+
        &\big[
            - ih(p_2,\pd_2) +
            \frac i2 \bar{\eta} (-1)^q H(p_2,p_2) \delta_{\bar{\cN}_2,0} +
            \frac i2 \eta (-1)^q \Hd(\pd_2,\pd_2) \delta_{\cN_2,0}
        \big]
        \times\\&\hspace{2cm}\times\theta(\cG_2 - 2\cS_2 + 1)\theta(2\cS_1 - \cG_1)
        +\\+
        &\big[
            - ih(p_2,\pd_2) -
            \frac i2 \bar{\eta} (-1)^q H(p_2,p_2) \delta_{\bar{\cN}_2,0} -
            \frac i2 \eta (-1)^q \Hd(\pd_2,\pd_2) \delta_{\cN_2,0}
        \big]
        \times\\&\hspace{2cm}\times\theta(\cG_2 - 2\cS_2 + 1)\theta(\cG_1 - 2\cS_1 + 1)\,.
\end{split}
\end{align}
Here $q$ is a form degree of a field on which $\Sigma^{(2)}_-$ acts; $\cS_{1,2}$ are analogues of $\cS$ \eqref{S_def} for $Y_1$ and $Y_2$, correspondingly.

\section{Calculations}\label{sec:calculations}

\subsection{Rank-one cohomology}\label{sec:rank-one_cohomology}

In this subsection we recall the structure of $H(\Sigma^{(1)}_-)$ discussed in \cite{Vasiliev:2007yc, Bychkov:2021zvd}. In \cite{Vasiliev:2007yc, Bychkov:2021zvd} it was shown that the $\Sigma^{(1)}_-$-cohomology corresponding to the rank-one fields, differential gauge parameters and dynamical equations at different spin values are spanned by the fields from the Table \ref{tab:cohom1}.

\begin{table}[t]
    \renewcommand{\arraystretch}{1.5}
    \centering
    \begin{tabular}{|c|c|c|c|}
        \hline
        Spin
            &$H^{0-\text{x}}(\Sigma^{(1)}_-)$
                &$H^{1-0}(\Sigma^{(1)}_-)$
                    &$H^{2-1}(\Sigma^{(1)}_-)$\\\hline
        $s = 0$
            &$0$   
                &$\phi_{0,0}$
                    &$h(y,\yd) j_{0,0}$\\\hline
        $s =  \frac12$
            &$0$
                &$\phi_{0,1} + c.c.$
                    &$h(y,\pd) j_{0,1} + c.c.$\\\hline
        $s = 1$
            &$\varepsilon_{0,0}$
                &$h(p,\pd) \phi_{1,1}$
                    &$[h(y,\pd) + h(p,\yd)] j_{1,1}$\\\hline
        $s = 2,\, 3,\, \dots$
           &$\varepsilon_{s-1,s-1}$
                &$\begin{array}{c}
                     h(p,\pd) \phi_{s,s}\\
                     h(y,\yd) \phi^{\tr}_{s-2,s-2}
                \end{array}$
                    &$\begin{array}{c}
                        \big[ H(p,p) + \Hd(\pd,\pd) \big] j_{s,s}\\
                        \big[ H(y,y) + \Hd(\yd,\yd) \big] j^{\tr}_{s-2,s-2}
                    \end{array}$\\\hline
        $s = \frac32,\, \frac52,\, \dots$
            &$\varepsilon_{s-1/2,s-3/2}  + c.c.$
                &$\begin{array}{c}
                    h(p,\pd) \phi_{s+1/2,s-1/2} + c.c.\\
                    h(y,\yd) \phi^{\tr}_{s-3/2,s-5/2} + c.c.\\
                    h(y,\pd) \phi^{\gamma\tr}_{s-3/2,s-1/2} + c.c.
                \end{array}$
                    &$\begin{array}{c}
                        H(p,p) j_{s+1/2,s-1/2} + c.c.\\
                        \Hd(\yd,\yd) j^{\tr}_{s-3/2,s-5/2} + c.c.\\
                        \Hd(\yd,\pd) j^{\gamma\tr}_{s-3/2,s-1/2} + c.c.
                    \end{array}$\\\hline
    \end{tabular}
    \caption{Structure of $\Sigma^{(1)}_-$-cohomology. Here $\varepsilon_{m,n}$, $\phi_{m,n}$, $j_{m,n}$, etc. is a reduced form of the notation \eqref{f_m,n} with an omitted argument $Y$. Note that $\varepsilon_{0,0} \equiv \varepsilon_{0,0}(x)$, $\phi_{0,0} \equiv \phi_{0,0}(x)$ and $j_{0,0} \equiv j_{0,0}(x)$ are functions of space-time coordinates.}
    \label{tab:cohom1}
\end{table}

\begin{table}[t]
    \renewcommand{\arraystretch}{1.5}
    \centering
    \begin{tabular}{|c|c|c|}
        \hline
        Spin
            &$H^{3-2}(\Sigma^{(1)}_-)$
                &$H^{4-3}(\Sigma^{(1)}_-)$\\ \hline
        $s = 0,\,\frac12$
            &$0$
                &$0$\\\hline
        $s = 1$
            &$[H(y,y) + \Hd(\yd,\yd)] \psi_{0,0}$
                &$0$\\\hline
        $s = 2,\, 3,\, \dots$
           &$\big[\cH(y,\pd) + \cH(p,\yd)\big] \psi_{s-1,s-1}$
            &$0$\\\hline
        $s = \frac32,\, \frac52,\, \dots$
            &$\cH(y,\pd) \psi_{s-3/2,s-1/2} + c.c.$
                &$0$\\\hline
    \end{tabular}
    \caption{Structure of $H^{3-2}(\Sigma^{(1)}_-)$ and $H^{4-3}(\Sigma^{(1)}_-)$.}
    \label{tab:cohom1_3-2}
\end{table}

Besides the cohomology groups presented in the Table \ref{tab:cohom1}, for the calculation of the $\Sigma^{(2)}_-$-cohomology we will need also $H^{3-2}(\Sigma^{(1)}_-)$ and $H^{4-3}(\Sigma^{(1)}_-)$. These can be easily found by direct computation, which we illustrate below. The final result is presented in Table \ref{tab:cohom1_3-2}.

Let us consider the case of $H^{3-2}(\Sigma^{(1)}_-)$. If $s \geqslant 3$ and $2 \leqslant G \leqslant 2s-4$ then for a 3-form $f\big|_G$ of degree $G$ and a 2-form $\epsilon\big|_{G+2}$ of degree $G+2$ one has
\begin{align*}
    f\big|_G + \Sigma^{(1)}_- \epsilon\big|_{G+2} \equiv
    &\theta(\cN-\bar{\cN}) \big[f_{m,n}(Y)+ih(p,\yd)\epsilon_{m+1,n-1}(Y)\big] +\\+
    &\theta(\bar{\cN}-\cN) \big[f_{n,m}(Y)+ih(y,\pd)\epsilon_{n-1,m+1}(Y)\big]\,,
\end{align*}
where $m=s-1+G/2,\,n=s-1-G/2$.
Let us decompose $\theta(\cN-\bar{\cN})$-part of this expression into a basis of 3-forms as explained in the Appendix \ref{app:technical_formulae}. The result is
\begin{align*}
    f_{m,n}(Y)+&ih(p,\yd)\epsilon_{m+1,n-1}(Y) =
        \frac{-1}{3(m+1)(n+1)}\times\\\times\big\{
            &\cH(p,\pd)\, [
                3 f_{m,n}(y,\yd|Y)
                - (m+1) \epsilon_{m+1,n-1}(\yd,\yd|Y)
            ] -\\-
            &\cH(p,\yd)\, [
                3 f_{m,n}(y,\pd|Y)
                - (m+1) \epsilon_{m+1,n-1}(\yd,\pd|Y)
                - (n+1) \epsilon_{m+1,n-1}(y,p|Y)
            ] -\\-
            &\cH(y,\pd)\, 
                3 f_{m,n}(p,\yd|Y)
            +\\+
            &\cH(y,\yd)\, [
                3 f_{m,n}(p,\pd|Y)
                - (n+1) \epsilon_{m+1,n-1}(p,p|Y)
            ]
        \big\}\,.
\end{align*}
Here, in accordance with the definitions from Appendix \ref{app:technical_formulae}, $f_{m,n}(y,\yd|Y)$ etc. are treated as independent coefficients of the decomposition. From this decomposition one can see that all components of $f_{m,n}(Y)$, except for $\cH(y,\pd)\, 3 f_{m,n}(p,\yd|Y)$, are $\Sigma^{(1)}_-$-exact. Next, one obtains that in this case the condition $\Sigma^{(1)}_- f\big|_G = 0$ becomes
\begin{equation*}
    0 = i h(p,\yd) f_{m,n}(Y) = \frac{-i}{4(m+1)(n+1)} h_{\ga\gad}\cH^{\ga\gad} (py)(\yd\pd)f_{m,n}(p,\yd|Y) = \frac{i}{4} h_{\ga\gad}\cH^{\ga\gad} f_{m,n}(p,\yd|Y)\,,
\end{equation*}
which gives $f_{m,n}(p,\yd|Y) = 0$. Thus $H^{3-2}(\Sigma^{(1)}_-)$ is trivial in the region $s\geqslant 3$, $2 \leqslant G \leqslant 2s-4$.

The other cases should be examined analogously, which gives the result presented in Table \ref{tab:cohom1_3-2}.

\subsection{Rank-two cohomology}\label{sec:rank-two_cohomology}

In this subsection the dynamical rank-two fields related to conserved bilinear HS currents are obtained by using Lemma \ref{lemma:gen_sigma-}. To this end, cohomology groups of $\Sigma^{(2)}_-$ are calculated. The method of calculation, known in homological algebra as bigraded spectral sequence\footnote{See \cite{bott1995differential} (p. 161 and below). Author thanks K.~Ushakov for the reference.}, is based on the following lemma.
\begin{lemma}\label{lemma:calc}
    Let the diagonalizable grading operators with non-negative integer eigenvalues $\cG_1,\,\cG_2$ and the operator $\sigma_- = \sigma_1 + \sigma_2$ such that $(\sigma_1)^2 = 0$, $(\sigma_2)^2 = 0$, $(\sigma_-)^2 = 0$ act on the space $\Lambda^\bullet \otimes V$;
    and let
    \begin{align}
    \begin{split}\label{comm_rels}
            [\cG_1,\cG_2] = 0\,,\qquad
                [\cG_1,\sigma_1] = -2\sigma_1\,,\qquad [\cG_1,\sigma_2] = 0\,,\\
                [\cG_2,\sigma_2] = -2\sigma_2\,,\qquad [\cG_2,\sigma_1] = 0\,.
    \end{split}
    \end{align}
    Then the following is true
    \begin{enumerate}
        \item
            $H(\sigma_-)$ is spanned by fields $J$ with definite $\cG_1 + \cG_2$ value:
                \begin{equation}\label{|G_def}
                    J = \sum_{\cG_1 + \cG_2 = G} J\big|_{G_1,G_2}\,,\qquad   
                        \cG_1 J\big|_{G_1,G_2} = G_1 J\big|_{G_1,G_2}\,,\quad
                        \cG_2 J\big|_{G_1,G_2} = G_2 J\big|_{G_1,G_2}\,.
                \end{equation}
        \item
            The component $J\big|_{G_1,G_2}$ of $J$ defined in \eqref{|G_def} that has maximal value of $G_1$, will be referred to as the \emph{base field} of $J$. The base field must satisfy the following conditions:
                \begin{enumerate}
                    \item $J\big|_{G_1,G_2} \in H(\sigma_2)/\sim$, where $A \sim B \Leftrightarrow A = B + \sigma_1 \xi\big|_{G_1+2,G_2}$ with arbitrary $\xi$ such that $\sigma_- \sum_{k\geqslant 1} \xi\big|_{G_1+2k,G_2-2k+2} = \sigma_1 \xi\big|_{G_1+2,G_2}$;
                    \item The chain of the following equations is consistent.
                        \begin{gather}
                            \sigma_1 J|_{G_1,G_2} = -\sigma_2 J|_{G_1-2,G_2+2}\,,\nonumber\\
                            \sigma_1 J|_{G_1-2,G_2+2} = -\sigma_2 J|_{G_1-4,G_2+4}\,,\label{chain}\\
                            \dots\nonumber
                        \end{gather}
                \end{enumerate}
        \item
            $H(\sigma_2)$ is spanned by fields with definite value of $\cG_2$. If this value is unique, i.e., if $\exists G_*: \forall f \in H(\sigma_2) \Longrightarrow \cG_2 f = G_* f$, then all the equations \eqref{chain} are consistent iff the equation $\sigma_1 J|_{G_1-2k,G_2+2k} = -\sigma_2 J|_{G_1-2k-2,G_2+2k+2}$ with $G_2+2k = G_*$ is consistent.
        \item 
            The base fields (solutions to the 2nd condition of the lemma) are in one-to-one correspondence with the elements of $H(\sigma_-)$.
    \end{enumerate}
\end{lemma}
\begin{proof}
    \begin{enumerate}
        \item
            From the commutation relations \eqref{comm_rels} it follows that $[\cG_1 + \cG_2,\sigma_-] = -2\sigma_-$. Then $[\cG_1 + \cG_2,\sigma_-] \big|_{H(\sigma_-)} = 0$ which immediately leads to the 1st point of the lemma.
        \item
            Let $J\big|_{G_1,G_2}$ be the base field of $J \in H(\sigma_-)$. Then, by definition of the base field, $J\big|_{G_1+2,G_2-2} = 0$. Hence, using that $\sigma_- J = 0$, one obtains that $\sigma_2 J\big|_{G_1,G_2} = 0$. Since one can add $\sigma_-$-exact elements to $J$, $J\big|_{G_1,G_2}$ is equivalent to $J\big|_{G_1,G_2} + \sigma_1\xi\big|_{G_1+2,G_2} + \sigma_2 \epsilon\big|_{G_1,G_2+2}$, where, as $G_1$ is the maximal $\cG_1$ value of $J$, $\xi$ must be such that $$\sigma_- \sum_{k\geqslant 1} \xi\big|_{G_1+2k,G_2-2k+2} = \sigma_1 \xi\big|_{G_1+2,G_2}\,.$$ Taking these facts together, one obtains the point {\it (a)}.
    
            Equations \eqref{chain} straightforwardly follow from the condition $\sigma_- J = 0$.
        \item
            The first statement of this point has the same origin as the point 1. Next, let there be such $G_*$ as demanded in the condition of the lemma. Note that if equations \eqref{chain} are consistent up to the $k-1$-th level then $\sigma_2\sigma_1 J|_{G_1-2k,G_2+2k} = 0$. Thus if $G_2+2k \neq G_*$ the $k$-th equation is also consistent, because the corresponding part of $H(\sigma_2)$ is trivial by definition of the $G_*$. Therefore, consistency could be violated only if $G_2+2k = G_*$.
        \item 
            Obviously, since equations \eqref{chain} are consistent, each solution to the 2nd condition of the lemma corresponds to some elements of $H(\sigma_-)$. Notice that if $J\big|_{G_1,G_2}$ is the base field of both $J \in H(\sigma_-)$ and $\Tilde{J} \in H(\sigma_-)$, such that $J - \Tilde{J} \neq \sigma_- \epsilon$, then the difference $J - \Tilde{J}$ belongs to $H(\sigma_-)$ too and has the base field with the $\cG_1$ value smaller than $G_1$. Thus the set of the fields built on all different solutions to the 2nd condition of the lemma forms a basis of $H(\sigma_-)$.
    \end{enumerate}
\end{proof}

\begin{figure}[t]
\begin{minipage}[t]{0.45\textwidth}
    \centering
    \begin{tikzpicture}
\tikzset{x=0.8cm,y=0.8cm}

\draw[black, ultra thick, ->] (0,0) -- (0,6) node[anchor=east]{$\cG_2$};
\draw[black, ultra thick, ->] (0,0) -- (6,0) node[anchor=north]{$\cG_1$};

\draw[gray, dashed] (3.5,0) -- (3.5,6);
\filldraw[black] (3.5,0) circle (2pt) node[anchor=north]{$G_1$};

\draw[gray, dashed] (0,1.5) -- (6,1.5);
\filldraw[black] (0,1.5) circle (2pt) node[anchor=east]{$G_2$};

\draw[color=lightgray] plot[mark=*] file {pic/plot1b.table};
\draw[color=black] plot[mark=*] file {pic/plot1a.table};
\filldraw[red] (3.5, 1.5) circle (2.5 pt) node[anchor=south west]{\color{black}$J\big|_{G_1,G_2}$};

\end{tikzpicture}
    \caption{Scheme of a $H(\sigma_-)$ element. Each point is a rank-two field with fixed values of $\cG_1$ and $\cG_2$. Fields at gray points are zero; $J\big|_{G_1,G_2}$ is the base field.}
    \label{fig:scheme1}
\end{minipage}\hfill
\begin{minipage}[t]{0.45\textwidth}
    \centering
    \begin{tikzpicture}







\tikzset{x=0.8cm,y=0.8cm}

\draw[black, ultra thick, ->] (0,0) -- (0,6) node[anchor=east]{$\cG_2$};
\draw[black, ultra thick, ->] (0,0) -- (6,0) node[anchor=north]{$\cG_1$};

\draw[gray, thick] (0,2) -- (6,2);
\filldraw[black] (0,2) circle (2pt) node[anchor=east]{$2s_2-2$};

\draw[gray, thick] (0,2.5) -- (6,2.5);
\filldraw[black] (0,2.5) circle (2pt) node[anchor=east]{$2s_2$};

\draw[gray, thick] (3.5,0) -- (3.5,6);
\filldraw[black] (3.5,0) circle (2pt) node[anchor=north east]{$2s_1-2$};

\draw[gray, thick] (4,0) -- (4,6);
\filldraw[black] (4,0) circle (2pt) node[anchor=north west]{$2s_1$};

\draw[color=lightgray] plot[mark=*] file {pic/plot2b.table};
\draw[color=black] plot[mark=*] file {pic/plot2a.table};
\filldraw[red] (3, 1.5) circle (2.5 pt); 

\draw (1.75,1) node{\huge$\omega\,\omega$};
\draw (1.75,4.25) node{\huge$\omega C$};
\draw (5,1) node{\huge$C \omega$};
\draw (5,4.25) node{\huge$CC$};
\end{tikzpicture}




    \caption{Sectors of rank-two fields with different form degree. The diagonal represents an example of a $H(\Sigma^{(2)}_-)$ element with a base field belonging to ``$\omega\omega$" sector.}
    \label{fig:scheme2}
\end{minipage}
\end{figure}

By virtue of Lemma \ref{lemma:calc}, each element of $H(\sigma_-)$ admits a graphical representation as shown in the Fig. \ref{fig:scheme1}. To describe $H(\sigma_-)$ one has to find a set of corresponding base fields. So, the main goal of our calculations is to resolve the 2nd condition of the Lemma \ref{lemma:calc} and obtain the base fields for the rank-two HS fields.

In the case of the HS theory, $\sigma_- \equiv \Sigma^{(2)}_-$ \eqref{Sigma2_-} which is naturally decomposed into $\sigma_1$ and $\sigma_2$ -- the parts depending solely on $Y_1$ or $Y_2$. In addition to the grading operators \eqref{G2_def} in the HS theory one has a pair of spin operators (analogues of \eqref{S_def} for $Y_1$ and $Y_2$) commuting with the grading and with the $\Sigma^{(2)}_-$. Hence we will apply Lemma \ref{lemma:calc} to fields with fixed spins $s_1$ and $s_2$ and without loss of generality assume that $s_1 \geqslant s_2$. As shown in the Fig. \ref{fig:scheme2}, values of spins determine the borders of the sectors ``$CC$", ``$\omega C$", etc., which correspond to different types of the rank-two fields \eqref{rk2_types} with different form degree.

Let $G_1$ and $G_2$ be the $\cG_1$ and $\cG_2$ degrees of the base field in consideration. The cases $G_1 \geqslant 2s_1$ and $G_1 < 2s_1$ are essentially different because the $H(\sigma_2)$ groups, which the respective base fields belong to due to the 2nd condition of Lemma \ref{lemma:calc}, are different. So if $G_1 \geqslant 2s_1$, i.e., if the base field lies in the ``$CC$" or the ``$C\omega$" sector of the diagram which in the case of the field-like cohomology represents 0- and 1-forms, respectively, such base field belongs to $H^{1-0}(\sigma_2)$ ($H^{q-(q-1)}(\sigma_2)$ is defined analogously to $H^{q-(q-1)}(\Sigma^{(1)}_-)$ of Section \ref{sec:definitions}). Similarly, the base fields for the gauge-like and equation-like cohomology belong to $H^{0-\text{x}}(\sigma_2)$ and $H^{2-1}(\sigma_2)$, respectively. If $G_1 < 2s_1$ the base fields belong to $H^{1-0}(\sigma_2)$, $H^{2-1}(\sigma_2)$ and $H^{3-2}(\sigma_2)$ for the gauge-like, field-like and equation-like cohomology, correspondingly. By construction of $\sigma_2$, these spaces are equivalent to the $\Sigma^{(1)}_-$ cohomology spaces discussed in Section \ref{sec:rank-one_cohomology}.

\begin{table}[t]
    \renewcommand{\arraystretch}{1.5}
    \centering
    \begin{tabular}{|m{6em}|c|}
        \hline
        Gauge-like $H(\Sigma^{(2)}_-)$ &
        $\begin{array}{c}
            h(p_2,\pd_2)\varepsilon^{\omega\omega}_{2s_1-2-\ad,\ad;\;\ceil{s_2},\floor{s_2};\;0,0}\\
            h(p_2,\pd_2)\varepsilon^{\omega\omega}_{2s_1-2s_2-2-\ad,\ad;\;0,0;\;\floor{s_2},\ceil{s_2}}\\
            \varepsilon^{C\omega}_{2s_1+\cd,0;\;\ceil{s_2}-1,\floor{s_2}-1-\cd;\;0,\cd}
        \end{array}$ \\\hline
        Field-like $H(\Sigma^{(2)}_-)$ &
        $\begin{array}{c}
            \{ \Hd(\pd_2,\pd_2) + \delta_{\{s_2\},0}\, H(p_2,p_2) \} \ \mathbf{j}^{\omega\omega}_{2s_1-2-\ad,\ad;\;\floor{s_2},\ceil{s_2};\;0,0}\\
            \{ H(p_2,p_2) + \delta_{\{s_2\},0}\, \Hd(\pd_2,\pd_2)\} \ \mathbf{j}^{\omega\omega}_{2s_1-2s_2-2-\ad,\ad;\;0,0;\;\ceil{s_2},\floor{s_2}}\\
            \{ \bar{\mathfrak{H}} + \delta_{\{s_2\},0}\, \mathfrak{H}\} \ \mathbf{j}^{\omega\omega}_{s+s_1-s_2-1,0;\;1,s-s_1+s_2;\;\ceil{s_2}-1,\floor{s_1-s}}\,,\ s_1>s+\{s_2\}\\
            h(p_2,\pd_2)\mathbf{j}^{C\omega}_{\floor{s\mp s_1},s\pm s_1-s_2;\;\ceil{s_2},0;\;0,\floor{s_2}}\\
            \mathfrak{h}\mathbf{j}^{C\omega}_{2s_1-1+\cd,0;\;\ceil{s_2}-1,\floor{s_2}-\cd;\;1,\cd}
        \end{array}$ \\\hline
        Equation-like $H(\Sigma^{(2)}_-)$ &
        $\begin{array}{c}
            \{ \cH(p_2,\yd_2) + \delta_{\{s_2\},0}\, \cH(y_2,\pd_2)\} \psi^{\omega\omega}_{s+s_1-s_2,0;\;0,s-s_1+s_2;\;\ceil{s_2}-1,\floor{s_1-s}-1}\\
            \{ \Hd(\pd_2,\pd_2) + \delta_{\{s_2\},0}\, H(p_2,p_2)\} \psi^{C\omega}_{\ceil{s\pm s_1},s \mp s_1 - s_2;\;\floor{s_2},0;\;0,\ceil{s_2}}
        \end{array}$ \\\hline
    \end{tabular}
    \caption{The base fields for the rank-two HS fields (up to complex conjugation). Here $\mathfrak{H}:=\Hd(\pd_2,\pd_2) +\# H(y_2,y_2) (y_1p_2)(p_1p_2)(\pd_1\pd_2)(\yd_1\pd_2)$, $\bar{\mathfrak{H}}$ is its c.c.; $\mathfrak{h}:=h(p_2,\pd_2) + \delta_{\{s_2\},1/2}\frac{s_2-1/2}{2s_2(2s_1 + \cd)} h(y_2,\pd_2)(y_1p_2)(p_1p_2)$. The coefficient in $\mathfrak{H}$ is defined in \eqref{sharp}. Here and below we use the standard notation $\{\dots\}$ for a fractional part of a number and $\floor{\dots}$ and $\ceil{\dots}$ for floor (integer part) and ceiling functions, respectively. Parameters in the subscripts are (half-)integer and are assumed to take any values that the subscripts are non-negative integers. The superscript refers to the sector of the diagram in Fig. \ref{fig:scheme2} which the base field belongs to.}
    \label{tab:rk2_fields}
\end{table}

\begin{table}[t]
    \renewcommand{\arraystretch}{1.5}
    \centering
    \begin{tabular}{|c|c|}
        \hline
        $G_1 = 0$ &
            $\{ \Hd(\pd_2,\pd_2) + H(p_2,p_2) \} \ 
                \mathbf{j}^{\omega\omega}_{s_1-c-1,s_1-c-1;\;s_2-c,s_2-c;\;c,c}\,,\qquad
                    1 \leqslant c \leqslant s_2-1$ \\\hline
        $G_1 = 1$ &
            $\begin{array}{c}
                \{ \Hd(\pd_2,\pd_2) + \delta_{\{s_2\},0}\, H(p_2,p_2) \} \ 
                    \mathbf{j}^{\omega\omega}_{\floor{s_1}-c,\ceil{s_1}-2-\cd;\;\floor{s_2}-c,\ceil{s_2}-\cd;\;c,\cd}
                        \,,\quad c \geqslant 1\,,\ \cd \leqslant \ceil{s_2}\\
                H(y_2,y_2) \ 
                    \mathbf{j}^{\omega\omega}_{\floor{s_1}-c,0;\;\floor{s_2}-c,0;\;c,s_1-3/2}\,,\quad s_1 = s_2\\
                H(y_2,p_2) \ 
                    \mathbf{j}^{\omega\omega}_{0,0;\;0,0;\;s_1+\frac12,s_1-\frac52}
                        \,,\quad s_1 = s_2
            \end{array}$ \\\hline
        $G_1 = 2$ &
            $\begin{array}{c}
                \mathfrak{H}_1 \ 
                    \mathbf{j}^{\omega\omega}_{s_1-c,s_1-2-\cd;\;\floor{s_2}-c,\ceil{s_2}-\cd;\;c,\cd}
                        \,,\quad c \geqslant 1\,,\ \cd \leqslant \ceil{s_2}
            \end{array}$ \\\hline
    \end{tabular}
    \caption{The base fields for the irregular field-like cocycles. Here $\mathfrak{H}_1 := \{ \Hd(\pd_2,\pd_2) + \delta_{\{s_2\},0} H(p_2,p_2) \}[ \delta_{\{s_1\},0} \#_1 (p_1p_2)(p_1y_2)(\pd_2\yd_1)(\yd_1\yd_2) + 1 ] + \#_2 \delta_{s_1,s_2}\delta_{a,0}\delta_{b,0} \{ H(y_2,y_2) + \Hd(\yd_2,\yd_2) \} (p_1p_2)^2(\yd_1\pd_2)^2$, $\#_1 = \frac{1-\ceil{s_2}}{(1+\cd)(s_1-c)(s_1+\ceil{s_2}-\cd)(\floor{s-2}-c+1)(\floor{s_2}-1)}$, $\#_2 = \frac{1}{s_1(s_1+1)(s_1-\cd)(s_1-\cd-1)}$. Parameters in the subscripts are (half-)integer and are assumed to take any values that the subscripts are non-negative integers. The superscript refers to the sector of the diagram in Fig. \ref{fig:scheme2} which the base field belongs to.}
    \label{tab:rk2_fields_irreg}
\end{table}

In our calculations we have checked the conditions of the Lemma \ref{lemma:calc} straightforwardly, working with the decompositions into a basis of differential forms, as we have done above for the rank-one cohomology case. As the intermediate expressions are quite long and cumbersome, thus we have put them into Appendix \ref{app:calculation_details}. We have not included there monotonous computations of the ``border effects" near the points $G_1 = 0$, $G_1 = 2s_1$ and $G_2 = 2s_2$; the details are presented only for the cases with $s_2 > 2$ and $2s_1+4 \leqslant G_1$ (Appendix \ref{app:highG}) or $3 \leqslant G_1 \leqslant 2s_1-4$ (Appendix \ref{app:lowG}). As it turns out, the answers for the exceptional cases near $G_1 = 2s_1$ and $G_2 = 2s_2$ can be obtained from the discussed ones by the parameters domain extension. For the near-zero-$G_1$ cases some comments are in order. At the points $G_1 = 0,1$, condition $2(b)$ of Lemma \ref{lemma:calc} is trivial because in that case $\sigma_1 J\big|_{G_1,G_2} = 0$ by definition of $\sigma_1$. At $G_1 = 2$ that condition is relaxed as well, but the reason is that the equation $\sigma_1 J\big|_{2,G_2} = -\sigma_2 J\big|_{0,G_2+2}$ (see eq. \eqref{chain}) includes three fields (with $\cN_1-\bar{\cN}_1 = 0,\pm 2$) instead of two as it is at $G_1>2$. Thus, as a result of relaxation of the 2nd condition of Lemma \ref{lemma:calc}, at $G_1 \leqslant 2$ the new field-like and gauge-like cocycles appear that have no analogues at $G_1>2$; we call such cocycles \emph{irregular}. Note that there are no irregular equation-like cocycles since the 2nd condition of Lemma \ref{lemma:calc} is trivial in that case, as is discussed in Appendix \ref{app:lowG}.

The main answer (the regular cocycles for the case of $s_2 \geqslant 2$) is presented in Table \ref{tab:rk2_fields}. The irregular field-like cocycles are listed in Table \ref{tab:rk2_fields_irreg}. 

\subsection{Remarkable properties}

As was mentioned in Section \ref{sec:general_idea}, not every element of $H(\Sigma^{(2)})$ is related to a conserved current. In order to single out the $H(\Sigma^{(2)})$ elements related to conserved currents we prove the following propositions.

\begin{prop}\label{prop:1}
    The base fields for the regular field-like $H(\Sigma^{(2)})$ listed in Table \ref{tab:dyn_rk2_fields} obey differential equations of the form
    \begin{equation}\label{tildej_eqn}
        D(p,\pd)\Tilde{j}^\bullet_{s+n,s-n}(Y) = (\text{descendants of the irregular cocycles})\,,
    \end{equation}
    where $\Tilde{j}^\bullet$ is defined with the help of \eqref{f_abc}:
    \begin{equation}\label{bfj-tildej}
        \mathbf{j}^\bullet_{a,\ad;\;b,\bd;\;c,\cd} = \cK_{a,\ad;\;b,\bd;\;c,\cd}\ \Tilde{j}^\bullet_{a+b,\ad+\bd}(Y)\,.
    \end{equation}
    For convenience, in the sequel these fields will be called \emph{relevant}.
\end{prop}
\begin{prop}\label{prop:2}
    The relevant fields are invariant under the rank-two gauge transformations generated by regular elements of gauge-like $H(\Sigma^{(2)})$, i.e., under the rank-two differential gauge transformations.
\end{prop}
\begin{prop}\label{prop:3}
    The base fields for the regular field-like $H(\Sigma^{(2)})$ from Table \ref{tab:rk2_fields} that are not included in Table \ref{tab:dyn_rk2_fields} are not governed by any differential equation. In the sequel they will be referred to as \emph{irrelevant}.
\end{prop}
\begin{prop}\label{prop:4}
    The rank-two fields built on $\mathbf{j}^{C\omega}_{s+s_1,s-s_1-s_2;\;s_2,0;\;0,s_2}$ and $\mathbf{j}^{C\omega}_{s-s_1-s_2,s+s_1;\;0,s_2;\;s_2,0}$ (and their fermionic analogues) from Table \ref{tab:dyn_rk2_fields} are goverened by differential equations that cannot be affected by rank-two fields of other types, including irregular ones (from Table \ref{tab:rk2_fields_irreg}).
\end{prop}

\begin{table}[t]
    \renewcommand{\arraystretch}{1.5}
    \centering
    \begin{tabular}{|c|}
        \hline
            $\begin{array}{c}
                 \{ \bar{\mathfrak{H}} + \delta_{\{s_2\},0}\, \mathfrak{H}\} \ \mathbf{j}^{\omega\omega}_{s+s_1-s_2-1,0;\;1,s-s_1+s_2;\;\ceil{s_2}-1,\floor{s_1-s}}\\
                  \{ \mathfrak{H} + \delta_{\{s_2\},0}\, \bar{\mathfrak{H}}\} \ \mathbf{j}^{\omega\omega}_{0,s+s_1-s_2-1;\;s-s_1+s_2,1;\;\floor{s_1-s},\ceil{s_2}-1}
            \end{array}
            \quad s_1>s+\{s_2\}$\\\hline
            $\begin{array}{c}
                 h(p_2,\pd_2)\mathbf{j}^{C\omega}_{\floor{s- s_1},s+ s_1-s_2;\;\ceil{s_2},0;\;0,\floor{s_2}}\\
                 h(p_2,\pd_2)\mathbf{j}^{C\omega}_{s+ s_1-s_2,\floor{s- s_1};\;0,\ceil{s_2};\;\floor{s_2},0}
            \end{array}$\\\hline
            $\begin{array}{c}
                 h(p_2,\pd_2)\mathbf{j}^{C\omega}_{\floor{s+ s_1},s- s_1-s_2;\;\ceil{s_2},0;\;0,\floor{s_2}}\\
                 h(p_2,\pd_2)\mathbf{j}^{C\omega}_{s- s_1-s_2,\floor{s+ s_1};\;0,\ceil{s_2};\;\floor{s_2},0}
            \end{array}
            \quad s>s_1+s_2$\\\hline
    \end{tabular}
    \caption{Structure of base fields for dynamical rank-two fields relevant to the conserved currents. Here $\mathfrak{H}:=\Hd(\pd_2,\pd_2) +\# H(y_2,y_2) (y_1p_2)(p_1p_2)(\pd_1\pd_2)(\yd_1\pd_2)$, $\bar{\mathfrak{H}}$ is its c.c. Coefficient is defined in \eqref{sharp}. Parameter $s$ is assumed to take any values that the subscripts are non-negative.}
    \label{tab:dyn_rk2_fields}
\end{table}

We start with discussion of the Proposition \ref{prop:1}, sketching the idea of appearance of $D(p,\pd)$ in \eqref{tildej_eqn}. For the detailed proof see Appendix \ref{app:on_rank-two_dynamical_equations}. For the general rank-two field which has the form $J+(\text{descendants})$ where $J \in H(\Sigma^{(2)}_-)$ the dynamical equation reads
\begin{equation}\label{rk2_dyn_eqn}
    \cP \big\{[D_L + \Sigma^{(2)}_- + \Sigma^{(2)}_+ ][J+(\text{descendants})]\big\} = 0\,,
\end{equation}
where $\cP$ is a formal projector onto the equation-like $H(\Sigma^{(2)}_-)$. The desired equation, \eqref{tildej_eqn}, is of the first order in space-time derivatives of $\Tilde{j}^\bullet_{s+n,s-n}(Y)$ as is necessary to interpret it as a conservation law. As noted in the Section \ref{sec:general_idea}, equation \eqref{rk2_dyn_eqn} is indeed the first-order differential equation if the corresponding element of the equation-like $H(\Sigma^{(2)}_-)$ has the same $\cG$-degree as $J$. Therefore, in the considered case the derivation of such equations amounts to acting by $D_L$ on $J\big|_{G_1,G_2}$ of the form presented in Table \ref{tab:dyn_rk2_fields} and singling out from the result only those components that correspond to elements of the equation-like $H(\Sigma^{(2)}_-)$ from Table \ref{tab:rk2_fields} with the degree $G_1 + G_2$.

Let us consider in detail the case of the field $h(p_2,\pd_2)\mathbf{j}^{C\omega}_{s\mp s_1-1/2,s\pm s_1-s_2;\;s_2+1/2,0;\;0,s_2-1/2}$ with half-integer $s_2$. Acting on it by $D_L$, one gets schematically the following
\begin{align*}
    D_L h(p_2,\pd_2)\mathbf{j}^{C\omega} =
    &\# \Hd(\pd_2,\pd_2) D(p_2,\yd_2)\mathbf{j}^{C\omega} +
    \# \Hd(\yd_2,\pd_2) D(p_2,\pd_2)\mathbf{j}^{C\omega} +\\+
    &\# H(p_2,p_2) D(y_2,\pd_2)\mathbf{j}^{C\omega} +
    \# H(y_2,p_2) D(p_2,\pd_2)\mathbf{j}^{C\omega}\,,
\end{align*}
where $\#$ are some non-zero coefficients, $D(u,\ud)$ is defined in \eqref{D_expansion}. Among these terms only that proportional to $\Hd(\pd_2,\pd_2)$ can belong to the equation-like cohomology (see Table \ref{tab:rk2_fields}).

Next, for $\mathbf{f}_{a,\ad;\;b,\bd;\;c,\cd}$ and $f_{a+b,\ad+\bd}$ from the formula \eqref{f_abc} one can obtain that
\begin{align*}
    D(p_2,\yd_2)  \mathbf{f}_{a,\ad;\;b,\bd;\;c,\cd} =
        &\# (D(p,\pd) f)_{a,\ad-1;\;b-1,\bd;\;c,\cd+1} +
        \# (D(p,\yd) f)_{a,\ad;\;b-1,\bd+1;\;c,\cd} +\\+
        &\# (D(y,\pd) f)_{a+1,\ad-1;\;b,\bd;\;c-1,\cd+1} +
        \# (D(y,\yd) f)_{a+1,\ad;\;b,\bd+1;\;c-1,\cd}\,,
\end{align*}
where we denoted $(D(p,\pd) f)_{a,\ad-1;\;b-1,\bd;\;c,\cd+1} := \cK_{a,\ad-1;\;b-1,\bd;\;c,\cd+1} D(p,\pd) f_{a+b,\ad+\bd}(Y)$, etc. Expanding $D(p_2,\yd_2)\mathbf{j}^{C\omega}$ via this formula, one can see that the $D(p,\pd)-$term has the form of $\psi^{C\omega}_{s\pm s_1+1/2,s \mp s_1 - s_2;\;s_2-1/2,0;\;0,s_2+1/2}$ from Table \ref{tab:rk2_fields} (where $s$ should be replaced by $s-1$). Hence, the considered field is indeed obeys the first-order dynamical equation of the form \eqref{tildej_eqn}. Let us emphasize that so far we have not specified the RHS of \eqref{tildej_eqn}, which means the other fields from Table \ref{tab:rk2_fields} could enter this equation. In Appendix \ref{app:on_rank-two_dynamical_equations} it is shown that the other relevant fields do not contribute \eqref{tildej_eqn}, and the analogous fact about irrelevant fields is stated in Proposition \ref{prop:3} which we are going to prove.

In the proof of Propositions \ref{prop:2}, \ref{prop:3} and \ref{prop:4} the following auxiliary lemma will be useful.
\begin{lemma}\label{lemma:eq_check}
    Let $J\big|_{g_1,g-g_1}$ and $\Psi\big|_{G_1,G-G_1}$ be the base fields for some elements of the field-like and equation-like $H(\Sigma^{(2)})$, correspondingly. Let $J\big|_{g_1,g-g_1} \propto \mathbf{j}_{a,\ad;\;m-a,\md-\ad;\;c,\cd}$ and $\Psi\big|_{G_1,G-G_1} \propto \psi_{A,\Ad;\;M-A,\Md-\Ad;\;C,\Cd}$ up to some vierbein-dependent factor. Then if $\Psi\big|_{G_1,G-G_1}$ represents the differential equation for $J\big|_{g_1,g-g_1}$ the following inequalities are true
    \begin{subequations}\label{eq_check_ineq}
    \begin{gather}
        |M-\Md| \leqslant m+\md\,,\qquad
        |m-\md| \leqslant M+\Md\,,\\
        |M-\Md-m+\md| + |m+\md-M-\Md| \leqslant G - g + 2\,.\label{eq_check_ineq_b}
    \end{gather}
    \end{subequations}
\end{lemma}
\begin{proof}
    Let $j_{m, \md}(Y)$ and $\psi_{M,\Md}(Y)$ be the counterparts of $\mathbf{j}_{a,\ad;\;m-a,\md-\ad;\;c,\cd}$ and $\psi_{A,\Ad;\;M-A,\Md-\Ad;\;C,\Cd}$ in the sense of formula \eqref{f_abc}. That the $J\big|_{g_1,g-g_1}$ obeys the equation associated with $\Psi\big|_{G_1,G-G_1}$ means that one can write
    \begin{equation}\label{jpsi_check}
        D(u_1,\ud_1)D(u_2,\ud_2)\dots D(u_k,\ud_k) j_{m, \md}(Y) = \psi_{M,\Md}(Y)\,,
    \end{equation}
    where $u_i=y,\,p$ and $\ud_i=\yd,\,\pd$. Note that if the difference between the numbers of $p$ and $y$ among $u_i$ exceeds $m$, the LHS is zero because one can commute the derivatives with $p$ to the right and eliminate $j_{m, \md}(Y)$. (The similar holds for $\yd$ and $\pd$.) To use this argument we define $N_{D(u,\ud)}$ as the number of $D(u,\ud)$ with certain $u,\,\ud$, in the LHS. It is easy to obtain from \eqref{jpsi_check} that
    \begin{align}\label{N_D_def}
        N_{D(y,\pd)} - N_{D(p,\yd)} =
            \frac{M-\Md-m+\md}{2}\,,\qquad
        N_{D(p,\pd)} - N_{D(y,\yd)} =
            \frac{m+\md-M-\Md}{2}\,.
    \end{align}
    Then to demand the LHS of equation \eqref{jpsi_check} to be non-trivial one has to set
    \begin{equation}\label{neq1}
        -m \leqslant N_{D(y,\pd)} - N_{D(p,\yd)} \leqslant \md\,,\qquad
        N_{D(p,\pd)} - N_{D(y,\yd)} \leqslant m\,,\qquad
        N_{D(p,\pd)} - N_{D(y,\yd)} \leqslant \md\,.
    \end{equation}
    
    Another condition comes from the viewpoint of grading. The total number of the derivatives in \eqref{jpsi_check} $k$ must not exceed $(G-g)/2+1$ in the $AdS$ theory and be equal to this number in the Minkowski theory. Hence,
    \begin{equation}\label{neq2}
        |N_{D(y,\pd)} - N_{D(p,\yd)}| + |N_{D(p,\pd)} - N_{D(y,\yd)}| \leqslant
            (G-g)/2+1\,.
    \end{equation}
    Inserting \eqref{N_D_def} into \eqref{neq1} and \eqref{neq2} one gets \eqref{eq_check_ineq}.
\end{proof}

Let us note that Lemma \ref{lemma:eq_check} can be used also for the field -- gauge parameter connection studying: replacement of ``field-like cohomology" and ``equation-like cohomology" by ``gauge-like cohomology" and ``field-like cohomology", correspondingly, in the condition of the lemma, does not change the conclusion.

Analysis of the $C\omega$-type fields from Table \ref{tab:rk2_fields} with the help of Lemma \ref{lemma:eq_check} is quite simple. One can check that for the irrelevant fields of this type taken with any element of the equation-like cohomology inequalities \eqref{eq_check_ineq} are false, hence these fields do not obey any differential equation. Using Lemma \ref{lemma:eq_check} from the gauge parameters perspective, one easily finds that none of the gauge-like cohomology elements can contribute to the transformation of $C\omega$-type relevant fields.

The irrelevant $\mathbf{j}^{C\omega}$ are gauge variant: repeating the same steps which we have made to obtain \eqref{tildej_eqn} one can show that the gauge transformation law for the $C\omega$-type irrelevant fields reads
\begin{equation}
    \delta_{\varepsilon} \tilde{j}^{C\omega}_{2s_1+\cd+\ceil{s_2}-2,\floor{s_2}-\cd}(Y) \propto D(p,\yd)\varepsilon^{C\omega}_{2s_1+\cd+\ceil{s_2}-1,\floor{s_2}-1-\cd}(Y)\,,
\end{equation}
where definition \eqref{f_abc} was again used. However these fields are not pure gauge, since the gauge variation obeys the identity $(D(p,\pd))^{\floor{s_2}-\cd} \delta_{\varepsilon} \tilde{j}^{C\omega}_{2s_1+\cd+\ceil{s_2}-2,\floor{s_2}-\cd}(Y) \equiv 0$, which is not true for general $\tilde{j}^{C\omega}_{2s_1+\cd+\ceil{s_2}-2,\floor{s_2}-\cd}(Y)$.

Analysis of the $\omega\omega$-type fields is more involved. Below we consider the case of integer $s_2$, the half-integer $s_2$ case is analogous. Lemma \ref{lemma:eq_check} gives that $\mathbf{j}^A \equiv \mathbf{j}^{\omega\omega}_{2s_1-2s_2-2-\ad,\ad;\;0,0;\;s_2,s_2}$ obeys no differential equations, while $\mathbf{j}^B \equiv \mathbf{j}^{\omega\omega}_{2s_1-2-\ad,\ad;\;s_2,s_2;\;0,0}$ may obey differential equations associated with $\psi^{C\omega}_{s-s_1,s+s_1-s_2;\;s_2,0;\;0,s_2}$ ($s \geqslant s_1$) and its complex conjugation. Actually, $\mathbf{j}^B$ is not governed by any differential equation as well, which is shown below using a gauge symmetry argument of Proposition \ref{prop:2}.

Firstly, let us notice that the regular gauge-like cocycles do not contribute rank-two gauge transformations of the irregular field-like cocycles. Indeed, for example, in $\cN_1 > \bar{\cN}_1$, $\cN_2 > \bar{\cN}_2$ sector, according to condition $2(b)$ of Lemma \ref{lemma:calc}, the regular fields satisfy
\begin{equation*}
    i h(p_1,\yd_1) \cE_{reg}^{\omega\omega}\big|_{G_1,2-2\{s_2\}} = i h(p_2,\yd_2) f(Y_1;Y_2)
\end{equation*}
with some function $f(Y_1;Y_2)$. Consequently,
\begin{equation*}
    i h(p_1,\yd_1) D_L\cE_{reg}^{\omega\omega}\big|_{G_1,2-2\{s_2\}} = i h(p_2,\yd_2) D_L f(Y_1;Y_2)\,.
\end{equation*}
For the irregular fields, relaxation of condition $2(b)$ of Lemma \ref{lemma:calc} means that the equation
\begin{equation*}
    i h(p_1,\yd_1) J_{irreg}^{\omega\omega}\big|_{G_1,2-2\{s_2\}} = i h(p_2,\yd_2) g(Y_1;Y_2)
\end{equation*}
is inconsistent for any $g(Y_1;Y_2)$. Therefore, in terms of \eqref{rk2_dyn_eqn},
\begin{equation*}
    \cP D_L\cE_{reg}^{\omega\omega}\big|_{G_1,2-2\{s_2\}} \neq J_{irreg}^{\omega\omega}\big|_{G_1,2-2\{s_2\}}\,,
\end{equation*}
which means that the regular gauge-like cocycles do not affect the irregular field-like cocycles. Hence, we will consider only the regular cocycles in the subsequent discussion.

Using Lemma \ref{lemma:eq_check} one obtains that $\varepsilon^A \equiv \varepsilon^{\omega\omega}_{2s_1-2s_2-2-\ad,\ad;\;0,0;\;s_2,s_2}$ can affect only\footnote{In this paragraph we do not assume $\ad$ in $\varepsilon^A$, $\mathbf{j}^A$, etc. to be consistent with each other.} $\mathbf{j}^A$, while $\varepsilon^B \equiv \varepsilon^{\omega\omega}_{2s_1-2-\ad,\ad;\;s_2,s_2;\;0,0}$ can affect $\mathbf{j}^B$ and $\mathbf{j}^C \equiv\mathbf{j}^{\omega\omega}_{s+s_1-s_2-1,0;\;1,s-s_1+s_2;\;s_2-1,s_1-s}$ at $s=s_1-1$, and $\mathbf{j}^D \equiv \mathbf{j}^{C\omega}_{2s_1-1+\cd,0;\;s_2-1,s_2-\cd;\;1,\cd}$. Let us show that $\mathbf{j}^C$ is actually invariant under gauge transformations associated with  $\varepsilon^B$. Indeed, on the one hand, $\mathbf{j}^C$ is relevant (see Table \ref{tab:dyn_rk2_fields}) and obeys differential equation of the form \eqref{tildej_eqn} associated with $\psi^{\omega\omega}_{2s_1-s_2-1,0;\;0,s_2-1;\;s_2-1,0}$, but the other fields that can be affected by $\varepsilon^B$, namely, $\mathbf{j}^B$ and $\mathbf{j}^D$, cannot contribute to this equation by virtue of Lemma \ref{lemma:eq_check}, as we have discussed above. On the other hand, the gauge variation of $\mathbf{j}^C$ cannot vanish by itself because, as the conditions of Lemma \ref{lemma:eq_check} are not violated, the resulting combination of derivatives does not cancel $\varepsilon^B$ out identically. Therefore, if $\mathbf{j}^C$ was gauge variant, $\varepsilon^B$ would has to satisfy a differential equation to provide gauge invariance of \eqref{tildej_eqn}. Notice that by proving the gauge invariance of $\mathbf{j}^C$ we completed the proof of Proposition \ref{prop:2}.

Now we can prove that $\mathbf{j}^B$ does not obey any differential equation. Since $\mathbf{j}^C$ is $\varepsilon^B$-invariant and $\mathbf{j}^D$ is not governed by differential equations, $\mathbf{j}^B\equiv \mathbf{j}^{\omega\omega}_{2s_1-2-\ad,\ad;\;s_2,s_2;\;0,0}$ with various $\ad$ are the only $\varepsilon^B$-variant fields that can contribute to the equation involving $\mathbf{j}^B$, if such exist. As this equation must be gauge invariant, the gauge variations of $\mathbf{j}^B$ with different $\ad$ must cancel each other out because here, as in the case of $\mathbf{j}^C$, the gauge variation does not vanish identically by itself. Let us fix the value $\ad \leqslant s_1-1$ of $\varepsilon^B$ in consideration (the case of $\ad > s_1-1$ is analogous); such fields we will denote by $\varepsilon^B\big|_\ad$ (and $\mathbf{j}^B\big|_\ad$). Let us define $\varepsilon^B\big|_{-1} := \varepsilon^{C\omega}_{2s_1,0;\;s_2-1,s_2-1;\;0,0}$. Then, similarly to obtaining \eqref{tildej_eqn}, one can show that
\begin{equation*}
    \delta_{\varepsilon^B}j^B_{s_2+2s_1-3-\ad,s_2+1+\ad}(Y) \sim D(p,\yd) \varepsilon^B_{s_2+2s_1-2-\ad,s_2+\ad}(Y)\,,
\end{equation*}
where $j^B_{\dots}(Y)$ and $\varepsilon^B_{\dots}(Y)$ are \eqref{f_abc}-counterparts of $\mathbf{j}^B\big|_{\ad+1}$ and $\varepsilon^B\big|_\ad$, correspondingly. Let us explore, at  which values of $\Ad$ field $\mathbf{j}^B\big|_{\Ad}$ can contribute to the equation on $\mathbf{j}^B\big|_{\ad+1}$ and cancel out its gauge variation with respect to $\varepsilon^B\big|_\ad$. Lemma \ref{lemma:eq_check} does not prohibit $\varepsilon^B\big|_\ad$ to contribute to $\mathbf{j}^B\big|_{\Ad}$ with $\Ad$ such that $\Ad \leqslant 1+\ad$ or simultaneously $\ad = s_1-2$ and $\Ad \geqslant s_1$. Let us hence confine ourselves to the case $\ad \leqslant s_1-3$. Therefore, only $\mathbf{j}^B\big|_{\Ad}$ with $\Ad \leqslant \ad$ can compensate gauge variation of $\mathbf{j}^B\big|_{\ad+1}$ in its equation. Obviously, for $\ad=-1$ there is no room for such $\mathbf{j}^B\big|_{\Ad}$. Therefore, $\mathbf{j}^B\big|_0$ indeed does not governed by differential equations. Consequently, at $\ad = 0$ the situation is the same: there are no any $\mathbf{j}^B\big|_{\Ad}$ to compensate the gauge variation of $\mathbf{j}^B\big|_1$, and hence it does not obey any differential equation as well. Next, using the mathematical induction method, one can extend this statement to all values of $\ad$. Thus, it is proven that $\mathbf{j}^B$ does not governed by any differential equation, hence the proof of Proposition \ref{prop:3} is complete.

Proposition \ref{prop:4} follows from Lemma \ref{lemma:eq_check} immediately.

At the end of the section, let us make some comments. Proposition \ref{prop:3} states that some $H(\Sigma^{(2)}_-)$ elements are non-dynamical, i.e., they are not subjects of any differential equation. Presence of non-dynamical fields among $\Sigma^{(2)}_-$-cocycles is not surprising, however, such fields are known as \emph{off-shell fields} (for instance, they were discussed in \cite{Spirin:2024zgy}). For off-shell fields, unfolded equations \eqref{gen_eqn} just express some fields via derivatives of the others. The second comment is related to Proposition \ref{prop:2}. In our proof we considered each relevant field separately. But since the result turned out to be general it would be interesting to explore if it has a general origin.

\section{Discussion}\label{sec:discussion}

In this section, we interpret the results of $H(\Sigma^{(2)}_-)$ calculation and, in particular, show how they are used for the bilinear HS currents classification problem.

\subsection{Currents from the rank-one fields perspective}\label{sec:rank-one_currents}

Let us recall some details of HS current analysis from the viewpoint of rank-one fields. In Table \ref{tab:cohom1} the structure of the field-like and equation-like cohomology is presented. That fields\footnote{Recall that this notation is based on the formula \eqref{f_m,n} with omitted argument $Y$. For the allowed values of  $m,\,n$ see Table \ref{tab:cohom1}.} $\phi_{m,n}$, $\phi^{\tr}_{m,n}$, $\phi^{\gamma\tr}_{m,n}$ and $j_{m,n}$, $j^{\tr}_{m,n}$ and $j^{\gamma\tr}_{m,n}$ from this table correspond to the dynamical fields and equations, respectively, should be understood as follows. The rank-one equations \eqref{rk1eq} are equivalent to a chain of equations expressing auxiliary fields via $\phi$ and an equation of the form $\hat{\partial} \phi = j$, where $\hat{\partial}$ denotes some differential operator and $\phi$ and $j$ stand for some $\phi$ and $j$ from Table \ref{tab:cohom1}. In  the HS theory $\phi$ is the Fronsdal field, $\hat{\partial}$ is the operator in Fronsdal equations, and $j$ is the current in the RHS of Fronsdal equations. More precisely, $\phi_{m,n}$ and $j_{m,n}$ are the traceless parts of the Fronsdal field and current, correspondingly; $\phi^{\tr}_{m,n}$ and $j^{\tr}_{m,n}$ are their traces and $\phi^{\gamma\tr}_{m,n}$ and $j^{\gamma\tr}_{m,n}$ are their $\gamma$-traces.

The spinor form of Fronsdal equations for integer $s$ reads
\begin{align}\label{Fr_eq_b}
\begin{split}
    &2s\Box \phi - D(y,\yd)D(p,\pd) \phi + D(y,\yd)D(y,\yd) \phi^{\tr}
        - 2s(s^2-2s-2) \phi =
            2s j_{s,s}(Y)\,,\\
    &2(2s-1)\Box \phi^{\tr} + D(y,\yd)D(p,\pd) \phi^{\tr} - D(p,\pd)D(p,\pd) \phi
        - 2(2s-1)s^2 \phi^{\tr} =\\&\pushright{=
            2(2s-1)j^{\tr}_{s-2,s-2}(Y)\,,}
\end{split}
\end{align}
where definition \eqref{D_expansion} is used and $\Box := -\frac12 (p_1p_2)(\pd_1\pd_2)D(Y_1)D(Y_2)$. If $s$ is half-integer then Fang-Fronsdal equations are
\begin{align}\label{Fr_eq_f}
\begin{split}
    D(y,\pd) \phi_{s-1/2,s+1/2}
        + D(y,\yd) \phi^{\gamma\tr}_{s-1/2,s-3/2}
        + i(s^2-1/4) \phi_{s+1/2,s-1/2} =
                j_{s+1/2,s-1/2}\,,\\
    \pushleft{D(p,\yd) \phi^{\gamma\tr}_{s-1/2,s-3/2}
        + \frac{(s-1/2)^2}{2s} D(p,\pd) \phi_{s-1/2,s+1/2}
        -}\\\pushright{- \frac{(s+1/2)^2}{2s} D(y,\yd) \phi^{\tr}_{s-5/2,s-3/2}
        - i(s^2-1/4)\phi^{\gamma\tr}_{s-3/2,s-1/2} =j^{\gamma\tr}_{s-3/2,s-1/2}}\,,\\
    D(y,\pd) \phi^{\tr}_{s-5/2,s-3/2}
        + D(p,\pd) \phi^{\gamma\tr}_{s-1/2,s-3/2}
        + i(s^2-1/4) \phi^{\tr}_{s-3/2,s-5/2} =
                j^{\tr}_{s-3/2,s-5/2}\,.
\end{split}
\end{align}
(There are also three complex conjugated equations.)

The conservation law for the currents in Fronsdal equations reads as
\begin{align}\label{cons_law_b}
    D(p, \pd) j_{s,s} - D(y,\yd) j^{\tr}_{s-2,s-2} = 0
\end{align}
for bosons and
\begin{align}\label{cons_law_f}
\begin{split}
    &\frac{1}{(2s+1)^2} D(p, \pd) j_{s+1/2,s-1/2} - \frac{1}{(2s-1)^2} D(y,\yd) j^{\tr}_{s-3/2,s-5/2} +\\&\hspace{2cm}+ \frac{8s}{(4s^2-1)^2} D(y,\pd) j^{\gamma\tr}_{s-3/2,s-1/2} + \frac{2is}{4s^2-1}j^{\gamma\tr}_{s-1/2,s-3/2} = 0
\end{split}
\end{align}
along with its complex conjugated for fermions. Note that $j_{m,n}$ makes no sense once $m$ or $n$ becomes negative, so $j_{m,n}$ is assumed to be zero in that case (analogously for $j^{\tr}_{m,n}$ and $j^{\gamma\tr}_{m,n}$). This makes formula \eqref{cons_law_f} correct even if some denominators are zero at $s = 1/2$.

\subsection{From cohomology to currents}\label{sec:from_the_cohom}

As explained in Section \ref{sec:general_idea}, from the viewpoint of the rank-two fields, bilinear currents are generated by the dynamical rank-two fields, which belong to the $\Sigma^{(2)}_-$-cohomology, according to Lemma \ref{lemma:gen_sigma-}. Lemma \ref{lemma:calc} states that each $H(\Sigma^{(2)}_-)$ element is determined by its \emph{``base field''}; the base fields are listed in Tables \ref{tab:rk2_fields} and \ref{tab:rk2_fields_irreg}.

Table \ref{tab:rk2_fields_irreg} contains the base fields for so-called \emph{irregular} cocycles -- specific cocycles of low degrees (equivalently, having low derivatives, $N_{der} = 0,\,1,\,2$) the appearance of which is due to the  ``border effects" near degree 0. As it will be discussed below, the irregular cocycles play important role in the construction because they generate traces of currents.

According to Proposition \ref{prop:3}, not every $\Sigma^{(2)}_-$-cocycle is related to a conserved current: the cocycles, called here \emph{irrelevant}, correspond to so-called off-shell fields (see e.g. \cite{Spirin:2024zgy}) not obeying any differential equations. The base fields for the \emph{relevant} regular cocycles are listed in Table \ref{tab:dyn_rk2_fields}. Lemma \ref{lemma:calc} and Proposition \ref{prop:1} imply the structure of these cocycles to be as follows
\begin{equation}\label{tildej}
    J(Y_1; Y_2) = \cR(Y_1; Y_2|p,\pd|h_{\ga\gad}) \Tilde{j}_{s+n,s-n}(Y)\,,
\end{equation}
where $\cR(Y_1; Y_2|p,\pd|h_{\ga\gad})$ is an operator constructed from the objects in its argument by resolving equations \eqref{chain} for base fields from Table \ref{tab:dyn_rk2_fields}; $s$ and $n$ are (half-)integer numbers such that $s\geqslant 0$, $-s\leqslant n \leqslant s$ and, following Table \ref{tab:dyn_rk2_fields}, $n$ is a combination of $s_1$ and $s_2$ which are two spins of the rank-two field, or, equivalently, the spins of the fields the current is formed by; $\Tilde{j}_{s+n,s-n}(Y)$ 
is determined by Table \ref{tab:dyn_rk2_fields} and formula \eqref{bfj-tildej}.

According to Proposition \ref{prop:1}, $\Tilde{j}_{s+n,s-n}(Y)$ satisfies the differential equation, reminiscent of the conservation law \eqref{cons_law_b} or \eqref{cons_law_f}:
\begin{equation}\label{tildej_eqn1}
        D(p,\pd)\Tilde{j}_{s+n,s-n}(Y) = (\text{descendants of the irregular cocycles})\,.
    \end{equation}
However, $\Tilde{j}_{s+n,s-n}(Y)$ are not currents yet at least because they have wrong degrees in $y$ and $\yd$ (cf. \eqref{Fr_eq_b}, \eqref{Fr_eq_f}) and the uncertainty in the RHS of their equations \eqref{tildej_eqn1}. The procedure of transition from $\Tilde{j}_{s+n,s-n}(Y)$ to currents is quite technical; the main properties of the resulting currents are formulated in the following proposition.

\begin{table}[t]
    \renewcommand{\arraystretch}{1.5}
    \centering
    \begin{tabular}{|c|c|}
        \hline
        Spin region
            &Maximal number of derivatives\\ \hline
        $s > s_1 + s_2$
            &$\floor{s} + \floor{s_1} + \floor{s_2}\,,\quad \floor{s} + \floor{s_1} + \floor{s_2} - 2\min\{s, s_1, s_2\}$\\
       $\begin{cases}
            s \leqslant s_1 + s_2\,,\\
            s_1 \leqslant s + s_2\,,\\
            s_2 \leqslant s + s_1
        \end{cases}$
            &$\floor{s} + \floor{s_1} + \floor{s_2} - 2\min\{s_1, s_2\}$\\\hline
    \end{tabular}
    \caption{Maximal number of derivatives in $s$-$s_1$-$s_2$ current. Each value corresponds to a pair of complex conjugated currents.}
    \label{tab:cur_der}
\end{table}

\begin{prop}\label{prop:5}
    Conserved currents $j_{s\pm\{s\},s\mp\{s\}}(Y)$ built from the fields of spins $s_1$ and $s_2$ have the maximal number of derivatives in accordance with Table \ref{tab:cur_der}; i.e., there are lower-derivative currents with $\floor{s} + \floor{s_1} + \floor{s_2} - 2 s_2$ derivatives and higher-derivative currents with $\floor{s} + \floor{s_1} + \floor{s_2}$ derivatives. (Recall that $\{\dots\}$ and $\floor{\dots}$ stand for a fractional and integer part of a number, respectively.)

    The lower-derivative currents obey
    \begin{align}\label{j_eqn_low}
        D(p, \pd) j_{s\pm\{s\},s\mp\{s\}}(Y) + \#_s D(y,\yd) j^{\tr}_{s\pm\{s\}-2,s\mp\{s\}-2}(Y) = 0\,,
    \end{align}
    where coefficient $\#_s$ is according to \eqref{cons_law_b} and \eqref{cons_law_f}.
    
    The higher-derivative currents obey
    \begin{equation}\label{j_eqn_high}
        D(p,\pd) j_{s\pm\{s\},s\mp\{s\}}(Y) = 0\,.
    \end{equation}
\end{prop}

\begin{proof}
    At first consider 
\begin{align}
\begin{split}\label{j-tildej}
    \tilde{\tilde{j}}_{s\pm\{s\},s\mp\{s\}}(Y) :=
        &\theta(n\mp\{s\}+1) (D(p,\yd))^{n\mp\{s\}}\Tilde{j}_{s+n,s-n}(Y) +\\+
        &\theta(-n\pm\{s\}) (D(y,\pd))^{-n\pm\{s\}}\Tilde{j}_{s+n,s-n}(Y)\,,
\end{split}
\end{align}
which has the proper $y$ and $\yd$ degrees. Let us count the number of derivatives inside $\tilde{\tilde{j}}_{s\pm\{s\},s\mp\{s\}}(Y)$. In the present setup this value is related to the degree: the maximal-derivative term in a rank-two field of degrees $G_1$ and $G_2$ contains $\floor{G_1/2}+\floor{G_2/2}$ derivatives of the Fronsdal fields. Hence, using Table \ref{tab:dyn_rk2_fields} and formulae \eqref{bfj-tildej}, \eqref{j-tildej} one can show that $\tilde{\tilde{j}}$ has the same number of derivatives as presented in Table \ref{tab:cur_der}.

Let us consider it in more detail. Without loss of generality we will assume that $s_1 \geqslant s_2$. In the $s > s_1+s_2$ case, there exist two pairs of complex conjugated $\tilde{\tilde{j}}$, which correspond to $\mathbf{j}^{C\omega}$-terms in Table \ref{tab:dyn_rk2_fields}. These terms have the degrees \eqref{G12_def} $G_1 = 2s-2\{s_2\}$ and $G_2 = 2\{s_2\}$, and the value of $n$ \eqref{tildej} $\pm s_1 + s_2$. This yields the first line of Table \ref{tab:cur_der}.

If $s_1 + s_2 \geqslant s > s_1$ the branch of $\mathbf{j}^{C\omega}$ corresponding to the higher-derivative $\tilde{\tilde{j}}$ vanishes; the lower-derivative $\tilde{\tilde{j}}$ is still generated by $\mathbf{j}^{C\omega}$. In the case of $s_1 \geqslant s$ and $s_1 \leqslant s + s_2$ $\tilde{\tilde{j}}$ corresponds to $\mathbf{j}^{\omega\omega}$-terms. It carries up to $\floor{s} + \floor{s_1} + \floor{s_2} - 2 s_2$ derivatives as presented in the second line of Table \ref{tab:cur_der}. In region $s_1 > s + s_2$ there are no relevant cocycles, according to Table \ref{tab:dyn_rk2_fields}.

The equations on $\tilde{\tilde{j}}$ result from those on $\Tilde{j}$ \eqref{tildej_eqn1} and $\tilde{\tilde{j}}$ definition \eqref{j-tildej}. Commuting Lorentz derivatives (see \eqref{D_comm2}) one obtains
\begin{equation}\label{tildetildej_eqn}
    D(p,\pd) \tilde{\tilde{j}}_{s\pm\{s\},s\mp\{s\}}(Y) = (\text{descendants of irregular cocycles})\,.
\end{equation}
Let us notice that by virtue of Proposition \ref{prop:4} for higher-derivative $\tilde{\tilde{j}}$ the RHS of \eqref{tildetildej_eqn} is zero. Therefore, simply stating that $j_{s\pm\{s\},s\mp\{s\}}(Y) = \tilde{\tilde{j}}_{s\pm\{s\},s\mp\{s\}}(Y)$ one immediately obtains \eqref{j_eqn_high}.

For the lower-derivative $\tilde{\tilde{j}}_{s\pm\{s\},s\mp\{s\}}(Y)$, the Ansatz for \eqref{tildetildej_eqn} takes the form
\begin{align}\label{tildetildej_eqn2}
\begin{split}
    D(p,\pd)& \tilde{\tilde{j}}_{s\pm\{s\},s\mp\{s\}}(Y) =\\=
        &\sum_{m,n,k} a_{m,n,k} D^{\alpha(m,n)-k}(p,\pd) D^{\beta(m,n)-k}(y,\yd) \times\\&\hspace{1.5cm}\times\big[
            \theta(\gamma(m,n)) D^{\gamma(m,n)}(y,\pd) + \theta(-\gamma(m,n)) D^{-\gamma(m,n)}(p,\yd) \big] \mathfrak{j}_{m,n}(Y) +\\+
        &\sum_{m,n,k} b_{m,n,k} D^{\beta(m,n)-k}(y,\yd) D^{\alpha(m,n)-k}(p,\pd) \times\\&\hspace{1.5cm}\times\big[
            \theta(\gamma(m,n)) D^{\gamma(m,n)}(y,\pd) + \theta(-\gamma(m,n)) D^{-\gamma(m,n)}(p,\yd) \big] \mathfrak{j}_{m,n}(Y)\,.
\end{split}
\end{align}
(Other combinations of Lorentz derivatives are expressed via the presented ones with the help of \eqref{D_comm_res}.) The RHS is the contribution of the irregular cocycles (see Section \ref{sec:rank-two_cohomology}) expressed here by $\mathfrak{j}_{m,n}(Y)$. At the assumption that $s_1 \geqslant s_2$, $\alpha(m,n) = \frac{\floor{s}+\floor{s_1}-\floor{s_2} - N^{\mathfrak{j}}_{der} -s\mp\{s\}+m+2}{2}$, $\beta(m,n) = \frac{\floor{s}+\floor{s_1}-\floor{s_2} - N^{\mathfrak{j}}_{der} +s\mp\{s\}-n}{2}$, $\gamma(m,n) = \frac{\pm2\{s\}-m+n}{2}$, where $N^{\mathfrak{j}}_{der} = 0,\,1,\,2$ is the number of derivatives of $\mathfrak{j}$.

To obtain equations \eqref{j_eqn_low} one has to absorb all the terms in \eqref{tildetildej_eqn2} of the form $D(p,\pd) f(Y)$ into the current $j_{s\pm\{s\},s\mp\{s\}}(Y):= \tilde{\tilde{j}}_{s\pm\{s\},s\mp\{s\}}(Y) - f(Y)$ and the terms of the form $D(y,\yd) g(Y)$ into its trace $j^{\tr}_{s\pm\{s\}-2,s\mp\{s\}-2}(Y):=-\frac{1}{\#_s}g(Y)$. Notice that the terms in the RHS with $\alpha(m,n)=\beta(m,n)=k$ (which carry neither $D(p,\pd)$ nor $D(y,\yd)$) at $k \geqslant 2$ express via combinations of the type (here $k=2$)
\begin{equation*}
    \big(\# D(y,\yd)^2D(p,\pd)^2 + \# D(y,\yd)D(p,\pd)D(y,\yd)D(p,\pd) + \#D(p,\pd)^2 D(y,\yd) \big)\big[\dots\big]\mathfrak{j}_{m,n}(Y)
\end{equation*} 
with the help of \eqref{D_comm_res}. Hence, these terms should be included into the definition of $j_{s\pm\{s\},s\mp\{s\}}(Y)$ and $j^{\tr}_{s\pm\{s\}-2,s\mp\{s\}-2}(Y)$.

The described transformations bring \eqref{tildetildej_eqn2} to the form
\begin{align}\label{tildetildej_eqn4}
\begin{split}
    &D(p, \pd) j_{s\pm\{s\},s\mp\{s\}}(Y) + \#_s D(y,\yd) j^{\tr}_{s\pm\{s\}-2,s\mp\{s\}-2}(Y) =\\&=
        \sum_{m,n}
        (\delta_{\alpha(m,n),0}\,\delta_{\beta(m,n),0} +
            \delta_{\alpha(m,n),1}\,\delta_{\beta(m,n),1})\, 
        (a_{m,n,k} + b_{m,n,k}) 
            \times\\&\hspace{1.5cm}\times
        \big[
            \theta(\gamma(m,n)) D^{\gamma(m,n)}(y,\pd) + \theta(-\gamma(m,n)) D^{-\gamma(m,n)}(p,\yd) \big] \mathfrak{j}_{m,n}(Y)\,.
\end{split}
\end{align}
The RHS is actually zero: if $\alpha(m,n) = \beta(m,n) = k = 0,\,1$ and $s_1 \geqslant s_2 + 3$ then $\min\{m,n\} = N^{\mathfrak{j}}_{der}-\floor{s_1}+\floor{s_2}+\{s\}\pm\{s\}+2k-2 < 0$ which makes no sense in view of definition \eqref{f_m,n}. In $s_2 \leqslant s_1 < s_2 + 3$ the absence of these terms has to be checked directly and separately, analogously to the discussion in Appendix \ref{app:on_rank-two_dynamical_equations}. We have not included these involved calculations into the paper.

Thus, \eqref{tildetildej_eqn4} is indeed equivalent to \eqref{j_eqn_low}, and the proposition is proven.
\end{proof}

Let us emphasize that our interpretation of the obtained $\Sigma^{(2)}_-$-cocycles as bilinear currents relies on the fact that only one $\tilde{j}$ affects \eqref{tildej_eqn1} and, consequently only one $\tilde{\tilde{j}}$ affects \eqref{tildetildej_eqn}. Due to this property the dynamical equations for the rank-two fields can be brought to the form matching the conservation laws that allows us to identify the cocycles with the currents. As  discussed in Appendix \ref{app:on_rank-two_dynamical_equations}, in the flat theory this fact is quite trivial, but in the $AdS_4$ case we have checked straightforwardly that unwanted terms of the sub-leading order in derivatives vanish. The latter result is not too  surprising however, as it has clear group-theoretical interpretation\footnote{Author is grateful to M.~Vasiliev for pointing out this fact.}: since free rank-two unfolded equations are invariant under the background $AdS$ symmetry, they must split into independent systems with different overall spins, as it happens in the rank-one case. We would avoid many cumbersome calculations once it is shown that $s$ in $\Tilde{j}_{s+n,s-n}(Y)$ indeed corresponds to overall spin of the rank-two field, for example, by constructing an operator $\cO$ with $s$-dependent eigenvalues, such that $[\cO,D_L + \Sigma^{(2)}_- + \Sigma^{(2)}_+] = 0$, $[\cO,\Sigma^{(2)}_-]\big|_{H(\Sigma^{(2)}_-)}=0$ and $[\cO,\cG]\big|_{H(\Sigma^{(2)}_-)}=0$. However, we did not manage to do so  leaving this problem for the future work.

\subsection{Currents and vertices}\label{sec:currents_and_vertices}

It is worth matching the currents \eqref{j-tildej} with the classification of cubic Lagrangian vertices in the HS theory. According to \cite{Bengtsson:1983pd,Bengtsson:1986kh,Metsaev:2005ar,Metsaev:2007rn,Manvelyan:2010jr,Joung:2011ww,Metsaev:2018xip}, three fields with arbitrary spins $s,\,s_1,\,s_2$ form two pairs of complex conjugated vertices with up to $\floor{s} + \floor{s_1} + \floor{s_2}$ and $\floor{s} + \floor{s_1} + \floor{s_2} - 2\min\{s,s_1,s_2\}$ derivatives.

From Table \ref{tab:cur_der} it follows that in the $s > s_1+s_2$ case the currents are straightforwardly related to the cubic vertices.

If the so-called \emph{triangle inequalities}
\begin{equation}\label{tri_ineq}
    s \leqslant s_1 + s_2\,,\qquad
    s_1 \leqslant s + s_2\,,\qquad
    s_2 \leqslant s + s_1
\end{equation}
are true, and $s \geqslant s_2$
(and without loss of generality $s_1 \geqslant s_2$) the currents carry up to $\floor{s} + \floor{s_1} + \floor{s_2} - 2 s_2$ derivatives where $s_2$ is in fact minimal among $s,\,s_1,\,s_2$. Thus, in the region $s \geqslant s_2$ inside of the triangle inequalities the currents correspond to the lower-derivative HS vertices. The absence of currents associated with the higher-derivative vertices in this case will be discussed below.

If the triangle inequalities are still true but $s < s_2$, the currents have $\floor{s} + \floor{s_1} + \floor{s_2} - 2s_2$ derivatives but $s_2$ is no longer the minimal spin. Since all the cubic Lagrangian vertices with these spins carry more derivatives, the relation between the currents and Lagrangian vertices is non-trivial in this case. Note that the stress-energy tensor for the spin-$s_*$ field (i.e., 2-derivative current with $s=2,\,s_1=s_2=s_*$) is of the discussed type. It is well-known after Fradkin and Vasiliev \cite{Fradkin:1987ks} that the corresponding Lagrangian vertex must be supplemented with the higher-derivative terms with up to $2s_*-2$ derivatives. The same should happen for other currents of this type, and these currents actually correspond to the lower-derivative vertices. (Recall that vertices directly related to the currents under discussion do not exist because these currents are not gauge invariant: for the stress-energy tensors this is the famous Weinberg-Witten theorem \cite{Weinberg:1980kq}.)

In the region $s_1 > s+ s_2$ our analysis have not revealed any currents, so this case is not present in Table \ref{tab:cur_der}. Also, above we have seen that inside of the triangle inequalities the higher-derivative vertices have no counterparts in our classification. These two facts have a simple explanation, in view of \cite{Joung:2013nma} \footnote{Author thanks K.~Mkrtchyan for pointing out this work.}: the respective currents are trivial so they do not manifest in the $\Sigma^{(2)}_-$-cohomology analysis. (Recall that trivial currents here are those which obey the conservation law identically, i.e., off-shell.) Indeed, a vertex produces a non-trivial spin-$s$ current once it deforms the spin-$s$ gauge transformation law. Inside of the triangle inequalities, the higher-derivative vertex is constructed from three HS Weyl tensors, i.e., it has the form
\begin{equation*}
    (pp_1)^{s+s_1-s_2}(pp_2)^{s+s_2-s_1}(p_1p_2)^{s_1+s_2-s} C_{2s,0}(Y) C_{2s_1,0}(Y_1) C_{2s_2,0}(Y_2)
\end{equation*}
(or its c.c.). Such \emph{Born-Infeld-type} vertices are deformationally-trivial, thus are not associated with any non-trivial conserved current. These are Class I vertices, in terms of \cite{Joung:2013nma}, and Abelian, in terms of \cite{Vasiliev:2011knf}. 

According to \cite{Joung:2013nma}, if $s_1 > s+ s_2$, the both higher- and lower-derivative vertices do not deform spin-$s$ gauge transformations, thus the corresponding currents are trivial as well. These vertices are of Class II, in terms of \cite{Joung:2013nma}, which means that they deform only spin-$s_1$ gauge transformation law thus giving rise to the first line of Table \ref{tab:cur_der} (with the redefinition $s \leftrightarrow s_1$). In \cite{Vasiliev:2011knf} such vertices were called ``current vertices", their structure, in terms of the unfolded fields, is $\omega C C$, where the HS gauge potential $\omega$ has spin $s_1$ and the HS Weyl tensors $C$ -- $s$ and $s_2$.

\subsection{Traces of the currents}

Proposition \ref{prop:5} states that the higher-derivative currents are traceless while the lower-derivative ones can have non-zero trace. Tracelessness of the higher-derivative currents is what we have expected to obtain, in view of \cite{Gelfond:2017wrh, Misuna:2017bjb, Tatarenko:2024csa}. Traces of the lower-derivative currents are generated by the irregular cocycles. Below we consider two examples, illustrating this fact: gravity and $s=s_1=3$, $s_2=2$ theory, both on Minkowski background. Let us emphasize that since we have found $\Sigma^{(2)}_-$-cocycles describing traces of currents, the traces are not removable by local field redefinitions.

In the case of gravity, the bilinear current is given by $J_{\mu\nu} := -(G^{(2)})_{\mu\nu}$, where $(G^{(2)})_{\mu\nu}$ is the quadratic part of the Einstein tensor; $\mu,\nu = 0,\dots 3$. From the rank-two field perspective, its trace is produced by the irregular cocycle $\{ \Hd(\pd_2,\pd_2) + H(p_2,p_2) \} \ \mathbf{j}^{\omega\omega}_{0,0;\;1,1;\;1,1}$ (see Table \ref{tab:rk2_fields_irreg}) with $G_1 = 0$ and $G_2 = 2$, hence carrying 1 derivative. And indeed, direct computation shows that
\begin{multline}\label{Jtr_grav}
    J_\mu{}^\mu \equiv -(G^{(2)})_\mu{}^\mu \approx
    \partial_\mu \mathfrak{j^\mu}\,,\\
    \mathfrak{j^\mu} = 
    \frac32 h_\mu{}^\nu \partial_\nu h_\rho{}^\rho
    -2 h_{\mu\nu} \partial_\rho h^{\nu\rho}
    -h^{\nu\rho} \partial_\nu h_{\mu\rho}
    +\frac12 h_\nu{}^\nu \partial^\rho h_{\mu\rho}
    +\frac32 h^{\nu\rho} \partial_\mu h_{\nu\rho}
    -\frac12 h_\nu{}^\nu \partial_\mu h_\rho{}^\rho
    \,,
\end{multline}
where $\approx$ means on-shell equality and $h_{\mu\nu}$ is the first-order perturbation of metric. By examining all possible quadratic local field redefinitions, one can see that $J_\mu{}^\mu$ \eqref{Jtr_grav} indeed cannot be eliminated.

It is remarkable that the rank-two field corresponding to the trace of the Einstein tensor \eqref{Jtr_grav} is affected by rank-two gauge transformations induced by the irregular cocycle $h(p_2,\pd_2) \varepsilon^{\omega\omega}_{1,0;\;1,2;\;1,0}$ (we have not presented here the full list of the irregular gauge-like cocycles). In the tensor language this corresponds to the transformation of the type
\begin{equation}\label{delta_frakj}
    \delta \mathfrak{j}^\mu = \partial_\nu \varepsilon^{\mu\nu} \Longrightarrow
    \delta J_\mu{}^\mu = \partial_\mu\partial_\nu \varepsilon^{\mu\nu}
\end{equation}
with symmetric and traceless parameter $\varepsilon^{\mu\nu}$. These transformations can be identified with the usual redefinitions of the traceless part of the metric. Indeed, since the trace part of the Einstein equations reads as $\Box (h^{(2)})_\mu{}^\mu - \partial^\mu\partial^\nu (h^{(2)})_{\mu\nu} = J_\mu{}^\mu$, shifts of the metric of the form $(h^{(2)})_{\mu\nu} \to (h^{(2)})_{\mu\nu} + \varepsilon^{\mu\nu}$, with $\varepsilon^{\mu\nu}$ as above, produce transformations of the current of the form \eqref{delta_frakj}.

In the $s=s_1=3$, $s_2=2$ theory the trace of the current is
\begin{equation}\label{Jtr_332_spinor}
    j^{\tr}_{1,1} \propto D(p,\pd) (p_1y)^2(\pd_2\yd)^2(\pd_1\pd_2)^2 \omega_{3,1}(p_1,\yd_1|Y_1) C_{0,4}(Y_2)
\end{equation}
in the spinor formalism (using definition \eqref{1-form_decomp}) and
\begin{multline}\label{Jtr_332_tensor}
    J_{\mu\nu}{}^{\nu} \approx
    \left(
    \frac{40}{3}
    (\partial_\ga \partial_\gb \phi_{\gg\gd}{}^\gd - \partial_\gb \partial_\gg \phi_{\ga\gd}{}^\gd) -
    16
    (\partial_\ga \partial_\gd \phi_{\gb\gg}{}^\gd - \partial_\gg \partial_\gd \phi_{\ga\gb}{}^\gd)
    \right)
    (\partial^\gb \partial^\gg h_\mu{}^\ga - \partial_\mu \partial^\gg h^{\ga\gb})
\end{multline}
in the tensor formalism. Here $\phi_{\ga\gb\gg}$ is the spin-3 field. We used results of \cite{Tatarenko:2024csa} to obtain \eqref{Jtr_332_spinor} and \cite{Zinoviev:2008ck} to obtain \eqref{Jtr_332_tensor}. It is easy to check that both expressions are equivalent. As for gravity, we have checked the possibility of cancelling out the trace by making a quadratic local field redefinition, and with the help of computer calculations we have seen that the trace cannot be eliminated in this case as well.

The trace \eqref{Jtr_332_spinor}, \eqref{Jtr_332_tensor} is connected with the irregular field-like cocycle $\{ \Hd(\pd_2,\pd_2) + H(p_2,p_2) \} \ \mathbf{j}^{\omega\omega}_{1,1;\;1,1;\;1,1}$ (and, possibly, other irregular cocycles; unfortunately we have not obtained the explicit expression analogous to \eqref{Jtr_grav}). One can see that this cocycle is variant under the rank-two gauge transformations generated by the irregular gauge-like cocycle $h(p_2,\pd_2) \varepsilon^{\omega\omega}_{2,1;\;1,2;\;1,0}$. In tensor language,
\begin{gather*}
    J_{\mu\nu}{}^{\nu} \approx
    \partial_\nu (
        c_1\, \eta_{\mu\rho} \Box +
        c_2\, \partial_\mu \partial_\rho
    )\, \mathfrak{j}^{\nu\rho} + \dots\,,
    \\
    \delta J_{\mu\nu}{}^{\nu} \approx
    \partial_\nu (
        c_1\, \eta_{\mu\rho} \Box +
        c_2\, \partial_\mu \partial_\rho
    )\, \delta \mathfrak{j}^{\nu\rho}
    =
    \partial_\nu\partial_\lambda (
        c_1\, \eta_{\mu\rho} \Box +
        c_2\, \partial_\mu \partial_\rho
    )\, \varepsilon^{\nu\rho\lambda}\,,
\end{gather*}
where $c_{1,2}$ are constants and $\varepsilon^{\nu\rho\lambda}$ is symmetric and traceless gauge parameter. Thus, similarly to the case of gravity, we treat such transformations as shifts of the spin-3 field with parameter $( c_1\, \eta_{\mu\rho} \Box + c_2\, \partial_\mu \partial_\rho )\, \varepsilon^{\nu\rho\lambda}$.

Let us note that the traceless parts of the currents are also affected by the considered gauge transformations in such a way that the conservation laws are gauge invariant. For example, the traceless part of the current in gravity $\tilde{J}_{\mu\nu}$ transforms as
\begin{equation*}
    \delta \tilde{J}_{\mu\nu} =
    - \Box \varepsilon_{\mu\nu}
    + \lp
        \delta^\alpha_\nu \partial_\mu \partial^\beta
        +\delta^\alpha_\mu \partial_\nu \partial^\beta
        -\frac12 g_{\mu\nu}\partial^\alpha \partial^\beta
    \rp \varepsilon_{\alpha\beta}
\end{equation*}
with $\varepsilon_{\mu\nu}$ from \eqref{delta_frakj}. The role of these transformations is somewhat analogous to that of the usual gauge transformations of the Fronsdal fields\footnote{Author thanks M.~Vasiliev for the useful discussion of this question.}. From the cohomological point of view, the traceless and  traceful parts of the Fronsdal field are different $\Sigma^{(1)}_-$-cocycles, but they are affected by the same gauge parameter. The
traceless and traceful parts of the Fronsdal field contribute to the Fronsdal equations together 
so that the gauge invariance of the equations is due to mutual cancellation of the gauge variations of the parts of the Fronsdal field.
Hence, the absence of one of the parts of the Fronsdal field is impossible since it would break the gauge symmetry. In the case of the bilinear currents, as we have seen, the trace and the traceless part of each current are generated by different $\Sigma^{(2)}_-$-cocycles, which, however are affected by the same gauge parameter and obey the conservation law only together. The gauge variations of the traceful and traceless parts of the current cancel each other out in the conservation law, making it gauge invariant. Thus in some sense the gauge symmetry glues together different $\sigma_-$-cocycles, in the both cases of rank-one (Fronsdal) fields and rank-two (bilinear currents) fields. In other words, within the symmetry approach, implying that properties of a theory are to be deduced from its symmetries, the traces of the Fronsdal fields and currents have to appear since the Fronsdal equations and the conservation laws must be gauge invariant. (Let us note that the logic of the present work is quite different: we obtain equations and gauge transformations independently, from the analysis of the $\Sigma^{(2)}_-$-cohomology. However, the afore-mentioned viewpoint is also reasonable.)

\section{Conclusion}\label{sec:conclusion}

In this paper, non-trivial primary bilinear conserved currents, built from massless fields of arbitrary integer or half-integer spins on 4d $AdS$ or Minkowski background, were analysed with the help of the $\sigma_-$-cohomology technique. A classification of the currents was worked out. It was shown that the current in the equation for a spin-$s$ field constructed from spin-$s_1$ and spin-$s_2$ field (all spins are assumed to be not less than $2$) carries $\floor{s} + \floor{s_1} + \floor{s_2} - 2\min\{s_1, s_2\}$ derivatives if
\begin{equation*}
    s \leqslant s_1 + s_2\,,\qquad
    s_1 \leqslant s + s_2\,,\qquad
    s_2 \leqslant s + s_1\,,
\end{equation*}
and $\floor{s} + \floor{s_1} + \floor{s_2}$ or $\floor{s} + \floor{s_1} + \floor{s_2} - 2\min\{s, s_1, s_2\}$ derivatives if $s > s_1 + s_2$, while in the regions $s_1 > s + s_2$ and $s_2 > s + s_1$ there are no non-trivial conserved currents.

The connection between these currents and the Lagrangian vertices, which have been classified in \cite{Bengtsson:1983pd,Bengtsson:1986kh,Metsaev:2005ar,Metsaev:2007rn,Manvelyan:2010jr,Metsaev:2018xip} was established in Section \ref{sec:currents_and_vertices}. So, some known vertices are not connected with any conserved current, while other are connected to the current directly or to its descendants. The later possibility realizes in the region $s < \min\{s_1, s_2\}$, and it generalizes the situation with the HS energy-momentum tensors, which are well-known \cite{Weinberg:1980kq,Fradkin:1987ks} to be not directly related to the Lagrangian vertices: the vertices include the higher-derivative terms. The vertices not corresponding to the currents in our classification are those that, according to \cite{Joung:2013nma}, do not deform the respective gauge transformation law, hence they do not produce non-trivial currents. A particular case of such vertices are the Born-Infeld-type vertices.

It is shown that the higher-derivative currents appearing in the region $s \geqslant s_1 + s_2$ are traceless, in agreement with \cite{Gelfond:2017wrh, Misuna:2017bjb, Tatarenko:2024csa}. The other currents can have non-zero traces, which are produced by elements of $\Sigma^{(2)}_-$-cohomology, called in Section \ref{sec:rank-two_cohomology} \emph{irregular}, and thus are not cancellable by local field redefinitions. 

Present analysis is based on a statement that bilinear conserved HS currents correspond to certain rank-two fields that belong to the $\Sigma^{(2)}_-$-cohomology (see Section \ref{sec:general_idea}). Therefore, groups of the so-called gauge-like, field-like and equation-like $\Sigma^{(2)}_-$-cohomology were found. As is discussed in Section \ref{sec:calculations}, $H(\Sigma^{(2)}_-)$ also contains elements that cannot be treated as conserved currents. An interesting result is that all such irrelevant cocycles represent non-dynamical rank-two fields (so-called off-shell fields), that are not governed by any differential equation with respect to the space-time coordinates. All rank-two fields that we identify with conserved currents are dynamical and satisfy the first-order differential equations treated as conservation laws.

The $\sigma_-$-cohomology technique used in this paper is valid for HS theory both on anti-de Sitter space and Minkowski space. Thus our results are true in both cases.

\acknowledgments{
Author is grateful to M.~Vasiliev who motivated this work and gave many valuable advices and comments. Author acknowledges A.~Tarusov and K.~Ushakov for careful reading of the draft and useful comments. Author thanks K.~Mkrtchyan for the correspondence. Author also wishes to thank O.~Gelfond and N.~Misuna for discussions and I.~Starostin for language corrections. This work was supported by Theoretical Physics and
Mathematics Advancement Foundation “BASIS” Grant No 24-1-1-9-5.
}

\begin{appendices}
\section{Technical formulae}\label{app:technical_formulae}

\paragraph{Differential forms}

Here the formulae that help to obtain expressions like \eqref{highG_eps_chain_pre_eq} are collected.

Schouten identity $(uv)(yp) + (vy)(up) + (yu)(vp) \equiv 0$ has the following simple corollary
\begin{equation}\label{Sch_id}
    (uv) F_n(y) =
        \frac{i}{n+1} [(up) (vy) - (uy) (vp)] F_n(y)\,,
\end{equation}
where $F_n(y)$ is a homogeneous polynomial in $y$ of degree $n$. With the help of this identity one can decompose any 1-form as follows:
\begin{align}
\begin{split}
    \omega_{m,n}(Y) = -&h(p_1,\pd_1) \omega_{m,n}(Y_1|Y) =\\=
        \frac{1}{(m+1)(n+1)} \big[
            &h(y,\yd) (p_1p)(\pd_1\pd) -
            h(y,\pd) (p_1p)(\pd_1\yd) -\\-
            &h(p,\yd) (p_1y)(\pd_1\pd) +
            h(p,\pd) (p_1y)(\pd_1\yd)
        \big] \omega_{m,n}(Y_1|Y)
        \equiv\\\equiv
        -\frac{1}{(m+1)(n+1)} \big[
            &h(y,\yd) \omega_{m,n}(p,\pd|Y) -
            h(y,\pd) \omega_{m,n}(p,\yd|Y) -\\-
            &h(p,\yd) \omega_{m,n}(y,\pd|Y) +
            h(p,\pd) \omega_{m,n}(y,\yd|Y)
        \big] \,.
\end{split}\label{1-form_decomp}
\end{align}
Here $\omega_{m,n}(Y_1|Y)$ is bilinear in $y_1$ and $\bar y_1$ with the labels $m$ and $n$ referring to the degrees in $y$ and $\bar y$, respectively.

Application of the identity \eqref{Sch_id} to 2- and 3-forms yields:
\begin{align}
\begin{split}\label{2-form_decomp}
    \Upsilon_{m,n}(Y) =
        -\frac12 &H(p_1,p_1)\Upsilon_{m,n}(y_1,y_1|Y)
        -\frac12 \Hd(\pd_1,\pd_1)\Upsilon_{m,n}(\yd_1,\yd_1|Y)
        =\\=
        \bigg[
            -\frac{1}{m(m+1)} &H(y,y) \Upsilon_{m,n}(p,p|Y)
            +\frac{2}{m(m+2)} H(y,p) \Upsilon_{m,n}(y,p|Y) -\\
            -\frac{1}{(m+1)(m+2)} &H(p,p) \Upsilon_{m,n}(y,y|Y)
            -\frac{1}{n(n+1)} \Hd(\yd,\yd) \Upsilon_{m,n}(\pd,\pd|Y) +\\
            +\frac{2}{n(n+2)} &\Hd(\yd,\pd) \Upsilon_{m,n}(\yd,\pd|Y)
            -\frac{1}{(n+1)(n+2)} \Hd(\pd,\pd) \Upsilon_{m,n}(\yd,\yd|Y)
        \bigg]\,;
\end{split}
\end{align}
\begin{align}
\begin{split}
    \Psi_{m,n}(Y) = -&\cH(p_1,\pd_1) \Psi_{m,n}(Y_1|Y) =\\=
        -\frac{1}{(m+1)(n+1)} \big[
            &\cH(y,\yd) \Psi_{m,n}(p,\pd|Y) -
            \cH(y,\pd) \Psi_{m,n}(p,\yd|Y) -\\-
            &\cH(p,\yd) \Psi_{m,n}(y,\pd|Y) +
            \cH(p,\pd) \Psi_{m,n}(y,\yd|Y)
        \big] \,.
\end{split}\label{3-form_decomp}
\end{align}

One can treat the formulae \eqref{1-form_decomp} -- \eqref{3-form_decomp} as definitions for $\omega_{m,n}(a,\ad|Y)$ etc. Equivalently they can be defined with the help of operators of formal differentiation with respect to $h(a,\ad)$, $H(a,b)$, etc.:
\begin{gather}
    \omega_{m,n}(a,\ad|Y) = \partial_h(a,\ad) \omega(Y)\,,\\
    J_{m,n}(a,b|Y) = \partial_H(a,b) J(Y)\,,\qquad
    J_{m,n}(\ad,\bd|Y) = \partial_\Hd (\ad,\bd) J(Y)\,,\\
    \Psi_{m,n}(a,\ad|Y) = \partial_\cH (a,\ad) \Psi(Y)\,.
\end{gather}

\paragraph{Lorentz derivative commutation relations}

As is well-known, in $AdS_4$ space with cosmological constant $\Lambda$ a commutator of Lorentz covariant derivatives is proportional to $\Lambda$. This also can be derived from the nilpotency of the $AdS$ background derivative \eqref{D}. Decomposing the equation $D_\Omega^2 f(Y) = 0$ with some zero-form $f(Y)$ into a basis of two-forms, one can obtain commutation relations for the Lorentz derivatives in the spinor language
\begin{align}
\begin{split}
    (\pd_1\pd_2)(p_1y_3)(p_2y_3)D(Y_1)D(Y_2) f(Y) = 2 \Lambda i (yy_3)(py_3) f(Y) \,,\\
    (p_1p_2)(\pd_1\yd_3)(\pd_2\yd_3)D(Y_1)D(Y_2) f(Y) = 2 \Lambda i (\yd\yd_3)(\pd\yd_3) f(Y) \,.\\
\end{split}\label{D_comm}
\end{align}
Here $D(Y_n) \equiv D(y_n,\yd_n)$, see definition \eqref{D_expansion}; $y_n$ and $\yd_n$, $n=1,2,3$, are auxiliary spinor variables. In particular, formulae \eqref{D_comm} lead to the following formulae used in the proof of Proposition \ref{prop:5}:
\begin{align}\label{D_comm2}
    &D(y,\pd)D(y,\yd) f(Y) = D(y,\yd)D(y,\pd) f(Y)\,,
        &&D(p,\yd)D(y,\yd) f(Y) = D(y,\yd)D(p,\yd) f(Y)\,,\\
    &D(y,\pd)D(p,\pd) f(Y) = D(p,\pd)D(y,\pd) f(Y)\,,
        &&D(p,\yd)D(p,\pd) f(Y) = D(p,\pd)D(p,\yd) f(Y)\,.
\end{align}

Another set of useful consequences of \eqref{D_comm} is
\begin{align}\label{D_comm_res}
\begin{split}
    D(p,\yd)D(y,\pd) f(Y)
    =& \Big[ D(y,\yd)D(p,\pd)
        - (\cN+1)\bar{\cN} \Box
        + \Lambda (\cN+1)\bar{\cN}(\bar{\cN}+2) \Big]f(Y)\,;\\
    D(y,\pd)D(p,\yd) f(Y)
    =& \Big[ D(y,\yd)D(p,\pd)
        - (\bar{\cN}+1)\cN \Box
        + \Lambda (\bar{\cN}+1)\cN(\cN+2) \Big]f(Y)\,;\\
    D(p,\pd)D(y,\yd) f(Y)
    =& \Big[ D(y,\yd)D(p,\pd)
        + (\bar{\cN}+\cN+2) \Box
        +\\&\hspace{1cm}+ \Lambda \big(\cN(\cN+2)(\bar{\cN}+1)+\bar{\cN}(\bar{\cN}+2)(\cN+1)\big) \Big]f(Y)\,,
\end{split}
\end{align}
where $\Box := -\frac12 (p_1p_2)(\pd_1\pd_2)D(Y_1)D(Y_2)$. Through the paper $\Lambda$ is set to $-1$.

\section{Calculation details}\label{app:calculation_details}

\subsection{High $G_1$ region}\label{app:highG}

In this subsection the base fields for the gauge-like, the field-like and the equation-like $H(\Sigma^{(2)}_-)$ are calculated in ``high $G_1$ region", i.e., under assumption that $s_2 > 2$ and $2s_1+4 \leqslant G_1$. Without loss of generality we consider only those base fields for which the value of $\cN_2$ is not less than the value of $\bar{\cN}_2$. The dependence of the fields on the spinor variables $Y_1$ and $Y_2$ is implicit in this subsection.

\paragraph{Gauge-like cohomology}

In accordance with Lemma \ref{lemma:calc}, the base field $\cE|_{G_1,G_2}$ belongs to the degenerate case of $H^{0-\text{x}}(\sigma_2)$ which is equivalent to $H^{0-\text{x}}(\Sigma^{(1)})$ from Table \ref{tab:cohom1}. Thus, in terms of definition \eqref{f_abc}, the relevant components of the rank-two field are
\begin{subequations}\label{highG_init_Eps}
\begin{align}
    &\cE\big|_{G_1,0} =
        \varepsilon_{a,\ad;\;s_2-1-c,s_2-1-\cd;\;c,\cd}
            &&\text{if $s_2$ is integer}\,;\\
    &\cE\big|_{G_1,1} =
        (\varepsilon_+)_{a,\ad;\;s_2-1/2-c,s_2-3/2-\cd;\;c,\cd} + \dots
            &&\text{if $s_2$ is half-integer}\,.
\end{align}
\end{subequations}
Here ellipsis denotes the part of $\cE\big|_{G_1,1}$ with $\cN_2 = \bar{\cN}_2-1$, which is analogous to the considered one. Notice that the point 2{\it (a)} of Lemma \ref{lemma:calc} is trivial in this case because the form degree of the base field is zero.

According to the point 2{\it (b)} of Lemma \ref{lemma:calc} the base field must obey \eqref{chain}. This can be taken into account by using the 3rd point of Lemma \ref{lemma:calc}. Since all non-trivial elements of $H^{1-0}(\sigma_2)$ have $G_2 = 0$ for integer $s_2$ or $G_2 = 1$ for half-integer $s_2$ (see Table \ref{tab:cohom1}), one has to check the consistency of
\begin{subequations}
\begin{align}
    &\sigma_1 \cE\big|_{G_1,0} = -\sigma_2 \cE\big|_{G_1-2,2}
        &&\text{if $s_2$ is integer}\,;\\
    &\sigma_1 \cE\big|_{G_1,1} = -\sigma_2 \cE\big|_{G_1-2,3}
        &&\text{if $s_2$ is half-integer}\,,
\end{align}
\end{subequations}
where $\sigma_{1,2}$ are $Y_{1,2}$ parts of $\Sigma^{(2)}_-$ \eqref{Sigma2_-}, correspondingly. Substitution of $\sigma_{1,2}$ yields
\begin{subequations}
\begin{align}
    &h(p_1,\pd_1) \cE\big|_{G_1,0} = h(p_2,\yd_2) \cE\big|_{G_1-2,2}
        &&\text{if $s_2$ is integer}\,;\\
    &h(p_1,\pd_1) \cE\big|_{G_1,1} = h(p_2,\yd_2) \cE\big|_{G_1-2,3}
        &&\text{if $s_2$ is half-integer}\,.
\end{align}
\end{subequations}

The decomposition of these equations into a basis of 1-forms (see \eqref{basic_frame_forms}, \eqref{indexless_notation} and Appendix \ref{app:technical_formulae}) reads:
\begin{subequations}\label{highG_eps_chain_pre_eq}
\begin{multline}
    h(y_2,\yd_2) (p_1p_2)(\pd_1\pd_2) \cE\big|_{G_1,0} + 
    h(p_2,\pd_2) (p_1y_2)(\pd_1\yd_2) \cE\big|_{G_1,0} -\\- 
    h(p_2,\yd_2) \big[(p_1y_2)(\pd_1\pd_2) \cE\big|_{G_1,0}
        + s_2^2 \delta_{\cN_2,\bar{\cN}_2+2}\cE\big|_{G_1-2,2}\big] -\\
    h(y_2,\pd_2) \big[(p_1p_2)(\pd_1\yd_2) \cE\big|_{G_1,0}
        + s_2^2 \delta_{\cN_2+2,\bar{\cN}_2}\cE\big|_{G_1-2,2}\big] = 0\,,
\end{multline}
\begin{multline}
    h(y_2,\yd_2) (p_1p_2)(\pd_1\pd_2) \delta_{\cN_2,\bar{\cN}_2+1}\cE\big|_{G_1,1} + 
    h(p_2,\pd_2) (p_1y_2)(\pd_1\yd_2) \delta_{\cN_2,\bar{\cN}_2+1}\cE\big|_{G_1,1} -\\- 
    h(p_2,\yd_2) \big[(p_1y_2)(\pd_1\pd_2) \delta_{\cN_2,\bar{\cN}_2+1}\cE\big|_{G_1,1}
        + s_2^2 \delta_{\cN_2,\bar{\cN}_2+2}\cE\big|_{G_1-2,3}\big] -\\
    h(y_2,\pd_2) (p_1p_2)(\pd_1\yd_2) \delta_{\cN_2,\bar{\cN}_2+1}\cE\big|_{G_1,1} = 0\,.
\end{multline}
\end{subequations}

One can easily check that it gives the following conditions:
\begin{align}\label{highG_eps_chain_eq}
&\begin{split}
    (p_1y_2)(\pd_1\yd_2)\varepsilon_{a,\ad;\;s_2-1-c,s_2-1-\cd;\;c,\cd} = 0\,,\\
    (p_1p_2)(\pd_1\pd_2)\varepsilon_{a,\ad;\;s_2-1-c,s_2-1-\cd;\;c,\cd} = 0\,;
\end{split}
&&\begin{split}
    (p_1y_2)(\pd_1\yd_2)(\varepsilon_+)_{a,\ad;\;s_2-1/2-c,s_2-3/2-\cd;\;c,\cd} = 0\,,\\
    (p_1p_2)(\varepsilon_+)_{a,\ad;\;s_2-1/2-c,s_2-3/2-\cd;\;c,\cd} = 0\,.
\end{split}
\end{align}
Combining formulae \eqref{highG_init_Eps} and \eqref{highG_eps_chain_eq}, one obtains that by virtue of Lemma \ref{lemma:calc} in the case under consideration the full list of the base fields is as presented in Table \ref{tab:highG_gauge}.
\begin{table}[H]
    \renewcommand{\arraystretch}{1.5}
    \centering
    \begin{tabular}{|c|c|}
        \hline
        \multirow{1}{7em}{Integer $s_2$}
            &$\varepsilon_{2s_1+\cd,0;\;s_2-1,s_2-1-\cd;\;0,\cd}$\\\hline
        \multirow{1}{7em}{Half-integer $s_2$}
            &$(\varepsilon_+)_{2s_1+\cd,0;\;s_2-1/2,s_2-3/2-\cd;\;0,\cd}$\\\hline
    \end{tabular}
    \caption{The base fields for gauge-like $H(\Sigma^{(2)}_-)$ in high $G_1$ region (up to complex conjugation).}
    \label{tab:highG_gauge}
\end{table}

\paragraph{Field-like cohomology}

In this case the base field $J\big|_{G_1,G_2} \in H^{1-0}(\sigma_2) \cong H^{1-0}(\Sigma^{(1)}_-)$ which is presented in Table \ref{tab:cohom1}. Consequently, $G_2 = 0$ for integer $s_2$ and $G_2 = 1$ for half-integer $s_2$:
\begin{subequations}\label{highG_j_init}
\begin{align}
    J\big|_{G_1,0} =
        &\delta_{\cN_2,\bar{\cN}_2} \big[
            h(p_2,\pd_2) \mathbf{j} +
            h(y_2,\yd_2) \mathbf{j}^{\tr}
        \big]\,,
            &&\text{if $s_2$ is integer}\,,\\
    J\big|_{G_1,1} =
        &\delta_{\cN_2,\bar{\cN}_2+1} \big[
            h(p_2,\pd_2) \mathbf{j}_+ +
            h(y_2,\yd_2) \mathbf{j}_+^{\tr} +
            h(y_2,\pd_2) \mathbf{j}_+^{\gamma\tr}
        \big] +\nonumber\\+
        &\delta_{\cN_2+1,\bar{\cN}_2} \big[\dots\big]\,,
            &&\text{if $s_2$ is half-integer}\,.
\end{align}
\end{subequations}

According to Lemma \ref{lemma:calc}, one has to factorize $J\big|_{G_1,G_2}$ by the equivalence relation defined in this lemma. Since $G_2 = 0,1$ the afore-mentioned equivalence relation becomes just $A \sim A + \sigma_1 \xi\big|_{G_1+2,G_2}$ with no restrictions on $\xi$ because $\sigma_2 \xi\big|_{G_1+2,G_2} = 0$ at $G_2 = 0,1$ by definition. Then the decomposition of the formula $J\big|_{G_1,0} \sim J\big|_{G_1,0} + \sigma_1 \xi\big|_{G_1+2,0} + \sigma_2 \epsilon\big|_{G_1,2}$ for integer $s_2$ and its analogue for half-integer $s_2$ into the basis of 1-frame forms reads as
\begin{subequations}\label{highG_j_fact_eqn}
\begin{align}
\begin{split}\label{highG_j_fact_eqn_bos}
    J\big|_{G_1,0} \sim
    \delta_{\cN_2,\bar{\cN}_2} \big\{
        &h(p_2,\pd_2)
            [\# (p_1y_2)(\pd_1\yd_2)\xi + \mathbf{j}] +
        h(p_2,\yd_2)
            [i \epsilon_+ - \# (p_1y_2)(\pd_1\pd_2)\xi] +\\+
        &h(y_2,\pd_2)
            [i \epsilon_- - \# (p_1p_2)(\pd_1\yd_2)\xi] +
        h(y_2,\yd_2)
            [\# (p_1p_2)(\pd_1\pd_2)\xi + \mathbf{j}^{\tr}]
    \big\}\,,
\end{split}\\
\begin{split}\label{highG_j_fact_eqn_ferm}
    J\big|_{G_1,1} \sim
    \delta_{\cN_2,\bar{\cN}_2+1} \big\{
        &h(p_2,\pd_2)
            [\# (p_1y_2)(\pd_1\yd_2)\xi_+ + \mathbf{j}_+] +
        h(p_2,\yd_2)
            [i \epsilon_+ - \# (p_1y_2)(\pd_1\pd_2)\xi_+] +\\+
        &h(y_2,\pd_2)
            [- \# (p_1p_2)(\pd_1\yd_2)\xi_+ + \mathbf{j}^{\gamma\tr}_+] +
        h(y_2,\yd_2)
            [\# (p_1p_2)(\pd_1\pd_2)\xi_+ + \mathbf{j}^{\tr}_+]
    \big\} +\\+
    \delta_{\cN_2+1,\bar{\cN}_2}\{&\dots\}\,.
\end{split}
\end{align}
\end{subequations}
Here $\# = \frac{i}{\floor{s_2}\ceil{s_2}}$; $\xi_+ \equiv \delta_{\cN_2,\bar{\cN}_2 + 1} \xi\big|_{G_1+2,1}$; $\epsilon_\pm \equiv \delta_{\cN_2,\bar{\cN}_2 \pm 2} \epsilon\big|_{G_1,2}$ in \eqref{highG_j_fact_eqn_bos} and $\epsilon_+ \equiv \delta_{\cN_2,\bar{\cN}_2 + 3} \epsilon\big|_{G_1,2}$ in \eqref{highG_j_fact_eqn_ferm}. One can check that using $\xi$ and $\epsilon$ all the components of $J\big|_{G_1,G_2}$ can be eliminated, except for
\begin{gather}
    \mathbf{j}_{a,\ad;\;s_2,0;\;0,s_2}\,,\qquad
    \mathbf{j}_{a,\ad;\;0,s_2;\;s_2,0}\,,\qquad
    \mathbf{j}_{a,\ad;\;b,\bd;\;c>0,\cd>0}\,,\nonumber\\
    (\mathbf{j}_+)_{a,\ad;\;s_2+1/2,0;\;0,s_2-1/2}\,,\qquad
    (\mathbf{j}_+)_{a,\ad;\;b,\bd;\;c>0,\cd}\,,\qquad
    (\mathbf{j}^{\gamma\tr}_+)_{a,\ad;\;b,\bd;\;c,\cd>0}\,.\label{highG_j_factorized}
\end{gather}

Next, the 3rd point of Lemma \ref{lemma:calc} should be applied. The corresponding subspace of $H(\sigma_2)$ is $H^{2-1}(\sigma_2)$ equivalent to $H^{2-1}(\Sigma^{(1)}_-)$ of Table \ref{tab:cohom1}; its non-trivial elements have $\cG_2$ value 1 for half-integer $s_2$ and 2 for integer $s_2$. Therefore, for half-integer $s_2$ the consistency of
\begin{equation}
    \sigma_1 J\big|_{G_1,1} = -\sigma_2 J\big|_{G_1-2,3}
\end{equation}
with some $J\big|_{G_1-2,3}$ should be checked, while for integer $s_2$ one has to find $J\big|_{G_1-2,2}$ from the equation
\begin{equation}
    \sigma_1 J\big|_{G_1,0} = -\sigma_2 J\big|_{G_1-2,2}
\end{equation}
and check the consistency of
\begin{equation}
    \sigma_1 J\big|_{G_1-2,2} = -\sigma_2 J\big|_{G_1-4,4}\,.
\end{equation}
The resulting restrictions on the base fields can be obtained in the same manner as those for the gauge-like cohomology and read as
\begin{align}\label{highG_j_step_eqn}
\begin{split}
    &(p_1p_2)(p_1y_2)(\pd_1\yd_2)(\pd_1\pd_2) \mathbf{j} = 0\,,\\
    &(p_1p_2)^2(\pd_1\pd_2)^2 \mathbf{j} = 0\,;
\end{split}
\begin{split}
    &(p_1y_2)(\pd_1\pd_2) \mathbf{j}_+^{\gamma\tr} =
        \frac{(2s_2-1)^2}{8s_2} (p_1p_2)(\pd_1\pd_2) \mathbf{j}_+\,,\\
    &(p_1y_2)(\pd_1\yd_2) \mathbf{j}_+^{\gamma\tr} = - (p_1p_2)(\pd_1\yd_2) \mathbf{j}_+\,,\\
    &(p_1p_2)(\pd_1\pd_2) \mathbf{j}_+^{\gamma\tr} = 0\,.
\end{split}
\end{align}

The final answer is the result of imposing \eqref{highG_j_step_eqn} on \eqref{highG_j_factorized}. It is presented in Table \ref{tab:highG_field}.
\begin{table}[H]
    \renewcommand{\arraystretch}{1.5}
    \centering
    \begin{tabular}{|c|c|}
        \hline
        \multirow{1}{7em}{Integer $s_2$}
            &$h(p_2,\pd_2)\mathbf{j}_{a,\ad;\;s_2,0;\;0,s_2}\,,\qquad
                |a-\ad-s_2|=2s_1$\\
            &$h(p_2,\pd_2)\mathbf{j}_{2s_1-1+\cd,0;\;s_2-1,s_2-\cd;\;1,\cd}$\\\hline
        \multirow{1}{7em}{Half-integer $s_2$}
            &$h(p_2,\pd_2)(\mathbf{j}_+)_{a,\ad;\;s_2+1/2,0;\;0,s_2-1/2}\,,\qquad
                |a-\ad-s_2|=2s_1$\\
            &$\big[h(p_2,\pd_2) + \frac{s_2-1/2}{2s_2(2s_1 + \cd)} h(y_2,\pd_2)(y_1p_2)(p_1p_2)\big](\mathbf{j}_+)_{2s_1-1+\cd,0;\;s_2-1/2,s_2-1/2-\cd;\;1,\cd}$\\\hline
    \end{tabular}
    \caption{The base fields for field-like $H(\Sigma^{(2)}_-)$ in high $G_1$ region (up to complex conjugation).}
    \label{tab:highG_field}
\end{table}

\paragraph{Equation-like cohomology}

In this case the base field $\Psi\big|_{G_1,G_2} \in H^{2-1}(\sigma_2)$ corresponding to $H^{2-1}(\Sigma^{(1)}_-)$ presented in Table \ref{tab:cohom1}. If $s_2$ is half-integer, then from Table \ref{tab:cohom1} it follows that
\begin{align}\label{highG_init_Psi_ferm}
\begin{split}
    \Psi\big|_{G_1,G_2} \equiv \Psi\big|_{G_1,1} =
    &\delta_{\cN_2,\bar{\cN}_2+1}\big\{
        \Hd(\pd_2,\pd_2)\mathbf{\psi}_+ +
        H(y_2,y_2)\mathbf{\psi}^{\tr}_+ +
        H(y_2,p_2)\mathbf{\psi}^{\gamma\tr}_+
    \big\}+\\+
    &\delta_{\cN_2+1,\bar{\cN}_2}\big\{\dots\big\}\,.
\end{split}
\end{align}

If $s_2$ is integer, $G_2 = 2$ and
\begin{align}
\begin{split}\label{highG_init_Psi_bos}
    \Psi\big|_{G_1,2} =
    \delta_{\cN_2,\bar{\cN}_2+2}\big\{
        \Hd(\pd_2,\pd_2)\mathbf{\psi} +
        H(y_2,y_2)\mathbf{\psi}^{\tr}
    \big\}+
    \delta_{\cN_2+2,\bar{\cN}_2}\big\{
        H(p_2,p_2)\mathbf{\psi} +
        \Hd(\yd_2,\yd_2)\mathbf{\psi}^{\tr}
    \big\}\,.
\end{split}
\end{align}

The equivalence relation of the 2nd point of Lemma \ref{lemma:calc} reads ($\xi_+(u,\bar{v})$ and $\epsilon_+(u,\bar{v})$ are coefficients of the decomposition of $\xi_+$ and $\epsilon_+$ into 1-frame forms -- see formula \eqref{1-form_decomp})
\begin{align}\label{highG_Psi_fact_eqn}
\begin{split}
    \pushleft{\Psi\big|_{G_1,2-2\{s_2\}} \sim}\\\sim
    \delta_{\cN_2,\bar{\cN}_2+2-2\{s_2\}}\big\{
        &H(p_2,p_2)
            [\#_{11} \epsilon_+(y_2,\yd_2) +
            \#_{12} (p_1y_2)^2 \xi_+(y_1,\pd_1) + \#_{13} (p_1y_2)(y_1y_2) \xi_+(p_1,\pd_1)] +\\+
        &H(p_2,y_2)
            [\#_{21} \epsilon_+(p_2,\yd_2) +
            \#_{22} (p_1y_2)(p_1p_2) \xi_+(y_1,\pd_1) + \#_{23} \xi_+(p_1,\pd_1) +
            \delta_{\{s_2\},\frac12} \psi^{\gamma\tr}_+] +\\+
        &H(y_2,y_2)
            [\#_{31} (p_1p_2)^2 \xi_+(y_1,\pd_1) + \#_{32} (p_1p_2)(y_1p_2) \xi_+(p_1,\pd_1) + 
            \delta_{\{s_2\},0} \psi^{\tr} + \delta_{\{s_2\},\frac12} \psi^{\tr}_+] +\\+
        &\Hd(\pd_2,\pd_2)
            [\#_{41} (\pd_1\yd_2)^2 \xi_+(p_1,\yd_1) + \#_{42} (\pd_1\yd_2)(\yd_1\yd_2) \xi_+(p_1,\pd_1) +
             \delta_{\{s_2\},0} \psi + \delta_{\{s_2\},\frac12} \psi_+] +\\+
        &\Hd(\pd_2,\yd_2)
            [\#_{51} \epsilon_+(p_2,\yd_2) +
            \#_{52} (\pd_1\yd_2)(\pd_1\pd_2) \xi_+(p_1,\yd_1) + \#_{53} \xi_+(p_1,\pd_1)] +\\+
        &\Hd(\yd_2,\yd_2)
            [\#_{61} \epsilon_+(p_2,\pd_2) +
            \#_{62} (\pd_1\pd_2)^2 \xi_+(p_1,\yd_1) + \#_{63} (\pd_1\pd_2)(\yd_1\pd_2) \xi_+(p_1,\pd_1)]
    \big\} +\\+
    \delta_{\cN_2+2-2\{s_2\},\bar{\cN}_2}\big\{
        &H(p_2,p_2)
            [\#_{41} (p_1y_2)^2 \xi_-(y_1,\pd_1) + \#_{42} (p_1y_2)(y_1y_2) \xi_-(p_1,\pd_1) + 
            \delta_{\{s_2\},0} \psi + \delta_{\{s_2\},\frac12} \psi_-] +\\+
        &H(p_2,y_2)
            [\#_{51} \epsilon_-(y_2,\pd_2) +
            \#_{52} (p_1y_2)(p_1p_2) \xi_-(y_1,\pd_1) + \#_{53} \xi_-(p_1,\pd_1)] +\\+
        &H(y_2,y_2)
            [\#_{61} \epsilon_-(p_2,\pd_2) +
            \#_{62} (p_1p_2)^2 \xi_-(y_1,\pd_1) + \#_{63} (p_1p_2)(y_1p_2) \xi_-(p_1,\pd_1)] +\\+
        &\Hd(\pd_2,\pd_2)
            [\#_{11} \epsilon_-(y_2,\yd_2) +
            \#_{12} (\pd_1\yd_2)^2 \xi_-(p_1,\yd_1) + \#_{13} (\pd_1\yd_2)(\yd_1\yd_2) \xi_-(p_1,\pd_1)] +\\+
        &\Hd(\pd_2,\yd_2)
            [\#_{21} \epsilon_-(y_2,\pd_2) +
            \#_{22} (\pd_1\yd_2)(\pd_1\pd_2) \xi_-(p_1,\yd_1) + \#_{23} \xi_-(p_1,\pd_1) + 
            \delta_{\{s_2\},\frac12} \psi^{\gamma\tr}_-] +\\+
        &\Hd(\yd_2,\yd_2)
            [\#_{31} (\pd_1\pd_2)^2 \xi_-(p_1,\yd_1) + \#_{32} (\pd_1\pd_2)(\yd_1\pd_2) \xi_-(p_1,\pd_1) +
            \delta_{\{s_2\},0} \psi^{\tr} + \delta_{\{s_2\},\frac12} \psi^{\tr}_-]
    \big\}\,.
\end{split}
\end{align}

Here
\begin{gather*}
    \#_{11} = -\#_{21}=\frac{1}{\floor{s_2}+2}\,,\qquad
        \#_{12} = -\#_{13} = \frac{1}{(\cN_1+2)(\floor{s_2}+1)(\floor{s_2}+2)}\,,\\
    \#_{22} = -\frac{2}{(\cN_1+2)\floor{s_2}(\floor{s_2}+2)}\,,\qquad
        \#_{23} = \frac{\cN_1\floor{s_2}+(y_1y_2)(p_1p_2)}{(\cN_1+2)\floor{s_2}(\floor{s_2}+2)}\,,\\
    \#_{31} = -\#_{32} = \frac{1}{(\cN_1+2)\floor{s_2}(\floor{s_2}+1)}\,,\qquad
        \#_{41} = -\#_{42} = \frac{1}{(\bar{\cN}_1+2)\ceil{s_2}(\ceil{s_2}-1)}\,,\\
    \#_{51} = -\#_{61} = \frac{1}{\bar{\cN}_1-2}\,,\qquad
        \#_{52} = -\frac{2}{(\bar{\cN}_1+2)\ceil{s_2}(\ceil{s_2}-2)}\,,\\
    \#_{53} = \frac{\ceil{s_2}(\bar{\cN}_1+2) + (\yd_1\yd_2)(\pd_1\pd_2)}{(\bar{\cN}_1+2)\ceil{s_2}(\ceil{s_2}-2)}\,,\qquad
        \#_{62} = -\#_{63} = \frac{1}{(\bar{\cN}_1+2)(\ceil{s_2}-1)(\ceil{s_2}-2)}\,.
\end{gather*}

According to Lemma \ref{lemma:calc}, in the case of integer $s_2$, the equation $\sigma_2 (\xi_+ + \xi_-) = \sigma_1 \xi\big|_{G_1+4,0}$ must be consistent with some $\xi\big|_{G_1+4,0}\,$. This yields
\begin{align}\label{highG_Psi_xi_cond}
\begin{split}
    (p_1p_2)^2 \xi_+(y_1,\pd_1) - (p_1p_2)(y_1p_2) \xi_+(p_1,\pd_1) =
        (\pd_1\pd_2)^2 \xi_-(p_1,\yd_1) - (\pd_1\pd_2)(\yd_1\pd_2) \xi_-(p_1,\pd_1)\,,\\
    (\pd_1\yd_2)^2 \xi_+(p_1,\yd_1) - (\pd_1\yd_2)(\yd_1\yd_2) \xi_+(p_1,\pd_1)
        = (p_1y_2)^2 \xi_-(y_1,\pd_1) - (p_1y_2)(y_1y_2) \xi_-(p_1,\pd_1)\,.
\end{split}
\end{align}

To use the 3rd point of Lemma \ref{lemma:calc} one has to know $H^{3-2}(\sigma_2)$.
As follows from Table \ref{tab:cohom1_3-2}, for integer $s_2$, elements of $H^{3-2}(\sigma_2)$ are non-trivial at $G_2 = 2$, and for half-integer $s_2$ -- at $G_2 = 1$. Then, the resulting restrictions are
\begin{gather}\label{highG_Psi_step_eqn}
    (s_2-1) (p_1p_2)(\pd_1\pd_2) \psi + (s_2+1) (p_1y_2)(\pd_1\yd_2) \psi^{\tr} = 0\,,\\
    (2s_2-1) (p_1p_2)(\pd_1\pd_2) \psi_+ + (2s_2+3) (p_1y_2)(\pd_1\yd_2) \psi^{\tr}_+
        - (s_2+3/2) (p_1p_2)(\pd_1\yd_2) \psi^{\gamma\tr}_+= 0\,.\nonumber
\end{gather}

The final result obtained from \eqref{highG_Psi_fact_eqn}, \eqref{highG_Psi_xi_cond} and \eqref{highG_Psi_step_eqn} is presented in Table \ref{tab:highG_eqn}.
\begin{table}[H]
    \renewcommand{\arraystretch}{1.5}
    \centering
    \begin{tabular}{|c|c|}
        \hline
        \multirow{1}{7em}{Integer $s_2$}
            &$[\Hd(\pd_2,\pd_2) + H(p_2,p_2)] \psi_{a,\ad;\;s_2,0;\;0,s_2}\,,\qquad
                |a-\ad-s_2|=2s_1$\\\hline
        \multirow{1}{7em}{Half-integer $s_2$}
            &$\Hd(\pd_2,\pd_2)(\psi_+)_{a,\ad;\;s_2-1/2,0;\;0,s_2+1/2}\,,\qquad
                |a-\ad-s_2|=2s_1$\\\hline
    \end{tabular}
    \caption{The base fields for equation-like $H(\Sigma^{(2)}_-)$ in high $G_1$ region (up to complex conjugation).}
    \label{tab:highG_eqn}
\end{table}

\subsection{Low $G_1$ region}\label{app:lowG}

In this subsection, we find the base fields for the gauge-like, the field-like and the equation-like $\Sigma^{(2)}_-$-cohomology, assuming that $s_2 > 2$ and $2 \leqslant G_1 \leqslant 2s_1-4$. The method used here is identical to that of Appendix \ref{app:highG}, so we will give less comments. In Appendix \ref{app:highG} value of $\cN_2$ was assumed to be greater than the value of $\bar{\cN}_2$, while here we for technical reasons (without loss of generality) suppose that $\cN_1 \geqslant \bar{\cN}_1$. As in Appendix \ref{app:highG}, argument $(Y_1; Y_2)$ is implicit here. 

\paragraph{Gauge-like cohomology}

The base fields for the gauge-like $H(\Sigma^{(2)}_-)$ are elements of $H^{1-0}(\sigma_2)$ as for the field-like $H(\Sigma^{(2)}_-)$ in high $G_1$ region. The equivalence relation from Lemma \ref{lemma:calc} reads
\begin{align}\label{lowG_eps_fact_eqn_bos}
\begin{split}
    \cE\big|_{G_1,0} \sim
    \delta_{\cN_2,\bar{\cN}_2} \big\{
        &h(p_2,\pd_2)
            [-\#(p_1y_2)(\yd_1\yd_2)\xi + \varepsilon] +
        h(p_2,\yd_2)
            [i \epsilon_+ + \#(p_1y_2)(\yd_1\pd_2)\xi] +\\+
        &h(y_2,\pd_2)
            [i \epsilon_- + \#(p_1p_2)(\yd_1\yd_2)\xi] +
        h(y_2,\yd_2)
            [-\#(p_1p_2)(\yd_1\pd_2)\xi + \varepsilon^{\tr}]
    \big\}
\end{split}
\end{align}
for integer $s_2$ and
\begin{align}\label{lowG_eps_fact_eqn_ferm}
\begin{split}
    \cE\big|_{G_1,1} \sim
    \delta_{\cN_2,\bar{\cN}_2+1} \big\{
        &h(p_2,\pd_2)
            [-\#(p_1y_2)(\yd_1\yd_2)\xi_+ + \varepsilon_+] +
        h(p_2,\yd_2)
            [i \epsilon_+ + \#(p_1y_2)(\yd_1\pd_2)\xi_+] +\\+
        &h(y_2,\pd_2)
            [\#(p_1p_2)(\yd_1\yd_2)\xi_+ + \varepsilon^{\gamma\tr}_+] +
        h(y_2,\yd_2)
            [-\#(p_1p_2)(\yd_1\pd_2)\xi_+ + \varepsilon^{\tr}_+]
    \big\}
    +\\+
    \delta_{\cN_2+1,\bar{\cN}_2} \big\{
        &h(p_2,\pd_2)
            [\#(p_1y_2)(\yd_1\yd_2)\xi_- + \varepsilon_-] +
        h(p_2,\yd_2)
            [\#(p_1y_2)(\yd_1\pd_2)\xi_- + \varepsilon^{\gamma\tr}_-] +\\+
        &h(y_2,\pd_2)
            [i \epsilon_- + \#(p_1p_2)(\yd_1\yd_2)\xi_-] +
        h(y_2,\yd_2)
            [\#(p_1p_2)(\yd_1\pd_2)\xi_- + \varepsilon^{\tr}_-]
    \big\}
\end{split}
\end{align}
for half-integer $s_2$. Here $\# = \frac{i}{\floor{s_2}\ceil{s_2}}$; $\xi$ and $\xi_\pm$ are unrestricted.

The conditions on the base fields from the 3rd point of Lemma \ref{lemma:calc} are
\begin{subequations}\label{lowG_eps_step_eqn}
\begin{align}
\begin{split}
    &(p_1p_2)(p_1y_2)(\yd_1\yd_2)(\yd_1\pd_2) \varepsilon =
        (p_1y_2)^2(\yd_1\yd_2)^2 \varepsilon^{\tr}\,,\\
    &(p_1p_2)^2(\yd_1\pd_2)^2 \varepsilon =
        (p_1p_2)(p_1y_2)(\yd_1\yd_2)(\yd_1\pd_2) \varepsilon^{\tr}\,;
\end{split}\\
\begin{split}
    &(p_1y_2)(\yd_1\pd_2) \varepsilon_+^{\gamma\tr} =
        \frac{\floor{s_2}^2}{2s_2} (p_1p_2)(\yd_1\pd_2) \varepsilon_+
        -\frac{\ceil{s_2}^2}{2s_2} (p_1y_2)(\yd_1\yd_2) \varepsilon_+^{\tr}\,,\\
    &(p_1y_2)(\yd_1\yd_2) \varepsilon_+^{\gamma\tr} = - (p_1p_2)(\yd_1\yd_2) \varepsilon_+\,,\\
    &(p_1p_2)(\yd_1\pd_2) \varepsilon_+^{\gamma\tr} = - (p_1p_2)(\yd_1\yd_2) \varepsilon_+^{\tr}\,;
\end{split}\\
\begin{split}
    &(p_1p_2)(\yd_1\yd_2) \varepsilon_-^{\gamma\tr} =
        \frac{\floor{s_2}^2}{2s_2} (p_1p_2)(\yd_1\pd_2) \varepsilon_-
        -\frac{\ceil{s_2}^2}{2s_2} (p_1y_2)(\yd_1\yd_2) \varepsilon_-^{\tr}\,,\\
    &(p_1y_2)(\yd_1\yd_2) \varepsilon_-^{\gamma\tr} = - (p_1y_2)(\yd_1\pd_2) \varepsilon_-\,,\\
    &(p_1p_2)(\yd_1\pd_2) \varepsilon_-^{\gamma\tr} = - (p_1y_2)(\yd_1\pd_2) \varepsilon_-^{\tr}\,.
\end{split}
\end{align}
\end{subequations}

Under these conditions \eqref{lowG_eps_fact_eqn_bos} and \eqref{lowG_eps_fact_eqn_ferm} leave non-trivial only fields in Table \ref{tab:lowG_gauge}.
\begin{table}[H]
    \renewcommand{\arraystretch}{1.5}
    \centering
    \begin{tabular}{|c|c|}
        \hline
        \multirow{1}{7em}{Integer $s_2$}
            &$h(p_2,\pd_2)\varepsilon_{a,\ad;\;s_2,s_2;\;0,0}\,,\qquad
                a+\ad=2s_1-2$\\
            &$h(p_2,\pd_2)\varepsilon_{a,\ad;\;0,0;\;s_2,s_2}\,,\qquad
                a+\ad=2s_1-2s_2-2$\\\hline
        \multirow{1}{7em}{Half-integer $s_2$}
            &$h(p_2,\pd_2)(\varepsilon_+)_{a,\ad;\;s_2+1/2,s_2-1/2;\;0,0}\,,\qquad
                a+\ad=2s_1-2$\\
            &$h(p_2,\pd_2)(\varepsilon_-)_{a,\ad;\;0,0;\;s_2-1/2,s_2+1/2}\,,\qquad
                a+\ad=2s_1-2s_2-2$\\\hline
    \end{tabular}
    \caption{The base fields for gauge-like $H(\Sigma^{(2)}_-)$ in low $G_1$ region (up to complex conjugation).}
    \label{tab:lowG_gauge}
\end{table}

\paragraph{Field-like cohomology}

The base field $J\big|_{G_1,G_2} \in H^{2-1}(\sigma_2)$, which makes this case analogous to that of equation-like cohomology in the high $G_1$ region. The equivalence relation from Lemma \ref{lemma:calc} is
\begin{align}
\begin{split}\label{lowG_fact_eqn}
    J\big|_{G_1,2-2\{s_2\}} \sim \\\sim
    \delta_{\cN_2,\bar{\cN}_2+2-2\{s_2\}}\big\{
        &H(p_2,p_2)
            [\#_{11} \epsilon_+(y_2,\yd_2) +
            \#_{12} (p_1y_2)^2 \xi_+(y_1,\yd_1) + \#_{13} (p_1y_2)(y_1y_2) \xi_+(p_1,\yd_1)] +\\+
        &H(p_2,y_2)
            [\#_{21} \epsilon_+(p_2,\yd_2) +
            \#_{22} (p_1y_2)(p_1p_2) \xi_+(y_1,\yd_1) + \#_{23} \xi_+(p_1,\yd_1) +
            \delta_{\{s_2\},\frac12} \mathbf{j}^{\gamma\tr}_+] +\\+
        &H(y_2,y_2)
            [\#_{31} (p_1p_2)^2 \xi_+(y_1,\yd_1) + \#_{32} (p_1p_2)(y_1p_2) \xi_+(p_1,\yd_1) + 
            \delta_{\{s_2\},0} \mathbf{j}^{\tr} + \delta_{\{s_2\},\frac12} \mathbf{j}^{\tr}_+] +\\+
        &\Hd(\pd_2,\pd_2)
            [\#_{41} (\yd_1\yd_2)^2 \xi_+(p_1,\pd_1) + \#_{42} (\pd_1\yd_2)(\yd_1\yd_2) \xi_+(p_1,\yd_1) +
             \delta_{\{s_2\},0} \mathbf{j} + \delta_{\{s_2\},\frac12} \mathbf{j}_+] +\\+
        &\Hd(\pd_2,\yd_2)
            [\#_{51} \epsilon_+(p_2,\yd_2) +
            \#_{52} (\yd_1\yd_2)(\yd_1\pd_2) \xi_+(p_1,\pd_1) + \#_{53} \xi_+(p_1,\yd_1)] +\\+
        &\Hd(\yd_2,\yd_2)
            [\#_{61} \epsilon_+(p_2,\pd_2) +
            \#_{62} (\yd_1\pd_2)^2 \xi_+(p_1,\pd_1) + \#_{63} (\pd_1\pd_2)(\yd_1\pd_2) \xi_+(p_1,\yd_1)]
    \big\} +\\+
    \delta_{\cN_2+2-2\{s_2\},\bar{\cN}_2}\big\{
        &H(p_2,p_2)
            [\#^{11} (p_1y_2)^2 \xi_-(y_1,\yd_1) + \#^{12} (p_1y_2)(y_1y_2) \xi_-(p_1,\yd_1) + 
            \delta_{\{s_2\},0} \mathbf{j} + \delta_{\{s_2\},\frac12} \mathbf{j}_-] +\\+
        &H(p_2,y_2)
            [\#^{21} \epsilon_-(y_2,\pd_2) +
            \#^{22} (p_1y_2)(p_1p_2) \xi_-(y_1,\yd_1) + \#^{23} \xi_-(p_1,\yd_1)] +\\+
        &H(y_2,y_2)
            [\#^{31} \epsilon_-(p_2,\pd_2) +
            \#^{32} (p_1p_2)^2 \xi_-(y_1,\yd_1) + \#^{33} (p_1p_2)(y_1p_2) \xi_-(p_1,\yd_1)] +\\+
        &\Hd(\pd_2,\pd_2)
            [\#^{41} \epsilon_-(y_2,\yd_2) +
            \#^{42} (\yd_1\yd_2)^2 \xi_-(p_1,\pd_1) + \#^{43} (\pd_1\yd_2)(\yd_1\yd_2) \xi_-(p_1,\yd_1)] +\\+
        &\Hd(\pd_2,\yd_2)
            [\#^{51} \epsilon_-(y_2,\pd_2) +
            \#^{52} (\yd_1\yd_2)(\yd_1\pd_2) \xi_-(p_1,\pd_1) + \#^{53} \xi_-(p_1,\yd_1) + 
            \delta_{\{s_2\},\frac12} \mathbf{j}^{\gamma\tr}_-] +\\+
        &\Hd(\yd_2,\yd_2)
            [\#^{61} (\yd_1\pd_2)^2 \xi_-(p_1,\pd_1) + \#^{62} (\pd_1\pd_2)(\yd_1\pd_2) \xi_-(p_1,\yd_1) +
            \delta_{\{s_2\},0} \mathbf{j}^{\tr} + \delta_{\{s_2\},\frac12} \mathbf{j}^{\tr}_-]
    \big\}\,.
\end{split}
\end{align}
Here
\begin{gather*}
    \#_{11} = \#_{21} = \frac{1}{\floor{s_2}+2}\,,\qquad
        \#_{12} = -\#_{13} = \frac{-1}{(\cN_1+2)(\floor{s_2}+1)(\floor{s_2}+2)}\,,\\
    \#_{22} = - \#_{31} = \#_{32} = \frac{1}{(\cN_1+2)(\floor{s_2}+1)\floor{s_2}}\,,\qquad
        \#_{22} = \frac{2}{(\cN_1+2)(\floor{s_2}+2)\floor{s_2}}\,,\\
    \#_{23} = -\frac{\cN_1\floor{s_2}+2(y_1y_2)(p_1p_2)}{(\cN_1+2)(\floor{s_2}+2)\floor{s_2}}\,,\qquad
        \#_{41} = -\#_{42} = \frac{1}{\bar{\cN}_1\ceil{s_2}(\ceil{s_2}-1)}\,,\\
    \#_{51} = -\#_{61} = \frac{1}{\ceil{s_2}-2}\,,\qquad
        \#_{52} = \frac{-2}{\bar{\cN}_1\ceil{s_2}(\ceil{s_2}-2)}\,,\\
    \#_{53} = \frac{(\bar{\cN}_1+2)\ceil{s_2}+2(\yd_1\yd_2)(\pd_1\pd_2)}{\bar{\cN}_1\ceil{s_2}(\ceil{s_2}-2)}\,,\qquad
        \#_{62} = -\#_{63} = \frac{1}{\bar{\cN}_1(\ceil{s_2}-1)(\ceil{s_2}-2)}\,;\\
    \#^{11} = -\#^{12} = \frac{-1}{(\cN_1+2)(\ceil{s_2}-1)\ceil{s_2}}\,,\qquad
        \#^{21} = -\#^{31} = \frac{1}{\ceil{s_2}-2}\,,\\
    \#^{22} = \frac{2}{(\cN_1+2)(\ceil{s_2}-2)\ceil{s_2}}\,,\qquad
        \#^{23} = -\frac{(\cN_1+2)\ceil{s_2}+2(y_1y_2)(p_1p_2)}{(\cN_1+2)(\ceil{s_2}-2)\ceil{s_2}}\,,\\
    \#^{32} = -\#^{33} = \frac{-1}{(\cN_1+2)(\ceil{s_2}-1)(\ceil{s_2}-2)}\,,\qquad
        \#^{41} = - \#^{51} = \frac{1}{\floor{s_2}+2}\,,\\
    \#^{42} = -\#^{43} = \frac{1}{\bar{\cN}_1(\floor{s_2}+1)(\floor{s_2}+2)}\,,\qquad
        \#^{52} = \frac{-2}{\bar{\cN}_1\floor{s_2}(\floor{s_2}+2)}\,,\\
    \#^{53} = \frac{\bar{\cN}_1\floor{s_2}+2(\yd_1\yd_2)(\pd_1\pd_2)}{\bar{\cN}_1\floor{s_2}(\floor{s_2}+2)}\,,\qquad
        \#^{61} = -\#^{62} = \frac{1}{\bar{\cN}_1\floor{s_2}(\floor{s_2}+1)}\,.
\end{gather*}

As follows from Lemma \ref{lemma:calc}, for integer $s_2$, $\xi_\pm$ must obey
\begin{align}
\begin{split}
    \frac{\bar{\cN}_1}{(\cN_1+2)} [ (p_1p_2)^2 \xi_+(y_1,\yd_1) -  (p_1p_2)(y_1p_2) \xi_+(p_1,\yd_1) ] =
        (\pd_1\pd_2)(\yd_1\pd_2) \xi_-(p_1,\yd_1) - (\yd_1\pd_2)^2 \xi_-(p_1,\pd_1)\,,\\
    \frac{\bar{\cN}_1}{(\cN_1+2)} [(p_1y_2)^2 \xi_-(y_1,\yd_1) - (p_1y_2)(y_1y_2) \xi_-(p_1,\yd_1)] =
        (\pd_1\yd_2)(\yd_1\yd_2) \xi_+(p_1,\yd_1) - (\yd_1\yd_2)^2 \xi_+(p_1,\pd_1)\,.
\end{split}
\end{align}

The analogue of conditions \eqref{highG_Psi_step_eqn} is
\begin{align}\label{lowG_step_eqn}
\begin{split}
    (\yd_2\pd_2) (p_2p_1)(\yd_1\pd_2)\mathbf{j} -
    (p_2y_2) (\yd_2\yd_1)(p_1y_2)\mathbf{j}^{\tr} = 0\,,\\
    2(\yd_2\pd_2) (p_2p_1)(\yd_1\pd_2)\mathbf{j}_\pm -
    2(p_2y_2) (\yd_2\yd_1)(p_1y_2)\mathbf{j}_\pm^{\tr}-
    (p_2y_2) (p_1p_2)(\yd_2\yd_1)\mathbf{j}_\pm^{\gamma\tr} = 0\,.
\end{split}
\end{align}

The final solution in this case is presented in Table \ref{tab:lowG_field}.
\begin{table}[H]
    \renewcommand{\arraystretch}{1.5}
    \centering
    \begin{tabular}{|c|c|}
        \hline
        \multirow{3}{7em}{Integer $s_2$}
            &$[\Hd(\pd_2,\pd_2) + H(p_2,p_2)]\ \mathbf{j}_{a,\ad;\;s_2,s_2;\;0,0}\,,\qquad
                a+\ad=2s_1-2$\\
            &$[\Hd(\pd_2,\pd_2) + H(p_2,p_2)]\ \mathbf{j}_{a,\ad;\;0,0;\;s_2,s_2}\,,\qquad
                a+\ad=2s_1-2s_2-2$\\
            &$[
                \mathfrak{H} + \bar{\mathfrak{H}}
            ]\ \mathbf{j}_{s+s_1-s_2-1,0;\;1,s-s_1+s_2;\;s_2-1,s_1-s}$\\\hline
        \multirow{3}{7em}{Half-integer $s_2$}
            &$\Hd(\pd_2,\pd_2)\ (\mathbf{j}_+)_{a,\ad;\;s_2-1/2,s_2+1/2;\;0,0}\,,\qquad
                a+\ad=2s_1-2$\\
            &$H(p_2,p_2)\ (\mathbf{j}_-)_{a,\ad;\;0,0;\;s_2+1/2,s_2-1/2}\,,\qquad
                a+\ad=2s_1-2s_2-2$\\
            &$\bar{\mathfrak{H}}
            \ (\mathbf{j}_-)_{s+s_1-s_2-1,0;\;1,s-s_1+s_2;\;s_2-1/2,s_1-s-1/2}$\\\hline
    \end{tabular}
    \caption{The base fields for field-like $H(\Sigma^{(2)}_-)$ in low $G_1$ region (up to complex conjugation).}
    \label{tab:lowG_field}
\end{table}
Here $s$ is a free (half-)integer parameter, which is restricted by $s+s_1 \geqslant s_2+1$, $s+s_2 \geqslant s_1+1$ and $s_1-s-\{s_2\} \geqslant 1$; $\mathfrak{H}$ is defined as follows ($\bar{\mathfrak{H}}$ is its complex conjugation):
\begin{equation}
\begin{split}
    \mathfrak{H} :=
        \Hd(\pd_2,\pd_2) +
        \# H(y_2,y_2) (y_1p_2)(p_1p_2)(\pd_1\pd_2)(\yd_1\pd_2)\,,
\end{split}
\end{equation}
\begin{align}\label{sharp}
    \# = \frac{\floor{s_2}-1}{\floor{s_1-s}(s+s_1-s_2)(\floor{s_2}+1)(\ceil{s_2}+1)}\,.
\end{align}

\paragraph{Equation-like cohomology}

The base fields of the equation-like $H(\Sigma^{(2)}_-)$ belong to $H^{3-2}(\sigma_2) \cong H^{3-2}(\Sigma^{(1)}_-)$ presented in Table \ref{tab:cohom1_3-2}. Then the analogue of \eqref{highG_j_init} is
\begin{subequations}\label{H3-2}
\begin{align}
    &\Psi\big|_{G_1,2} = [\cH(y_2,\pd_2) + \cH(p_2,\yd_2)] \delta_{\cN_2,\bar{\cN}_2}\psi\,,
        &&\text{if $s_2$ is integer}\,,\\
    &\Psi\big|_{G_1,1} = \cH(y_2,\pd_2) \delta_{\cN_2+1,\bar{\cN}_2}\psi_+ +
    \cH(p_2,\yd_2) \delta_{\cN_2,\bar{\cN}_2+1}\psi_-\,,
        &&\text{if $s_2$ is half-integer}\,.
\end{align}
\end{subequations}
The analogue of \eqref{highG_j_fact_eqn} is
\begin{align}
\begin{split}
    \Psi\big|_{G_1,1} \sim
    \delta_{\cN_2,\bar{\cN}_2+1}\big\{
        &\cH(p_2,\pd_2)
            [\#\epsilon_+(\yd_2,\yd_2) + \dots] +\\+
        &\cH(p_2,\yd_2)
            [\#\epsilon_+(y_2,p_2) + \#\epsilon_+(\yd_2,\pd_2) + \dots] +\\+
        &\cH(y_2,\pd_2)
            [\#_1 (y_1p_2)(\yd_1\yd_2) \xi_+(p_1,p_1) 
            -\#_1 (p_1p_2)(\yd_1\yd_2) \xi_+(p_1,y_1) -\\&\qquad
            -\#_2 (p_1p_2)(\yd_1\yd_2) \xi_+(\pd_1,\yd_1) + 
            \#_2 (p_1p_2)(\pd_1\yd_2) \xi_+(\yd_1,\yd_1) +
            \psi_+] +\\+
        &\cH(y_2,\yd_2)
            [\#\epsilon_+(p_2,p_2) + \dots]
    \big\} +\\+
    \delta_{\cN_2+1,\bar{\cN}_2}\big\{
        &\cH(p_2,\pd_2)
            [\#\epsilon_-(y_2,y_2) + \dots] +\\+
        &\cH(p_2,\yd_2)
            [\#_1 (y_1y_2)(\yd_1\pd_2) \xi_-(p_1,p_1) 
            -\#_1 (p_1y_2)(\yd_1\pd_2) \xi_-(p_1,y_1) -\\&\qquad
            -\#_2 (p_1y_2)(\yd_1\pd_2) \xi_-(\pd_1,\yd_1) + 
            \#_2 (p_1y_2)(\pd_1\pd_2) \xi_-(\yd_1,\yd_1) +
            \psi_-] +\\+
        &\cH(y_2,\pd_2)
            [\#\epsilon_-(y_2,p_2) + \#\epsilon_-(\yd_2,\pd_2) + \dots] +\\+
        &\cH(y_2,\yd_2)
            [\#\epsilon_-(\pd_2,\pd_2) + \dots]
    \big\}\,.
\end{split}
\end{align}
Here ellipsis denotes terms that can be eliminated by $\epsilon$, $\#$ -- some non-zero coefficients; $\#_1 = \frac{1}{3(\cN_1+1)(\ceil{s_2}+1)(\floor{s_2}-1)}$, $\#_2 = \frac{1}{3(\bar{\cN}_1+1)(\ceil{s_2}+1)(\floor{s_2}-1)}$; if $s_2$ is integer $\xi_\pm$ obey
\begin{align}
\begin{split}
    \frac{\bar{\cN}_1+1}{\cN_1+1} \big[
        (y_1p_2)(\yd_1\yd_2) \xi_+(p_1,p_1) -
        &(p_1p_2)(\yd_1\yd_2) \xi_+(p_1,y_1) -\\-
        &(y_1y_2)(\yd_1\pd_2) \xi_-(p_1,p_1) +
        (p_1y_2)(\yd_1\pd_2) \xi_-(p_1,y_1)
        \big] =\\=
    (p_1p_2)(\yd_1\yd_2) \xi_+(\pd_1,\yd_1) -
    &(p_1p_2)(\pd_1\yd_2) \xi_+(\yd_1,\yd_1) -\\-
    &(p_1y_2)(\yd_1\pd_2) \xi_-(\pd_1,\yd_1) +
    (p_1y_2)(\pd_1\pd_2) \xi_-(\yd_1,\yd_1)\,.
\end{split}
\end{align}

The conditions from the 3rd point of Lemma \ref{lemma:calc} are trivial because $H^{4-3}(\sigma_2)=0$; the final answer for this case is in Table \ref{tab:lowG_eqn}
\begin{table}[H]
    \renewcommand{\arraystretch}{1.5}
    \centering
    \begin{tabular}{|c|c|}
        \hline
        Integer $s_2$
            &$[\cH(y_2,\pd_2) + \cH(p_2,\yd_2)] \psi_{s+s_1-s_2,0;\;0,s-s_1+s_2;\;s_2-1,s_1-s-1}$\\\hline
        Half-integer $s_2$
            &$\cH(p_2,\yd_2) (\psi_-)_{s+s_1-s_2,0;\;0,s-s_1+s_2;\;s_2-1/2,s_1-s-3/2}$\\\hline
    \end{tabular}
    \caption{The base fields for equation-like $H(\Sigma^{(2)}_-)$ in low $G_1$ region (up to complex conjugation).}
    \label{tab:lowG_eqn}
\end{table}

\section{On rank-two dynamical equations}\label{app:on_rank-two_dynamical_equations}

The purpose of this appendix is to obtain formula \eqref{tildej_eqn} and, in particular, to show that it has no contribution from the other regular cocycles. Note that here we will keep track only of the regular cocycles (listed in Table \ref{tab:rk2_fields}), omitting the irregular ones.

Let us consider a rank-two field $J = J(Y_1; Y_2) = J\big|_G + J\big|_{G+2} + \dots$, where $J\big|_G \in H(\Sigma^{(2)}_-)$ is built from $\tilde{j}^\bullet_{s+n,s-n}$ according to Proposition \ref{prop:1}, $J\big|_{G+2}$, etc. are the descendants of $J\big|_G$. Our goal is to show that if the equation
\begin{equation}\label{rk2_eqn_k}
    (D_L J + \Sigma^{(2)}_- J + \Sigma^{(2)}_+ J)\big|_{G+2k} = 0
\end{equation}
is true for $k=0$, it is satisfied identically for $k \geqslant 1$. Thus different $\tilde{j}^\bullet_{s+n,s-n}$ and their descendants cannot interact; this substantiates their absence in \eqref{tildej_eqn}.

There are 4 cases that have to be discussed separately, depending on $\tilde{j}^\bullet_{s+n,s-n}$ parameters: $\bullet = \omega\omega$ or $\bullet = C\omega$ and $s_2$ is integer or half-integer ($s_1$ and $s_2$ values are fixed according to Section \ref{sec:rank-two_cohomology}). The general line of the discussion in all the cases is as follows.
\begin{enumerate}
    \item Consider the projection of equation \eqref{rk2_eqn_k} onto $H(\Sigma^{(2)}_-)$:
    \begin{equation}\label{rk2_dyn_eqn_k}
        \cP \big\{D_L J\big|_{G+2k} +
            \big(
                \Sigma^{(2)}_+ J\big|_{G+2k-2} +
                \Sigma^{(2)}_+ J\big|_{G+2k}
            \big)\big|_{G+2k} \big\} = 0\,,
    \end{equation}
    where $\cP$ is the projector. Show that at $k\geqslant 1$ equation \eqref{rk2_dyn_eqn_k} is equivalent to
    \begin{equation}\label{j_alpha,beta_eqn}
        D(p,\pd) j^{(k)}_\alpha + D(y,\pd) j^{(k)}_\beta + D(p,\yd) j^{(k)}_\gamma + \#_k\, j^{(k-1)}_\alpha + j^{(k)}_\delta = 0\,,
    \end{equation}
    where $j^{(k)}_\alpha = j^{(k)}_\alpha(Y)$, etc. are certain fields, $\#_k$ is a numerical coefficient. At $k=0$ \eqref{rk2_dyn_eqn_k} gives $D(p,\pd)\tilde{j}^\bullet_{s+n,s-n} = 0$.

    \item By counting degrees of spinor variables (analogously to Lemma \ref{lemma:eq_check}) show that
    \begin{align}
        &j^{(k)}_\alpha = c^{(k)}_\alpha\, D^k(y,\yd) \tilde{j}^\bullet_{s+n,s-n}\,,
        &&j^{(k)}_\beta = c^{(k)}_\beta\, D(p,\yd)D^{k-1}(y,\yd) \tilde{j}^\bullet_{s+n,s-n}\,,\nonumber\\
        &j^{(k)}_\gamma = c^{(k)}_\gamma\, D(y,\pd)D^{k-1}(y,\yd) \tilde{j}^\bullet_{s+n,s-n}\,,
        &&j^{(k)}_\delta = c^{(k)}_\delta\, D^{k-1}(y,\yd) \tilde{j}^\bullet_{s+n,s-n}\,,\label{c_alpha,beta_def}
    \end{align}
    where $c^{(k)}_\alpha$, etc. are numerical coefficients. Define $\vec{c}_{k} = (c^{(k)}_\alpha,\ c^{(k)}_\beta,\ c^{(k)}_\gamma,\ c^{(k)}_\delta)^T$.
    
    \item Account for the formula
    \begin{multline}
        D(y,\pd)D(p,\yd)D^{k-1}(y,\yd) \tilde{j}^\bullet_{s+n,s-n} =
            \mu_k(s,n) D(p,\pd) D^k(y,\yd) \tilde{j}^\bullet_{s+n,s-n} +\\+
            \nu_k(s,n) D^k(y,\yd) D(p,\pd) \tilde{j}^\bullet_{s+n,s-n} +
            \chi_k(s,n) D^{k-1}(y,\yd) \tilde{j}^\bullet_{s+n,s-n}\,,\label{DypDpy_comm}
    \end{multline}
    obtained with the help of \eqref{D_comm_res}; here
    \begin{gather*}
        \mu_k(s,n) = -\frac{(s-n+1) (s+n)}{k (2s+k+1)}\,,\qquad
        \nu_k(s,n) = \frac{(s-n+k+1) (s+n+k)}{k (2s+k+1)}\,,\\
        \chi_k(s,n) = -(s-n+1) (s+n) (s-n+k+1) (s+n+k)\,.
    \end{gather*}
    By virtue of \eqref{c_alpha,beta_def} and \eqref{DypDpy_comm}, \eqref{j_alpha,beta_eqn} amounts to
    \begin{align}
        &\vec{a}_k \cdot \vec{c}_k=0\,,
            &&\vec{a}_k = (1,\ \mu_k(s,n),\ \mu_k(s,-n),\ 0)\,;\label{ac_eqn}\\
        &\vec{b}_k \cdot \vec{c}_k + \#_k c^{(k-1)}_\alpha=0\,,
            &&\vec{b}_k = (0,\ \chi_k(s,n),\ \chi_k(s,-n),\ 1)\,,\label{bc_eqn}
    \end{align}
    where it is accounted that $D(p,\pd)\tilde{j}^\bullet_{s+n,s-n} = 0$. Equations \eqref{ac_eqn} and \eqref{bc_eqn} are vanishing conditions for the coefficients in front of $D(p,\pd) D^k(y,\yd) \tilde{j}^\bullet_{s+n,s-n}$ and $D^{k-1}(y,\yd) \tilde{j}^\bullet_{s+n,s-n}$, correspondingly.

    \item Find matrix $M_k$ such that $\vec{c}_{k+1} = M_k \vec{c}_{k}$. To this end one has to resolve the equation
    \begin{equation}\label{step_eqn}
        \cP_{H(\sigma_2)} \big\{D_L J\big|_{G+2k} +
            \big(
                \Sigma^{(2)}_+ J\big|_{G+2k-2} +
                \Sigma^{(2)}_+ J\big|_{G+2k}
            \big)\big|_{G+2k} + \sigma_1 J\big|_{G+2k+2} \big\} = 0\,,
    \end{equation}
    where $\cP_{H(\sigma_2)}$ is the projector onto $H(\sigma_2)$, to find $c^{(k+1)}_\beta,\ c^{(k+1)}_\gamma,\ c^{(k+1)}_\delta$, and equation \eqref{ac_eqn} to find $c^{(k+1)}_\alpha$.

    \item Using $M_k$ check that \eqref{bc_eqn} is true at $(k+1)$-th level if \eqref{ac_eqn} and \eqref{bc_eqn} were true at $k$-th.
\end{enumerate}

The non-trivial steps are the 1st and the 4th, they have to be done for each case separately. Unfortunately we cannot present here a detailed discussion of all the cases, since the intermediate formulae are quite long (see, e.g., \eqref{ww_bos_J_alpha,beta_def}, \eqref{ww_bos_Mk}), so let us consider only the case of $\tilde{j}^{\omega\omega}_{s+n,s-n}$ with integer $s_2$. According to Table \ref{tab:rk2_fields}, the corresponding equation-like base field is $[\cH(y_2,\pd_2) + \cH(p_2,\yd_2)] \psi_{s+k-1+s_1-s_2,0;\;0,s+k-1-s_1+s_2;\;s_2-1,s_1-s-k}$, therefore $\cP$ in \eqref{rk2_dyn_eqn_k} can be defined as
\begin{equation}\label{ww_bos_dyn_eqn}
    \cP\{\dots\} := (\partial_\cH(y_2,\pd_2) \dots)_{s+k-1+s_1-s_2,0;\;0,s+k-1-s_1+s_2;\;s_2-1,s_1-s-k}\,,
\end{equation}
where $\partial_\cH$ is defined in Appendix \ref{app:technical_formulae}. The corresponding $H(\sigma_2)$ is $H^{3-2}(\sigma_2)$, hence
\begin{equation}
    \cP_{H(\sigma_2)}\{\dots\} := \delta_{\cN_2,\bar{\cN}_2}\partial_\cH(y_2,\pd_2) \dots\,.
\end{equation}
One sees that $\cP\{\dots\} = (\cP_{H(\sigma_2)}\{\dots\})_{s+k-1+s_1-s_2,0;\;\dots}$, so \eqref{rk2_dyn_eqn_k} can be obtained from \eqref{step_eqn}. Let us consider \eqref{step_eqn} in this case: assuming that $\delta_{\cN_2+2,\bar{\cN}_2}J\big|_{G+2k} = H(p_2,p_2) \mathbf{j}_k + \Hd(\yd_2,\yd_2) \mathbf{j}^{\tr}_k$ one obtains
\begin{align}\label{ww_bos_step_eqn}
\begin{split}
    \eqref{step_eqn} \Leftrightarrow
        - &(y_2p_2) D(p_2,\pd_2) \mathbf{j}_k
        + (\pd_2\yd_2) \mathbf{j}^{\tr}_k -\\
        - &i (y_2p_2)(y_1p_2)(\pd_2\pd_1) \mathbf{j}_{k-1}
        + i (y_2y_1)(\pd_2\yd_2)(\pd_1\yd_2) \mathbf{j}^{\tr}_{k-1} -\\
        - &i (y_2p_2)(p_1p_2)(\pd_2\yd_1) \mathbf{j}_{k+1}
        + i (y_2p_1)(\pd_2\yd_2)(\yd_1\yd_2) \mathbf{j}^{\tr}_{k+1} 
        = 0\,.
\end{split}
\end{align}
From \eqref{ww_bos_dyn_eqn} and \eqref{ww_bos_step_eqn} one sees that in formula $\delta_{\cN_2+2,\bar{\cN}_2}J\big|_{G+2k} = J_\alpha + J_\beta + J_\gamma + J_\delta +  \dots$, where
\begin{align}\label{ww_bos_J_alpha,beta_def}
\begin{split}
    J_\alpha = [
        \kappa_{\alpha_1} &H(p_2,p_2)
        \cK_{s+s_1-s_2+k-1,0;\;1,s-s_1+s_2+k;\;s_2-1,s_1-s-k} +\\+
        \kappa_{\alpha_2} &\Hd(\yd_2,\yd_2)
        \cK_{s+s_1-s_2+k,1;\;0,s-s_1+s_2+k-1;\;s_2-2,s_1-s-k-1}
        ]\ j^{(k)}_\alpha\,,\\
    J_\beta = \kappa_\beta
        &H(p_2,p_2) \cK_{s+s_1-s_2+k-2,0;\;0,s-s_1+s_2+k;\;s_2,s_1-s-k}\ j^{(k)}_\beta\,,\\
    J_\gamma = \kappa_\gamma 
        &\Hd(\yd_2,\yd_2) \cK_{s+s_1-s_2+k,0;\;0,s-s_1+s_2+k-2;\;s_2-2,s_1-s-k}\ j^{(k)}_\gamma\,,\\
    J_\delta = [
        \kappa_{\delta_1} &H(p_2,p_2)
        \cK_{s+s_1-s_2+k-1,0;\;1,s-s_1+s_2+k;\;s_2-1,s_1-s-k} +\\+
        \kappa_{\delta_2} &\Hd(\yd_2,\yd_2)
        \cK_{s+s_1-s_2+k,1;\;0,s-s_1+s_2+k-1;\;s_2-2,s_1-s-k-1}
        ]\ j^{(k+1)}_\delta\,,
\end{split}
\end{align}
\begin{align}
\begin{split}
    \kappa_{\alpha_1} = -\frac{i (s_2^2-1) (k+s+s_1)
   (k+s-s_1+s_2)}{3 (k+s+s_1-s_2+1)
   (k+s-s_1+s_2+1)}\,,\\
    \kappa_{\alpha_2} = \frac{i (s_2+1)
   (k+s+s_1-s_2)}{3 (k+s+s_1-s_2+1)
   (k+s-s_1+s_2+1)}\,,\\
    \kappa_\beta = -\frac{3 (k+s+s_1-s_2-1)
   (k+s-s_1+s_2+1)}{s_2 (s_2^2-1) (k+s-s_1+s_2)}\,,\\
    \kappa_\gamma = -\frac{3 (k+s+s_1-s_2+1)
   (k+s-s_1+s_2-1)}{(s_2+1) (k+s+s_1-s_2)}\,,\\
    \kappa_{\delta_1} = \frac{i (s_2-1)^2 (s_2+1) (k+s)}{3
   (k+s+s_1-s_2)}\,,\\
    \kappa_{\delta_2} = \frac{i (s_2+1)^2 (k+s)}{3
   (k+s+s_1) (k+s-s_1+s_2)}\,,
\end{split}
\end{align}
$J_\alpha$, etc. are only parts of $\delta_{\cN_2+2,\bar{\cN}_2}J\big|_{G+2k}$ that contribute to \eqref{rk2_dyn_eqn_k}. (Recall that $\cK_{\dots}$ is defined in \eqref{K_abc}.) Substituting \eqref{ww_bos_J_alpha,beta_def} into \eqref{ww_bos_step_eqn} one indeed obtains equation \eqref{j_alpha,beta_eqn} with $\#_k = \frac{(k+s+s_1-s_2) (k+s-s_1+s_2) (k^2+k (2 s-3)+(s-3)s+s_1 (s_2-s_1)+2)}{(k+s+s_1-1)(k+s+s_1-s_2-1) (k+s-s_1+s_2-1)}$. Next one finds matrix $M_k$ from \eqref{ww_bos_step_eqn} and \eqref{ac_eqn}. The result is quite long so we present simpler expression for $\tilde{M}_k$ defined via
\begin{equation*}
    (c^{(k)}_\alpha,\ c^{(k)}_\beta,\ c^{(k)}_\gamma,\ c^{(k+1)}_\delta)^T = \tilde{M}_k (c^{(k-1)}_\alpha,\ c^{(k-1)}_\beta,\ c^{(k-1)}_\gamma,\ c^{(k)}_\delta)^T\,.
\end{equation*}
\begin{align}\label{ww_bos_Mk}
\begin{array}{cc}
    (\tilde{M}_k)_{11} = \frac{s^2 (k^2+k+2 s_1 (s_2-s_1)-3)+s (k^2-k (2
        (s_1-2 s_2) (s_1-s_2)+1)+4 s_2
        (s_2-s_1)-2)+2 k s^3
        }{k
        (k+2 s+1) (k+s+s_1-1) (k+s+s_1-s_2-1)
        (k+s+s_1-s_2+1) (k+s-s_1+s_2-1) (k+s-s_1+s_2+1)} +\\+\frac{
        (s_1-s_2) (-s_1 ((k-1)
        k+s_2^2+1)+s_2 (2 (k-1) k+s_2^2-1)+s_1^3-s_2 s_1^2)+s^4}{k
        (k+2 s+1) (k+s+s_1-1) (k+s+s_1-s_2-1)
        (k+s+s_1-s_2+1) (k+s-s_1+s_2-1) (k+s-s_1+s_2+1)}\,,\\
    (\tilde{M}_k)_{12} = \frac{(s+s_1-s_2) (s-s_1+s_2+1) (k+s-s_1-1)}{k
        (k+2 s+1) (k+s+s_1-1) (k+s+s_1-s_2-1)
        (k+s-s_1+s_2-1) (k+s-s_1+s_2+1)}\,,\\
    (\tilde{M}_k)_{13} = \frac{(s+s_1-s_2+1) (s-s_1+s_2)}{k (k+2 s+1)
        (k+s+s_1-s_2-1) (k+s+s_1-s_2+1) (k+s-s_1+s_2-1)}\,,\\
    (\tilde{M}_k)_{14} = \frac{k (k+2s+1) (s^2+s-(s_1-s_2)^2)+(s+s_1-s_2)
        (s+s_1-s_2+1) (s-s_1+s_2) (s-s_1+s_2+1)}{k
        (k+s-1) (k+2 s+1) (k+s+s_1-s_2) (k+s+s_1-s_2+1)
        (k+s-s_1+s_2) (k+s-s_1+s_2+1)}\,,\\
    (\tilde{M}_k)_{21} = \frac{(s_2+1) (k+s-s_1+s_2-2)}{2 (k+s+s_1-1)
        (k+s+s_1-s_2-1) (k+s-s_1+s_2-1) (k+s-s_1+s_2+1)}\,,\\
    (\tilde{M}_k)_{22} = \frac{k+s-s_1-1}{(k+s+s_1-1) (k+s+s_1-s_2-1)
        (k+s-s_1+s_2-1) (k+s-s_1+s_2+1)}\,,\\
    (\tilde{M}_k)_{23} = 0\,,\\
    (\tilde{M}_k)_{24} = \frac{1}{2 (k+s-1)
        (k+s+s_1-s_2) (k+s-s_1+s_2+1)}\,,\\
    (\tilde{M}_k)_{31} = -\frac{(s_2-1) (k+s+s_1-s_2-2)}{2 (k+s+s_1-1)
        (k+s+s_1-s_2-1) (k+s+s_1-s_2+1) (k+s-s_1+s_2-1)}\,,\\
    (\tilde{M}_k)_{32} = 0\,,\\
    (\tilde{M}_k)_{33} = \frac{1}{(k+s+s_1-s_2-1) (k+s+s_1-s_2+1)
        (k+s-s_1+s_2-1)}\,,\\
    (\tilde{M}_k)_{34} = \frac{1}{2 (k+s-1) (k+s+s_1-s_2+1)
        (k+s-s_1+s_2)}\,,\\
    (\tilde{M}_k)_{41} = \frac{(s_2-1) (s_2+1) (k+s-1) (k+s)}{(k+s+s_1-1)
        (k+s+s_1) (k+s+s_1-s_2-1) (k+s+s_1-s_2)
        (k+s-s_1+s_2-1) (k+s-s_1+s_2)}\,,\\
    (\tilde{M}_k)_{42} = 0\,,\\
    (\tilde{M}_k)_{43} = 0\,,\\
    (\tilde{M}_k)_{44} = \frac{(k+s)
        (k^2+k (2 s-1)+(s-1) s+s_1 (s_2-s_1))}{(k+s-1)
        (k+s+s_1) (k+s+s_1-s_2)^2 (k+s-s_1+s_2)^2}\,.
\end{array}
\end{align}
And one can check that the property from the 5th step is indeed true.

Let us note that $j^{(k-1)}_\alpha$ and $j^{(k)}_\delta$ appear in \eqref{j_alpha,beta_eqn} via $\Sigma^{(2)}_+ J\big|_{G+2k-2}$ and $\Sigma^{(2)}_+ J\big|_{G+2k}$. Therefore, in the flat limit these terms vanish, as well as $\chi_k(s,n)$, leading to that \eqref{bc_eqn} becomes trivial. Thus the discussion above is redundant in flat theory, but it is relevant in $AdS_4$ case.

\end{appendices}
\bibliographystyle{JHEP}
\bibliography{refs}
\end{document}